\documentclass[usenatbib]{mn2e}

\usepackage{graphicx}
\usepackage{amssymb}
\usepackage{amsmath}
\usepackage{color}
\usepackage{abbreviations}

\newcommand{\mtext}[1]{\quad\mbox{#1}\quad}

\newcommand{\be}{\begin{equation}}
\newcommand{\ee}{\end{equation}}

\newcommand{\pder}[2]{\frac{\partial #1}{\partial #2}}

\newcommand{\fracp}[2]{\left(\frac{#1}{#2}\right)}

\newcommand{\beq}{\begin{equation}}
\newcommand{\eeq}{\end{equation}} 
\newcommand{\ind}[1]{{\rm #1}}
\newcommand{\epsinf}{\epsilon_{\infty}}
\newcommand{\del}{\partial}
\newcommand{\fracb}[2]{\left(\frac{#1}{#2}\right)}

\graphicspath{ {./}{figures/} }

\renewcommand{\bf}[1][]{#1}

\begin{document}

\title[3D MHD simulations of the Crab Nebula]
{Three-Dimensional Magnetohydrodynamic Simulations of the Crab Nebula}

\author[Porth et al.]{Oliver Porth$^{1,2}$\thanks{E-mail: o.porth@leeds.ac.uk (OP)}, 
Serguei S. Komissarov$^{1}$\thanks{E-mail: serguei@maths.leeds.ac.uk (SSK)}, 
Rony Keppens$^{2}$\\
$^{1}$Department of Applied Mathematics, The University of Leeds, Leeds, LS2 9GT \\
$^{2}$Centre for mathematical Plasma Astrophysics, Department of Mathematics, KU Leuven,
 Celestijnenlaan 200B, 3001 Leuven, Belgium}

\date{Received/Accepted}
\maketitle
\begin{abstract} 
    In this paper we give a detailed account of the first three-dimensional (3D) 
relativistic magnetohydrodynamic (MHD) simulations of Pulsar Wind Nebulae (PWN), with 
parameters most suitable for the Crab Nebula. In order to clarify the new features 
specific to 3D models, reference 2D simulations have been carried out as well. 
Compared to the previous 2D simulations, we considered pulsar winds with much stronger magnetisation, up to $\sigma\simeq $ few,  and accounted more accurately for the anticipated 
magnetic dissipation in the striped zone of these winds. While the 3D models preserve 
the separation of the post termination shock flow into the equatorial and 
polar components, their relative strength and significance differ. While the 
highly magnetised 2D models produce highly coherent and well collimated polar jets 
capable of efficient ``drilling''  through the supernova shell, in the corresponding 
3D models the jets are disrupted by the kink mode current driven instability 
and ``dissolve'' into the main body of PWN after propagation of several shock radii. With the exception of the region 
near the termination shock, the 3D models do not preserve  the axisymmetry of their 
pulsar winds and do not exhibit the strong $z$-pinch configuration characteristic of the 
2D models. Our results show that contrary to the expectations based on 1D analytical 
and semi-analytical models, our numerical solutions with highly magnetized pulsar winds 
still produce termination shocks with radii comparable to those deduced from the 
observations. The reason for this is not only the randomization of PWN magnetic field
observed in the 3D solutions, but also the magnetic dissipation. Even in our 2D 
numerical solutions the dissipation in the turbulent nebula flow is sufficiently 
strong to yield particle-dominated 
PWN.  The synthetic synchrotron images of the simulated nebula retain the toroidal 
appearance of the inner nebulae, revealed in the previous 2D simulations, as well as 
finer features such as wisps and the inner knot. The polarization and variability of the 
synthetic knot is in excellent agreement with the recent optical observations of the 
Crab's HST knot 1. Overall the synthetic polarization of the inner nebulae also agrees 
with the observations. However, the polar jet, while still being a dynamical feature in the 3D simulations, is not as bright.  This suggests that in 
addition to the particle acceleration at the termination shock, additional in-situ 
acceleration mechanisms have to be invoked in future studies.        
\end{abstract}


\section{Introduction}

The Crab Nebula is one of the most iconic cosmic objects at present time.   
It has a very long and very eventful history of astronomical and astrophysical 
studies, with the discovery of the Crab pulsar and its connection to supernova explosions, 
in particular to the ``guest star'' observed by Chinese and Arab astronomers in the 
year 1054, being its most commented on highlights. The more recent discoveries include 
the dynamic jet and torus around the Crab pulsar and the still mysterious gamma ray 
flares.   
At the time of writing this manuscript, SAO/NASA ADS search engine returns 1598 hits 
for papers with words ``crab'' and  ``nebula'' in their titles and 4358 hits 
for papers with these words in their abstracts. 

The main interest to the Crab Nebula in modern astrophysics stems from
the fact that it is one of the brightest sources of non-thermal
emission in the sky throughout the whole observational range of photon
energies. This emission comes from the ultrarelativistic plasma
filling the interior of the nebula. The relatively small overall
linear size of the nebula, which is only several light years across,
allows direct observations of the global dynamics of this plasma,
whereas its large angular size (of 7 arcmin) ensures that its spatial
structure is well resolved, allowing studies of small-scale structural
variability. There are other types of objects in the Universe which
share many similarities with the Crab Nebula, e.g. radio galaxies,
quasars, gamma ray bursters, which are of great interest to
researchers of various specializations.  They all provide important
clues to the properties of relativistic plasma. Yet given its unique
parameters, the Crab Nebula remains the most productive testbed of
modern high-energy astrophysics.

The observational data, accumulated over many years, make a compelling
case for the relativistic plasma being continuously supplied into the
nebula by the Crab pulsar; in the form of an ultrarelativistic
magnetized wind \citep{rees-gunn-74,kc84a}.  The early quantitative
models of the interaction between this wind and the much slower
massive supernova shell were developed within the framework of ideal
relativistic magnetohydrodynamics (MHD), using one-dimensional (1D)
approximations. They fitted the observational parameters of the nebula
and its pulsar quite well, but required the Poynting flux of the
pulsar wind to be only a small fraction, $\sim10^{-3}$, of the total
wind power \citep{rees-gunn-74,kc84a,emm-che-87,begelman1992}.
Assuming that the electrons producing the observed non-thermal
synchrotron component of the nebula emission were accelerated at the
termination shock of the wind, these models could explain its spectrum
from the optical to gamma ray energies \citep{kennel1984}.  The even
more energetic inverse-Compton emission, which was discovered much
later, also fits within this scenario very well
\citep{atoyan1996,MH10}.

Some shortcomings of the 1D model were obvious from the start. For
example, the nebula is not spherical but shows significant elongation,
and it is filled with a complex network of line-emitting
filaments. However, these were ignored until the X-ray Chandra
observations revealed the spectacular axisymmetric ``torus'' in the
inner Crab Nebula as well as a weak jet apparently originated from the
pulsar \citep{weisskopf2000}.  This discovery prompted a revision of
the 1D model by taking into account the earlier theoretical studies of
magnetically dominated pulsar winds which concluded that they must be
highly aspherical, with the wind power scaling as $\sin^2\theta$,
where $\theta$ is the angle measured from the rotational axis of the
pulsar \citep{michel-73,bogovalov-99}. Recent 3D relativistic MHD
simulations of pulsar winds show a somewhat different
$\theta$-dependence, but generally confirm the theory
\citep{tch-12}. This implies that the wind termination shock must be
flattened at the poles, and that the freshly supplied plasma enters
the nebula predominantly as an equatorial flow, providing a nice
explanation of the Crab's torus
\citep{bogovalov-khan-02b,lyub-02}. The nature of its jet is less
obvious and remained controversial for a while. In particular,
\citet{lyub-02} proposed that the jet is formed downstream of the
termination because of the strong hoop stress of the azimuthal
magnetic field supplied by the wind, which pulls the shocked plasma
towards the wind symmetry axis and then squeezes it away along the
axis, very much like toothpaste is extracted from its container.

In order to check these ideas and study the 2D models in details, a
number of groups carried out axisymmetric relativistic MHD simulations
of pulsar wind nebulae
\citep{ssk-lyub-03,komissarov2004,del-zanna2004,bogovalov2005}. Although
they employed rather different computer codes, all these simulations
produced quite similar results.  For the pulsar wind parameters
similar to those of the 1D model of \cite{kc84a}, the numerical
solutions could reproduce the observed jet-torus structure of the Crab
Nebula.  In addition, even the rather crude initial attempts to model
the PWN synchrotron emission using such numerical solutions provided a
neat explanation for the most compact and bright feature of the
nebula, its mysterious HST knot 1, which is located within one
arcsecond from the Crab pulsar \citep{hester1995}. In these models,
the knot is a Doppler-beamed emission from the patch of the highly
oblique termination shock where the post-shock flow is still highly
relativistic and directed towards the observer
\citep{komissarov2004}. Later, the more advanced models, following the
approach of \citet{kennel1984}, confirmed that this feature was robust
\citep{del-zanna2006,komissarov2011}.  They also demonstrated that the
models could explain the observed spectrum and polarization of the
Crab Nebula \citep{bucc-05,volpi2008}.

These early numerical simulations had rather low resolution and hence
high numerical viscosity, damping short lengthscale motions in the
nebula.  Moreover, they also imposed equatorial symmetry and studied
the flow only in one hemisphere. When these limitations were eased in
the later study by \citet{camus2009}, a much more dynamic and unsteady
flow structure emerged. The main new feature was the highly distorted
termination shock, variable on the scale comparable to its light
crossing time and the highly inhomogeneous flow entering the
nebula. While the overall jet-torus structure still persisted, the
synthetic synchrotron images of the nebula also included bright fine
filaments moving away from the pulsar with the initial speed
approaching a large fraction of the speed of light.  These are highly
reminiscent of the so-called ``wisps'' of Crab Nebula, which were
discovered almost a century ago by \citet{lampland-21}, who also
reported their variability. This variability was carefully studied by
\citet{scargle-69}, who concluded that it is likely related to the
fast motion of the wisps. However, only much more recent observations
were able to deliver a firm evidence in favor of this interpretation
\citep{hester2002}.

In spite of all these successes of the MHD models, one issue, the
so-called $\sigma$-problem, has been overshadowing them almost from
the start. They all utilised the pulsar wind models, where the ratio
of the Poynting flux to the kinetic energy flux, the parameter
traditionally denoted as $\sigma$, was much less than unity. In fact,
the 1D-model would fit the observations only for $\sigma\sim
10^{-3}$. Moreover, for significantly higher values of $\sigma$, the
MHD solution could not be found, with the termination shock collapsing
onto the wind origin \citep{kc84a}. In contrast, the theory of pulsar
magnetospheres predicts winds which are strongly Poynting-dominated,
at least at their base \citep[see ][and references therein]{arons12}.

Several possible explanations of this $\sigma$-problem have been put
forward over the years. The simplest one is that the electromagnetic
energy of the pulsar wind is converted into kinetic energy of the wind
on its way from the pulsar to the termination shock. Although claims
have been made that ideal MHD acceleration mechanisms can provide the
required energy conversion \citep[e.g.][]{Vlahakis-04}, it has now
become clear that this is not the case
\citep[e.g.][]{kvkb-09,lyub-09,lyub-10}.

The acceleration can be facilitated via non-ideal processes, involving
magnetic dissipation. This is only relevant in the so called
``striped-wind'' zone of pulsar winds where alternations of the
magnetic field direction are expected on the length scale of the light
cylinder radius \citep{coroniti-90}.  For the wind of the Crab pulsar,
the dissipation length scale still significantly exceeds the radius of
the wind termination shock \citep{lyub-kirk-01}, unless the pulsar
produces much more plasma compared to the predictions of the current
models of pair-production in pulsar magnetospheres \citep[see ][and
  references therein]{arons12}.  Alternatively, the energy associated
with the alternating component of magnetic field of the striped wind
can be rapidly dissipated at the termination shock itself
\citep[e.g.][]{lyub-03,sironi2011,amano-13}.  Although the
striped-wind model allows conversion of a large fraction of the total
Poynting flux into the internal energy of PWN plasma, much more is
needed to approach the target value of $\sigma\sim10^{-3} \div
10^{-2}$. Indeed, the dissipation is confined to the striped zone of
the wind and only the alternating component of magnetic field
dissipates. Closer to the poles, the magnetization of the pulsar wind
plasma remains very high even after it crosses the termination shock,
and as a result, the overall magnetization of the plasma injected into
the nebula is much higher than that of the Kennel-Coroniti model,
unless the striped zone spreads over almost the entire wind, implying
that the pulsar is an almost orthogonal rotator
\citep{coroniti-90,komissarov2013}.

In contrast to these ideas, \citet{begelman1998} argues that the
axisymmetric models may be highly unrealistic when it comes to the
global structure of PWN. He has shown that the plasma configuration
assumed in these models is unstable to the magnetic kink instability
and speculated that the disrupted configuration may be less demanding
on the magnetization of pulsar winds. Indeed, one would expect the
magnetic pressure due to randomized magnetic field to dominate the
mean Maxwell stress tensor, and the adiabatic compression to have the
same effect on the magnetic pressure as on the thermodynamic pressure
of relativistic gas. Under such conditions, the global dynamics of PWN
produced by high-$\sigma$ winds may not be that much different from
those of PWN produced by particle-dominated winds.  These expectations
have received strong support from the recent numerical studies
\citep{mizuno2011b} of the magnetic kink-instability for the
cylindrical magnetostatic configuration, which was used in
\citet{begelman1992} to model PWN. These simulations have shown a
relaxation towards a quasi-uniform total pressure distribution inside
the computational domain on the dynamical time-scale. Although
important, these simulations do not claim to model PWN, simply because
the continuous injection of magnetic flux and energy into PWN by their
pulsar winds is not accounted for.

In addition to the dissipation of magnetic stripes in the wind or at
the termination shock, the magnetic dissipation could occur inside PWN
as well \citep{lyutikov2010c,komissarov2013}. In fact, the development
of the kink instability is bound to facilitate such dissipation, as
this has already been demonstrated in the simulations by
\citet{mizuno2011b}. In principle, simultaneous observations of both
the synchrotron and inverse-Compton emission allow to measure the
energy distribution between the magnetic fields and the emitting
electrons (and positrons). From a simple ``one-zone'' model of the
Crab Nebula, it follows that its magnetic energy is only a small
fraction, $\sim 1/30$, of the energy stored in the emitting particles
\citep{MH10,komissarov2013}. This shows that unless the striped wind
zone of the Crab's wind fills almost the entire wind volume, there
must be a substantial magnetic dissipation inside the nebula
\citep{komissarov2013}.  This is certainly the case for the magnetic
inclination angle of the Crab pulsar $\alpha\sim45^o$, found by
\citet{harding2008} in their modelling of the high-energy emission of
the pulsar.

The issue of magnetic dissipation is now gradually moving into the
focus of high-energy astrophysics.  For a long time, it was widely
accepted that the ultrarelativistic particles producing the observed
nonthermal emission in all astrophysical phenomena are mainly
accelerated at shocks. However, this is very problematic in the case
of highly magnetised flows, which are often invoked in theories of
relativistic galactic and extragalactic jets and gamma ray
bursts. Firstly, even strong, high Mach number shocks, in high
$\sigma$ plasma are much less dissipative compared to their low
$\sigma$ counterparts \citep[e.g.][]{kc84a,ssk-sd-12}. But what is
even more important is that the recent advanced particle-in-cell (PIC)
simulations show that the acceleration of non-thermal particles does
not operate at such shocks \citep{sironi2009,ss-11}.  Although the
ideal MHD conversion of magnetic energy into the bulk motion energy of
particles is less of an issue for collimated jets than for
uncollimated pulsar winds
\citep[e.g.][]{vk-04,kbvk-07,tnm-10,lyub-09}, it is still quite
impossible to reach $\sigma<10^{-2}$, required by the PIC simulations
in order to allow the particle acceleration.  This is why the magnetic
reconnection sites are now considered as more promising candidates for
the particle acceleration.

The recently discovered short intense gamma-ray flares in the Crab
Nebula \citep{agile-flare-11a,fermi-flare-11} seem to be the first
clear signatures of magnetic reconnection in relativistic plasma. The
very high photon energies of the flares, exceeding 100~MeV, require
strong Doppler boosting in order to overcome the radiation reaction
limit of shock acceleration. However during the flares, none of the
bright and compact, and hence presumably Doppler-boosted, emission
features in the Nebula show increased activity during the flares at
lower energies -- in conflict with what is expected in the case of
shock acceleration mechanisms \citep{weiss-12}. In contrast, the
electrostatic acceleration in reconnection current
sheets readily produces a quasi-monoenergetic spectrum of particle
energies \citep{cerr-uzd-12,cl12}.  

In this paper we provide a detailed account of the first
three-dimensional (3D) relativistic MHD simulations of the Crab
Nebula. Our main motivation was to study the effects which the
additional degree of freedom has on the structure and dynamics of
PWN. Mainly, we wanted to see (i) if the magnetic field would become
randomized just enough to resolve the $\sigma$-problem and yet to
agree with the high polarization typical for the inner Crab Nebula
\citep{hickson1990,MoranShearer2013}, (ii) if the jet-torus structure,
especially its jet component, will be preserved, and (iii) whether the
strong nonstationary dynamics of the termination shock discovered in
2D simulations and connected to the wisp production will carry on in
3D.  Preliminary results of the simulations have been reported in
\citet{porth-13}.

\section{Simulations Overview}

\subsection{Equations and numerical method}

We solve the equations of ideal relativistic magnetohydrodynamics 
in Minkowski space-time. These are the continuity equation

\beq
   \pder{}{t} (\rho\Gamma) + \nabla_i (\rho\Gamma v^i) = 0 \, , 
\label{eq:continuity}
\eeq
the energy equation 

\beq
   \pder{}{t} (w \Gamma^2 -P + \frac{B^2+E^2}{8\pi}) + 
   \nabla_i (w \Gamma^2 v^i+S^i) = 0 \, ,
\label{eq:energy}
\eeq
the momentum equations 
\beq
   \pder{}{t} (w \Gamma^2 v^j + \frac{S^j}{c^2}) + 
   \nabla_i (w \Gamma^2 v^iv^j+ P g^{ij} + T^{ij}) = 0 \, ,
\label{eq:momentum}\eeq
and the Faraday equation 

\beq
   \frac{1}{c} \pder{}{t} B^j +
    e^{jik} \pder{}{x^i} E_k = 0 \, , 
\label{eq:faraday}\eeq
where $\rho$ and $P$ are the rest mass density and thermodynamic
pressure respectively, $w=\rho c^2 + \gamma P/(\gamma-1)$ is the
relativistic enthalpy, $\gamma$ is the adiabatic index ($\gamma=4/3$
in the simulations), $v$ and $\Gamma$ are its velocity and Lorentz
factor, $B$ and $E$ are the magnetic and electric fields, $S^i
=e^{ijk}E_jB_k/4\pi c$ is the Poynting vector, $g^{ij}$, $e^{ijk}$ and
$\nabla_i$ are the metric tensor, the Levi-Civita tensor and the
covariant derivative operator of Euclidean space respectively, $T^{ij}
= -E^i E^j -B^i B^j + (B^2+E^2) g^{ij}/2$ is the Maxwell stress
tensor.  These equations are supplemented with the Ohm's law of ideal
MHD,

\beq
   E^i=-\frac{1}{c} e^{ijk}v_j B_k \, .
\eeq

The numerical simulations are performed with MPI-AMRVAC
\citep{amrvac}, which integrates the special relativistic
magneto-hydrodynamic conservation laws using finite volume
discretisation and adaptive mesh refinement (AMR). 
 In our simulations we use the following 
hybrid approach.  Outside of the termination shock, we apply 
the compact stencil third order reconstruction scheme
\citep[LIM03,][]{cada2009}, combined with an HLLC
Riemann solver \citep{Honkkila:2007:HSI:1232960.1233238}. 
In the pulsar wind and around the termination shock, we 
utilise a robust total variation diminishing Lax-Friedrich
scheme in combination with the minmod limiter.  The solution is
advanced in time with a third order Runge-Kutta integration.
The divergence-free condition for $\mathbf{B}$ is enforced using the
\cite{powell1994} source term approach.

Our 3D simulations are
performed in the Cartesian coordinates.  We employ a cubic domain with
the edge length of $2\times 10^{19}\rm cm$, large enough to contain
today's Crab Nebula.  The large box size also ensures that no
back-reaction from the outflow boundaries can influence the simulated
nebula bubble.  The base level of AMR includes $64^{3}$ cells.  
Four more levels are used to resolve the expanding nebula bubble, with
the cell-size of $\Delta x=1.95\times 10^{16} \rm cm$ at the fifth
level.  Higher resolution cases are achieved by increasing this nebula
refinement up to the ninth level.

A higher resolution is required to properly resolve the termination shock and
the flow near the origin.  To this end, additional grid levels
centered on the termination shock are automatically activated, the actual 
number being dependent on the shock size. For the simulations presented 
in this paper, normally $3-4$ extra grid-levels are employed, 
resulting in $8-9$ levels in total. 
Up to 20 grid-levels  in total are allowed in the vicinity of the origin 
and in order to  speed up the simulations, grid blocks that only hold the
stationary wind solution are normally de-activated from further
temporal updates and automatically re-activated when the termination
shock comes within their range.  

In order to study the impact of the axisymmetry condition, we also
performed 2D simulations whose  numerical setup was identical to 
that of corresponding 3D simulations in 
all respects, with the exception of cylindrical coordinates $\{r,z\}$.  
The standard reflective boundary conditions were applied at the symmetry 
axis $r=0$.

\subsection{Problem initialization}\label{sec:initialization}

Initially, the computational domain is split in two zones separated by
a spherical boundary of radius $r_i=10^{18}$cm. The  
outer zone describes a radially-expanding cold supernova shell. The
solution in this zone has Hubble's velocity profile \beq v=v_\ind{i}
\fracp{r}{r_\ind{i}} \label{eq:vhubble} \eeq and constant density
$\rho_\ind{e}$. The values of parameters $\rho_\ind{e}$ and
$v_\ind{i}$ are determined from the mass and energy of the shell as follows.

Assuming that the ejecta of total mass $M_{e}$ is distributed in a
shell of constant density $\rho_{e}$ between radii $r_{i}$ and
$r_{e}$, we have
\begin{align}
\rho_{e} = \frac{M_{e}}{\int_{r_{i}}^{r_{e}} 4\pi r^{2} dr}\,.
\end{align}
The velocity at the outer edge of the shell $v_{e}$ can be found as 

\begin{align}
v_{e} = \left[\frac{E_{e}}{\int_{r_{i}}^{r_{e}} 2\pi r^{4}/r_{e}^{2}
    \rho_{e} dr}\right]^{1/2} = \left( \frac{10}{3}
\frac{E_{e}}{M_{e}}\right)^{1/2} r_{e}\left (
\frac{r_{e}^{3}-r_{i}^{3}}{r_{e}^{5}-r_{i}^{5}} \right)^{1/2} \, ,
\end{align}
where $E_{e}$ is the total kinetic energy of the shell. 
In a self-similar evolution, the ratio of inner to outer shell radius
$x \equiv r_{i} /r_{e}$ remains constant during the entire expansion.
This allows us to eliminate $r_{i}$ and obtain

\begin{align}
v_{e} = \left( \frac{10}{3} \frac{E_{e}}{M_{e}} \right)^{1/2} \left(
\frac{1-x^{3}}{1-x^{5}} \right)^{1/2} \, ,  
\label{eq:ve}
\end{align}
which depends only mildly on $x$ when $x\ll1$.  E.g. for $x=1/5$, the
$x$-dependent factor of relation (\ref{eq:ve}) is $\simeq 0.996$ and
hence this dependence is dropped in the following, to give

\begin{align}
v_{e} = \left( \frac{10}{3} \frac{E_{e}}{M_{e}}
\right)^{1/2}\,. \label{eq:vereal}
\end{align}
The velocity $v_i$ of the initial expansion at the contact
discontinuity is simply $v_i=x v_e$.  For our simulations, we adopt
the parameters $M_e=3M_\odot=6\times10^{33}\rm\, g$ and
$E_e=10^{51}\rm\, erg$ in accordance with \cite{del-zanna2004}.  We
choose $x=1/5$ to obtain $v_i=1495 \rm\, km\,s^{-1}$.  Since we do not
study the dynamics at the forward supernova-blast in this work, the
outer radius $r_e$ has no further significance and we simply continue
this SNR solution up to the outer boundary of the computational domain.

The inner zone is initially filled with unshocked pulsar wind. In our
model of the wind, we assume that the alternating component of
magnetic field in its striped zone has completely dissipated along the
way from the pulsar to the PWN. Although this may not be the case and
most of the dissipation occurs instead at the termination shock,
dynamically this makes no difference \citep{lyub-03}. The total energy
flux density of the wind follows that of the monopole model
\citep{michel-73}

\begin{align}
f_{\rm tot} (r,\theta) = \frac{L}{L'}\left(\frac{1}{r^{2}}\right)
\left(\sin^{2}\theta+b\right) \, .
\label{eq:ftot}
\end{align}
The parameter $b=0.03$ is added for numerical reasons to avoid
vanishing energy flux at the poles. $L'=\pi (8/3+4b)$ simply gives the
normalization such that the total power in the pulsar wind corresponds
to the current spin down luminosity of the Crab pulsar, 
$L=5\times10^{38} \rm erg\,s^{-1}$ \cite[e.g.][and references therein]{hester2008}.  
This energy is divided between the magnetic, $f_{\rm m}$, and kinetic, $f_{\rm k}$, terms

\begin{align}
f_{\rm m}(r,\theta) = \sigma(\theta)\frac{f_{\rm
    tot}(\theta,r)}{1+\sigma(\theta)}, \quad
f_{\rm k}(r,\theta) = \frac{f_{\rm tot}(r,\theta)}{1+\sigma(\theta)} \,,
\label{eq:fs}
\end{align}
where $\sigma(\theta)$ is the latitude dependent wind magnetization.

\begin{figure}
\begin{center}
\includegraphics[width=80mm]{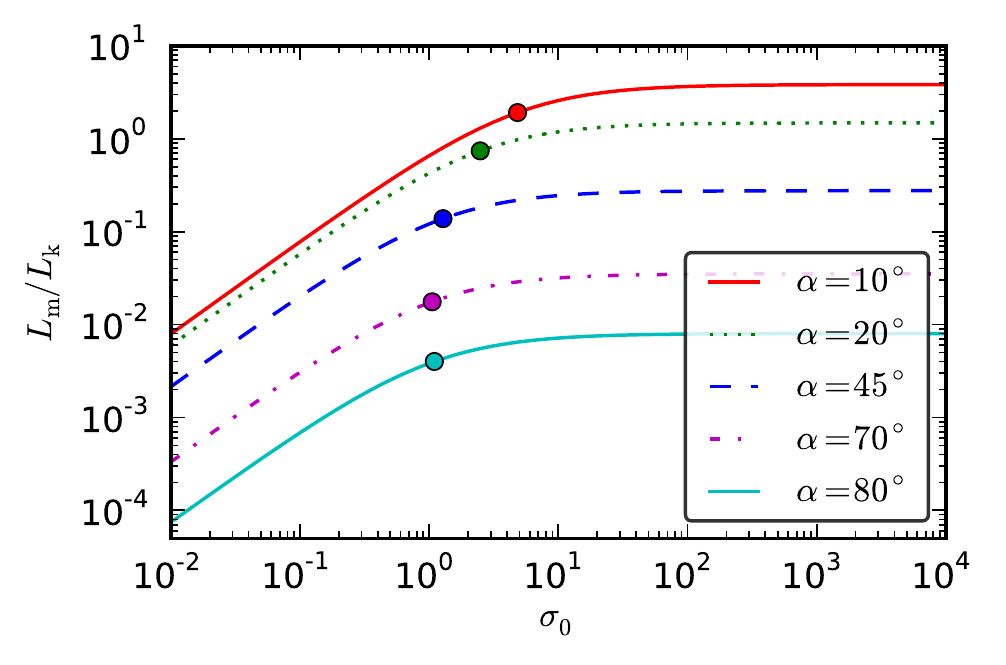}
\caption{The ratio of electromagnetic and kinetic luminosities injected
  into the PWN against the magnetization parameter  $\sigma_{0}$ of the 
  pulsar wind for various angles of obliqueness $\alpha$.  The circles mark 
  the locations where the luminosity ratio assumes half of the value 
corresponding to $\sigma_{0} = \infty$.  For $\alpha<45^{\circ}$, the
  saturation regime is obtained already for moderate values of
  $\sigma_{0}\simeq 1$.  This is why our dynamical simulations can
  provide reasonable models for the much higher magnetization, 
  $\sigma_{0}> 10^{3}$,  predicted in the theory of pulsar magnetospheres.  }
\label{fig:lmOverLk}
\end{center}
\end{figure}

For simplicity, we assume that before reaching the dissipation zone of magnetic 
stripes, the wind magnetisation saturates at
\begin{align}
\tilde{\sigma}_{0}(\theta) = \left\{
\begin{array}{ll}
\left(\theta/\theta_{\rm 0}\right)^{2}\ \sigma_{0}; &
\theta\le\theta_{\rm 0}\\ \sigma_{0}; & \theta > \theta_{\rm 0}
\end{array}
\right.
\label{eq:sig0}
\end{align}
where the additional (small) parameter $\theta_{0}=10^{\circ}$ was introduced
to ensure that the Poynting flux vanishes on the symmetry axis ($\sigma_0$ is 
one of the model parameters reported in table \ref{tab:simulations}.).    
The dissipation of magnetic stripes changes the magnetisation of the striped 
wind zone so that

\begin{align}
\sigma(\theta) =
\frac{\tilde{\sigma}_{0}(\theta)\chi_{\alpha}(\theta)}{1+\tilde{\sigma}_{0}(\theta)(1-\chi_{\alpha}(\theta))}\,
,\label{eq:sigmaprof}
\end{align}
where 

\beq \chi_\alpha(\theta) = \left\{
    \begin{array}{ll}
       (2\phi_\alpha(\theta)/\pi-1)^2, & \pi/2-\alpha<\theta<\pi/2
      +\alpha \\ 1, & \mbox{otherwise}
\end{array}
\right. , \label{eq:chi}
\eeq 
with $\phi_{\alpha} (\theta)\equiv \arccos(-\cot(\theta)\cot(\alpha))$.  
Here, $\alpha$ is the magnetic inclination angle of the
pulsar, which determines the size of the striped wind zone
\citep{komissarov2013}. 
{\bf Equations (\ref{eq:sigmaprof}) and (\ref{eq:chi}) can be read as follows: For polar angles $\theta < \pi/2-\alpha$ polarity reversals of the magnetic field are not present (unstriped region) and the initial magnetisation profile $\bar{\sigma}_0(\theta)$ is obtained  (with $\chi_\alpha=1$). Starting at $\theta \ge \pi/2-\alpha$ the striped region of the flow emerges, giving rise to a drop in $\chi_\alpha(\theta)$ and a decreasing effective magnetisation that ultimately vanishes on the equator (with $\chi_\alpha(\pi/2)=0$).}

Figure~\ref{fig:lmOverLk} shows the ratio of the
wind total electromagnetic luminosity 

\begin{align}
L_{\rm m} = 2\pi r^2 \int_{0}^{\pi}\sin\theta~f_{\rm m}(r,\theta) d\theta 
\label{eq:lm}
\end{align}
to its total kinetic luminosity

\begin{align}
L_{\rm k} &= 2\pi r^2 \int_{0}^{\pi}\sin\theta~f_{\rm k}(r,\theta) d\theta
\label{eq:lk}
\end{align}
as a function of $\sigma_0$ for various values of the magnetic 
inclination angle. 
In the simulations we use $\sigma_0=0.01,1,3$ and
$\alpha=10^{\circ},45^\circ$.  As one can see in
Fig.~\ref{fig:lmOverLk}, the ratio $L_{\rm m}/L_{\rm k}$
saturates for $\sigma_0 \gg 1$, with the asymptotic value being
determined by the extent of the striped wind zone.  In fact, for all
models, except those with very small $\alpha$, this ratio is already
very close to the asymptotic value when $\sigma_0=3$.  This is why
we expect our models with $\sigma_0=3$ not to be very different from
those with much higher $\sigma_0$, which are predicted in the theory of 
pulsar magnetospheres. One sees that in the high sigma limit, the obliqueness 
$\alpha$ has a much larger impact on the overall wind magnetization than 
$\sigma_0$.  
Usually, the parameter $\alpha$ is not very well constrained
observationally.  However, in the case of the Crab pulsar,
\cite{harding2008} found $\alpha\simeq 45^{\circ}$ by fitting the
high energy spectrum and the pulse profile. Therefore,  we  adopt this
value for our reference case.

The magnetization of the plasma injected into the PWN is also influenced by 
the termination shock.  As $B_{\phi} \beta_{n}= const.$ across
the shock, the Poynting flux changes according to the shock
compression by the factor $\eta=\beta_{n,1}/\beta_{n,2}$, 
where $\mathbf{\boldsymbol \beta \equiv v}/c$. 
For a cold, highly relativistic wind, the jump conditions for an
oblique shock \citep[see Appendix A of][]{komissarov2011} imply the  
compression

\begin{align}
\eta(\sigma) = 6 (1+\sigma)\left(1+2\sigma+\sqrt{1+16\sigma+16\sigma^{2}}\right)^{-1}
\end{align}
and we obtain a profile for the generalised local magnetisation 

\begin{align}
\sigma_{\rm s} \equiv \frac{f_{\rm m}}{f_{\rm tot}-f_{\rm m}} = \frac{\sigma(\theta) \eta(\sigma(\theta))}{1+\sigma(\theta)(1-\eta(\sigma(\theta)))}.  
\end{align}
where the denominator includes kinetic and thermal energy.  
The total downstream magnetic and kinetic plus thermal luminosities, 
$L_{\rm m,s},~L_{\rm k+t,s}$,  can be derived in the same way as in 
equations (\ref{eq:fs}),(\ref{eq:lm}) and (\ref{eq:lk}).  
The effective downstream magnetization
$\bar{\sigma}_{s}\equiv L_{\rm m,s}/L_{\rm k+t,s}$ differs from the
upstream value $\bar{\sigma}\equiv L_{\rm m}/L_{\rm k}$ by a factor, which
varies between one and three.  For reference, the values of $\bar{\sigma}$ and
$\bar{\sigma}_{s}$ in our simulations are listed in table
\ref{tab:simulations}.

The wind's magnetic field is purely azimuthal and changes direction at
the equatorial plane. Its strength, as measured in the pulsar frame,
is found via 

\beq 
B_{\phi}(r,\theta) = \pm\sqrt{4\pi f_{\rm m}
  (r,\theta)/v_{r}}\, , 
\eeq 
where $v_{r} \simeq c$ is the radial wind velocity.

The Lorentz factor of the Crab's pulsar wind, as well as its angular
dependence, is not known but is expected to be very high, in the range
$\Gamma=10^2-10^6$. Our 3D code simply cannot cope with such
high values. For this reason, we adopt the much lower value
$\Gamma=10$ for all its streamlines, like in the previous axisymmetric
studies \citep[e.g.][]{komissarov2004,camus2009}.

The wind rest mass density can be found via 

\beq
  \rho(r,\theta) = f_{\rm k}(r,\theta) /
  (\Gamma^{2}c^{2}v_{r}). \label{eq:rho} 
\eeq
In spite of the apparent dramatic mismatch between the real and 
simulated wind Lorentz factors, we do not expect 
this to make a significant impact on the PWN dynamics.  
Indeed, according to equation
(\ref{eq:rho}), the kinetic energy flux depends only on the product
$\Gamma^{2}\rho$.  Moreover, the magnetization $\sigma=B^{2}/(4\pi
\Gamma^{2}\rho c^{2})$ depends only on the term $\Gamma^{2}\rho$ too.
Hence, our models can be scaled to more realistic values of the wind
Lorentz factor without changing the dynamical properties of the wind
(ram pressure and magnetization) when the wind density is scaled
according to $\rho\propto\Gamma^{-2}$. At the same time, the position 
of the TS is determined by the ram pressure balance and will be the same  
as in our simulations. In order 
to verify this conclusion, we have performed low resolution 
2D test simulations with the wind Lorentz factor as high as $\Gamma=80$ 
and observed no significant change in the termination shock size.

\begin{table}
\caption{Overview of simulations.  Designation 2D* means 2D
  axisymmetric simulations with enforced equatorial symmetry, where  
  only the northern hemisphere is simulated and the condition  
  $[B_{\phi}]=2B_{\phi}$ imposed at the equator.  
  Designation 2D** is the same as 2D* but with the boundary condition
  $[B_{\phi}]=0$.  The column labeled $\Delta x$ shows the resolution
  in the post-shock nebula flow in the units of $10^{16}\rm cm$. }
\label{tab:simulations}
\begin{tabular}{@{}llllllll}
ID & $\alpha$ & $\sigma_{0}$ & Dim. & $\Delta x$ & $\bar{\sigma}$& $\bar{\sigma}_{\rm s}$ & $t_{\rm end}$\\
\hline
\hline
A3D & $45^{\circ}$ & 0.01 & 3D & 1.95 & 0.0021 & 0.0063 & 106\\
B3D & $45^{\circ}$ & 1    & 3D & 1.95 & 0.12   & 0.22   & 74 \\
B3Dhr & $45^{\circ}$ & 1    & 3D & 0.89 & 0.12   & 0.22   & 54\\
C3D & $45^{\circ}$ & 3    & 3D & 1.95 & 0.19   & 0.31   & 74\\
D3D & $10^{\circ}$ & 3    & 3D & 1.95 & 1.47   & 2.38   & 54\\
A2D & $45^{\circ}$ & 0.01 & 2D & 1.95 & 0.0021 & 0.0063 & 848\\
B2D & $45^{\circ}$ & 1    & 2D & 1.95 & 0.12   & 0.22   & 800\\
C2D & $45^{\circ}$ & 3    & 2D & 1.95 & 0.19   & 0.31   & 138\\
D2D & $10^{\circ}$ & 3    & 2D & 1.95 & 1.47   & 2.38   & 63\\
\\
B2Dhr & $45^{\circ}$ & 1  & 2D & 0.98 & 0.12   & 0.22   & 121\\
B2Dvhr & $45^{\circ}$ & 1 & 2D & 0.49 & 0.12   & 0.22   & 106\\
B2Duhr & $45^{\circ}$ & 1 & 2D & 0.24 & 0.12   & 0.22   & 74\\
B2Dehr & $45^{\circ}$ & 1 & 2D & 0.12 & 0.12   & 0.22   & 32\\
B2Deq & $45^{\circ}$ & 1 & 2D* & 1.95 & 0.12   & 0.22   & 112\\
B2Dhreq & $45^{\circ}$ & 1 & 2D* & 0.98 & 0.12   & 0.22   & 106\\ 
B2Dvhreq & $45^{\circ}$ & 1 & 2D* & 0.49 & 0.12   & 0.22   & 74\\
B2DhreqS & $45^{\circ}$ & 1 & 2D** & 0.98 & 0.12   & 0.22   & 106\\ 
\end{tabular}
\end{table}

\subsection{Modelling of synchrotron emission}
\label{sec:synchr-modell}

{\bf
For the production of synthetic synchrotron maps (see section
\ref{sec:synchrotron}), additional scalars are advanced with the flow.  
Following \cite{camus2009}, we assume the power-law distribution of
leptons injected at the termination shock

\begin{align}
f(\epsilon) = A\,n_0 \epsilon^{-p}\hspace{1cm} \mtext{for} \epsilon <
\epsilon_{\infty,0} \, ,
\label{inject-s}
\end{align}
where $\epsilon_{\infty,0}$ is the cutoff energy of the injected  
particles, chosen to be $1\rm PeV$ at every point of the shock,
$n_0$ is their number density and $A$ is the normalization constant.  We
set $p=2.2$, as this value was found best suited to approximate the
synchrotron spectrum of the Crab from the optical to X-ray frequencies
in a number of previous studies \citep{kennel1984,atoyan1996,volpi2008}.  
Assuming that downstream of the termination shock these particles are subject 
to the synchrotron and adiabatic energy losses only, their local energy spectrum 
is given by 

\begin{align}
f(\epsilon) = A\, n_0  \fracp{n_0}{n}^{-\frac{2+p}{3}} 
\left(1-\frac{\epsilon}{\epsilon_\infty}\right)^{p-2}\epsilon^{-p} 
\ ,\  \epsilon < \epsilon_{\infty}\,. 
\label{evolv-s}
\end{align}
In this equation, $\epsilon_{\infty}$ is the local cutoff energy and $n$ is the 
local number density of the particles. 
The evolution of these parameters is described by three additional 
conservation laws, which are integrated simultaneously with the main 
RMHD system. }

The first two are the continuity equation for $n$

\be
\pder{}{t} (n\Gamma) +  \nabla_{i}(n \Gamma v^{i}) = 0 \, ,
\label{eq:consn}
\ee
and the advection equation for $n_0$ 

\be
\pder{}{t} (n_0 n\Gamma) +  \nabla_{i}(n_{0} n \Gamma v^{i}) = 0 \, .
\label{eq:advectn0}
\ee
In terms of the co-moving total time derivative $d/dt' =
\Gamma\left(\del/\del_{t}+\mathbf{v}\cdot \nabla\right)$, the evolution of 
$\epsilon_{\infty}$ is governed by 

\begin{align}
\frac{d}{d t'} \ln \epsilon_{\infty} = \frac{d}{d t'} \ln n^{1/3} +
\frac{1}{\epsinf}\fracp{d \epsinf}{d t'}_{\rm S}
\label{eq:epsinf}
\end{align}
where the radiative loss term is given by

\begin{align}
\fracp{d \epsinf}{d t'}_{\rm S} = - \tilde{c}_{2} B'^{2}
\epsilon_{\infty}^{2}
\end{align}
\citep[e.g.][]{del-zanna2006}.  Here, $B'$ signifies the magnetic
field as measured in the co-moving frame and
$\tilde{c}_{2}=4e^{4}/(9m_{e}^{4}c^{7})$ denotes the second
synchrotron constant for an assumed isotropic pitch angle distribution
of particles \citep[e.g.][]{pacholczyk1970}.  As demonstrated by
\cite{camus2009}, equation (\ref{eq:epsinf}) can be cast in
conservation form as 

\be
\pder{}{t} (\epsinf n^{2/3}\Gamma) + \nabla_{i}(\epsinf n^{2/3} \Gamma v^{i}) = 
- \tilde{c}_{2} B'^{2} \epsilon_{\infty}^{2} n^{2/3} 
\label{eq:epsinf1}
\ee
This is the third additional conservation law. 

{\bf Since the termination shock is highly non-spherical and variable, the value of 
$n_0$ depends on the distance from the origin of the location 
where the fluid element crosses the shock. Since the pulsar wind is not 
spherically symmetric it should also depend on the polar angle of this 
location. The exact dependence is difficult to predict as it is influenced 
not only by the shock compression factor but also  
by the particle flux distribution in the pulsar wind and the efficiency of 
the particle acceleration at the termination shock, which are not  
well constrained at the moment. In the silmulations we introduce $n_0$ using the 
following prescription. In the pulsar wind zone, which is identified by its 
high Lorentz factor, we reset $n_0$ every time step to its prescribed value 
$n_\ind{w}(r,\theta)$. Everywhere else $n_0$ is computed according to
the aforesaid evolution equations. 

For $n_\ind{w}$ we investigate two recipies. In the recipe A

\begin{align}
n_\ind{w} \propto r_{\rm}^{-2}
\end{align}
Obviously, this recipe does not distinguish
between the striped wind zone and the polar zone, which is
stripe-free. In contrast, in the recipe B we assume that the synchrotron
electrons (and positrons) are only accelerated at the termination
shock of the striped wind:

\begin{align}
n_\ind{w}  \propto (1-\chi_\alpha(\theta))~ r^{-2}
\end{align}
This reflects the results of recent studies of collisionless
relativistic shocks, which indicate very low efficiency of particle
acceleration at such shocks when the angle between the shock normal
and the magnetic field is sufficiently large, the so called
superluminal shocks \citep{sironi2009}. } However in the case of striped
wind, the particle acceleration may still operate downstream of the
MHD shock, in the region where the alternating magnetic field of the
stripes is dissipated \citep{lyub-03,sironi2011}.

Note that this is a purely passive treatment in
the sense that we neglect the influence of radiative cooling on the
flow dynamics.  Normally, this is justified by the fact that the 
Crab Nebula total luminosity is only $\simeq 10\%$ of the spin down 
luminosity of the Crab pulsar \citep[e.g.][]{del-zanna2006}.  However, 
as discussed by \cite{Foy2007}, locally the impact of radiative losses can be 
quite high. For example, the Crab wisps may loose up to $30\%$ of their 
energy. We plan to explore
this further in a subsequent paper. Another shortcoming of this 
approach is that it ignores the potentially important particle acceleration 
inside the nebula, which is related to its turbulence and magnetic 
reconnection. 

Given the lepton spectrum (\ref{evolv-s}), the synchrotron emissivity in 
the observer frame is  

\begin{align}
  j_\nu \propto {\cal C} 
\left\{\begin{array}{lll}
   \nu^{\frac{1-p}{2}}
   \left(1-\sqrt{\frac{\nu}{\nu_\infty}}\right)^{p-2} & ; & 
   \nu < \nu_\infty\\
   0 &; & \nu > \nu_\infty
\end{array}
\right.
\end{align}
{\bf where
$$ 
{\cal C} = n_\ind{0} \mathcal{D}^{2+(p-1)/2} B_\perp'^{(p/2+1/2)} 
\fracp{n_0}{n}^{-\frac{p+2}{3}} \, , 
$$ 
where} 
\begin{align}
\mathcal{D} = \frac{1}{\Gamma(1-\mathbf{\boldsymbol \beta\cdot n})} 
\end{align}
is the Doppler factor. 
The cutoff-frequency is defined as $\nu_\infty\equiv c_1
B_\perp'\epsilon_\infty^2$ with the synchrotron constant $c_1=3e/(4\pi
m_e^3 c^5)$.  Note that the emissivity depends on the magnetic field
component perpendicular to the line of sight vector $\mathbf{n'}$ in
the co-moving system (indicated by dashes): $B_\perp' \equiv
|\mathbf{B'\times n'}|$ which is obtained via the Lorentz transformation.
Optically thin intensity maps are then obtained via the  straight forward
integration

\begin{align}
I_\nu(x,y)=\int_{-\infty}^\infty dz\,j_{\nu}(x,y,z)
\end{align}
in Cartesian coordinates where the $z$ direction is aligned with
the line of sight.  To obtain the linear polarization degree and
direction, we additionally solve the radiation transport equations for the 
Stokes parameters Q and U via

\begin{align}
I_\nu^p(x,y) &\equiv Q(x,y) + i U(x,y) \\
&= \frac{p+1}{p+7/3} \int_{-\infty}^\infty dz\,j_{\nu}(x,y,z)e^{2i\chi(x,y,z)} \, .
\end{align}
This is analogous to the method described by \cite{del-zanna2006}.
In this equation, $\chi$ is the local angle of the newly emitted photons
electric field $\mathbf{e}$, as measured in the observer frame with respect to 
the $y$ axis in the plane of the sky.  Theory of Relativity introduces a swing of
the polarization angle, which is similar to the relativistic 
aberration effect \citep{Blandford1979,porth2011}.  
The radiation electric field in the observer frame is given by  
\begin{align}
\mathbf{e\propto n\times q\ ; \hspace{1cm} q= B+n\times(\boldsymbol
  \beta\times B)} \, ,
\label{eq:swing}
\end{align}
where all quantities are measured in the observer frame \citep{Lyutikov:2003}.  
Hence $\chi$ is determined from

\begin{align}
\cos \chi = \mathbf{\hat{y}\cdot \hat{e}}\ ; \hspace{1cm} \sin \chi =
\mathbf{n\cdot (\hat{y}\times \hat{e}) },
\end{align}
where $\mathbf{\hat{v}}$ is the unit vector in direction of 
$\mathbf{v}$.  Note that $\chi$ increases counter-clockwise.  From the
above Stokes parameters, the  degree of linear polarization  is

\begin{align}
\Pi_\nu = \frac{\sqrt{Q_\nu^2+U_\nu^2}}{I_\nu} = |I_\nu^p|/I_\nu \, ,
\end{align}
which has a maximal value of $(p+1)/(p+7/3)\simeq 0.706\ (p=2.2)$.  In
this notation, the photon e-field direction is simply given by the
real and imaginary part of $\sqrt{I_\nu^p}$,  

\begin{align}
e_x &= - \Im\sqrt{I_\nu^p}\\
e_y &= \  \Re\sqrt{I_\nu^p}\,. 
\end{align}

\subsection{Analytical reference  model}
\label{sec:analyt-refer-model}

In the classical paper on the Crab Nebula by \citet{kennel1984}, one of the 
main observational constraints on the MHD model is the size of its
termination shock. During the phase of self-similar expansion, which is 
appropriate given the current age of the nebula, it is a {\bf roughly} constant fraction 
of the nebula size.    

Given the initial setup, all our solutions exhibit two distinct phases. 
During the first phase, the initial discontinuity at $r_\ind{i}$
splits into three waves and the termination shock rapidly
moves inwards, reflecting the build-up of the shocked 
PWN plasma.  During the second phase, the shock re-bounces and begins to
expand, at a much slower rate, towards its asymptotic
self-similar position. Unfortunately, the computational cost of actually 
reaching the self-similar phase in 3D simulations is prohibitive and this complicates 
testing of our numerical models against both the observations and the 
above mentioned analytical models. To overcome this problem, we decided to look 
for a simple 1D analytical solution describing the 
transition to the self-similar regime.  To this end, we considered the  
adiabatic spherically-symmetric expansion of PWN powered by an
unmagnetised wind. Since the analytical models suggest very low wind magnetization 
\citep{rees-gunn-74,kc84a}, the unmagnetized case turns out to be quite 
suitable here.  

Assuming the polytropic equation of state with $\gamma=4/3$ 
and 
{\bf a radial power-law for the expanding nebula radius $r_{\rm n}=r_{\rm i}(t/t_0)^{\alpha_{\rm r}}$,  we find that the 
evolution of the PWN energy $E$ is governed by the equation 
\begin{align}
   \dot{E} = L-\alpha\frac{E}{t} \, .
\end{align}
Given the initial condition $E(t_0) = 0$, the solution to this equation is 

\be
     E=\frac{Lt}{1+\alpha}\left(1-\fracb{t_0}{t}^{\alpha_{\rm r}+1}\right) \, .  
\label{eq:ehydro}
\ee
The velocity of the accelerating nebula boundary is then $v_{\rm n}=\alpha_{\rm r} r_{\rm n}/t$ and 
when we set $t=0$ as the time of the super nova, $t_0=\alpha_{\rm r} r_{\rm i}/v_{\rm i}$ corresponds to the initial time of the simulations.  

From the momentum conservation at the termination shock, one finds the 
characteristic shock size (see appendix \ref{sec:shape}) 

\begin{align}
r_{0} = \fracp{L}{4 \pi p c}^{1/2}
\end{align}
and hence 

\be
	r_0 = (1+\alpha_{\rm r})^{1/2}\alpha_{\rm r}^{-1}\, r_{\rm n} (v_{\rm n}/c)^{1/2} \left(1-\fracb{t_0}{t}^{\alpha_{\rm r}+1}\right)^{-1/2} \, . \label{eq:r0hydro}
\ee
This equation tells us that the time-scale of the transition to the self-similar 
expansion corresponds to a simulated time of $t_{0}$, which is $\simeq\alpha_{\rm r}\, 210~$yr for our 
parameters.  
}
For the angular dependence of the wind energy flux utilised in our simulations, 
$f_{\rm tot}(\theta) \propto \sin^{2}\theta$,  
the maximal radial and vertical extents of the termination shock in this 
reference model are  
\be 
   r_{max} \simeq \textcolor{black}{0.825}\, r_0\ \text{and}\ z_{max} \simeq \textcolor{black}{0.232}\, r_0 \, 
\label{eq:rmax}
\ee 
(see appendix \ref{sec:shape}). 
{\bf As argued by \cite{bucciantini2003, bucciantini2004}, the equation of motion for the PWN radius in fact depends only on the total pressure at the nebula boundary, rendering the expansion law $r_{\rm n}$ independent of magnetisation.  We can hence utilise the hydrodynamic results obtained by means of the thin shell approximation for the forward PWN shock \citep[e.g.][]{van-der-SwaluwAchterberg2001} and use the resulting index $\alpha_{\rm r}=6/5$ as our reference value.  
}

\citet{kennel1984} used $v_{\rm n}=2000\rm ~ km\,s^{-1}$ and deduced 
the radius of Crab's termination shock to be 
$r_{\rm s}=0.056\, r_{\rm n}$. Following the high resolution
Chandra observations reported by \cite{weisskopf2000}, it seems natural
to identify the equatorial size of the termination shock with 
that of the Crab's X-ray inner ring,  $r_\ind{ir} \sim0.14\,\rm pc$.  Adopting 
the volume averaged nebula radius of $r_{\rm n}=1.65\rm pc$
\citep{hester2008}, we obtain $r_{\rm s}=0.085\, r_{\rm n}$, 
slightly larger than the \cite{kennel1984} value.  
If we plug the above value for $v_{\rm n}$ into equations 
(\ref{eq:r0hydro}) and (\ref{eq:rmax}) for $t\gg t_{0}$, {\bf we obtain
 $r_{\rm max}/r_{n}=0.083$}, which is actually in a very good agreement 
with the observed ratio! This suggests that we may consider our 3D numerical 
magnetic models of PWN as being in agreement with the observations 
if they do not differ much from our non-magnetic analytical model.   
According to equation (\ref{eq:vhubble}), the initial expansion speed 
in our simulations is set to $v_{\rm i}\simeq1500\rm ~km\,s^{-1}$, which 
lets room for a moderate acceleration over the nebula history.

\section{Results}

\subsection{Global energy balance}
\label{sec:geb}

In our analysis of the energy balance of simulated PWN, we distinguish between 
the magnetic, kinetic and thermal energy densities which are defined as follows:
\begin{align}
e_{\rm m} &\equiv 
\frac{B^2+E^2}{8\pi}\label{eq:emag} \, ,\\
e_{\rm k} &\equiv \left(\Gamma-1\right)\Gamma\rho c^{2} \, , \\
e_{\rm t} &\equiv 4 \Gamma^{2}p - p \, .
\end{align}
The PWN volume is distinguished from that of the supernova shell by means of the 
passive scalar tracer $\tau\in[0,1]$, which is advected along with the flow. 
In the pulsar wind, we set $\tau=1$, and in the initial solution, 
$\tau=0$ for $r>r_\ind{i}$.  Using this tracer, the PWN volume is determined by 
the integral 

\begin{align}
V_{\rm PWN} = \int \Theta(\tau-10^{-3})~\Theta(9-\Gamma)~dV \, ,
\label{eq:mask}
\end{align}
where $\Theta(\cdot)$ is  the Heaviside step function.
The second step function in this equation is introduced in order to 
exclude the unshocked wind zone from the nebula test volume.

\begin{figure*}
\begin{center}
\includegraphics[width=80mm]{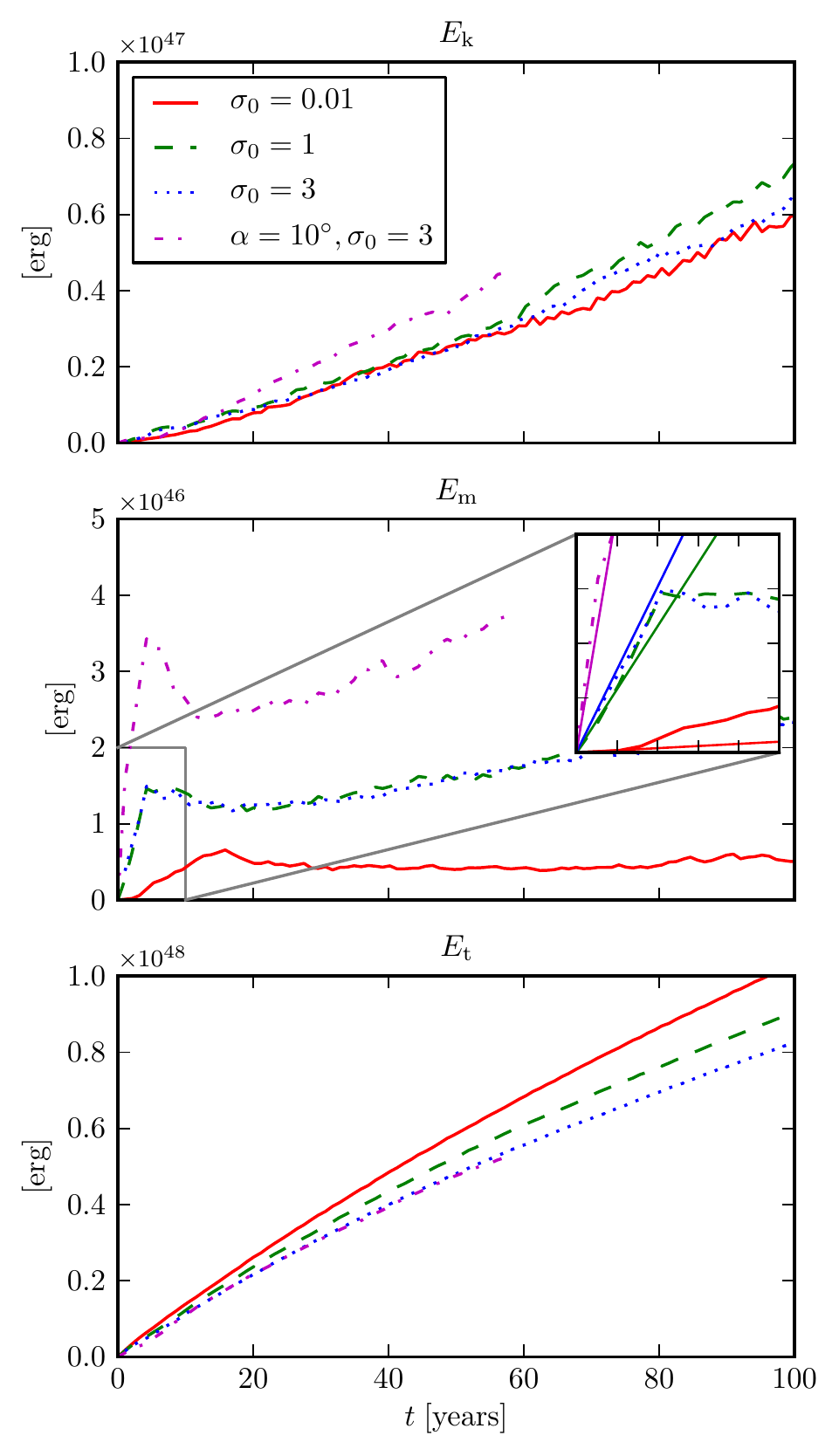}
\includegraphics[width=80mm]{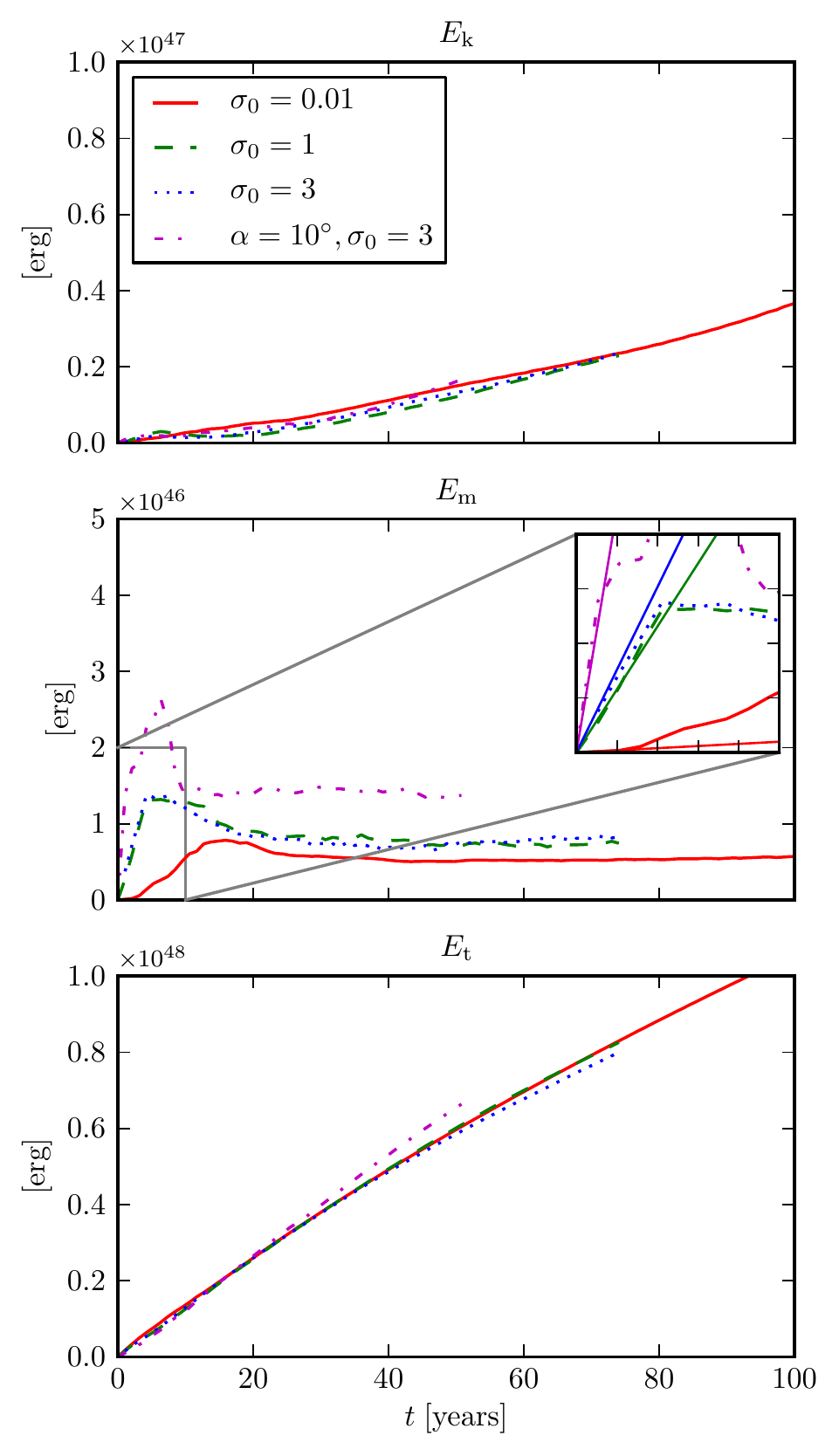}
\caption{Evolution of the total kinetic, $E_\ind{k}$ , magnetic,  $E_\ind{m}$  and
  thermal energy  $E_\ind{t}$ of the nebula.  The left panels show the results of 
  2D runs \{A2D, B2D, C2D, D2D\}, whereas the right panels show the 
  3D  results for the models with equivalent setups \{A3D, B3D, C3D, D3D\}.  
  The insets in the magnetic energy plots (second panel) compare the naive expectation 
  $L_{\rm m,s}t$  (straight lines ) with the actual data.  
  }
\label{fig:energetics}
\end{center}
\end{figure*}

The evolution of the energies in our simulations is
shown in figure \ref{fig:energetics}.
It is interesting to view the nebula magnetic energy content in
relation to the injected power $L_{\rm m,s}$.  Its initial 
evolution depends mainly on the wind magnetization. In the cases with
high magnetization, $\sigma_{0}\ge1$, we simply have 
$E_{\rm m}\simeq L_{\rm m,s}t$, whereas in the low sigma case $E_{\rm m}$ 
grows faster (see the inserts in Figure \ref{fig:energetics}).  
The possibility of such odd evolution for low sigma models has been discussed 
in \citet{rees-gunn-74} -- the magnetic energy grows at the expense of the 
thermal energy as work is carried out against the magnetic hoop stress.    

During this initial period the magnetic flux is conserved.  
Five to ten years after the start of the simulations, the turbulence in the 
PWN is sufficiently developed and the magnetic dissipation begins 
to influence the evolution of $E_{\rm m}$. 
At this stage, the magnetic energy in the 2D cases continues
to rise slowly, whereas it levels off in the 3D realisations\footnote{Since our 3D runs 
have not reached the phase of self-similar expansion, they cannot be used to deduce the 
long-term evolution on the scale of the Crab Nebula lifetime.}.
Extrapolating from the simulations, we
expect todays Crab Nebula to contain a total energy of $\simeq10^{49}\rm
erg$ in good agreement with the previous estimates
\citep[e.g.][]{hester2008}.

\subsection{Global structure}
\label{sec:gs}

\begin{figure*}
\begin{center}
\includegraphics[width=85mm]{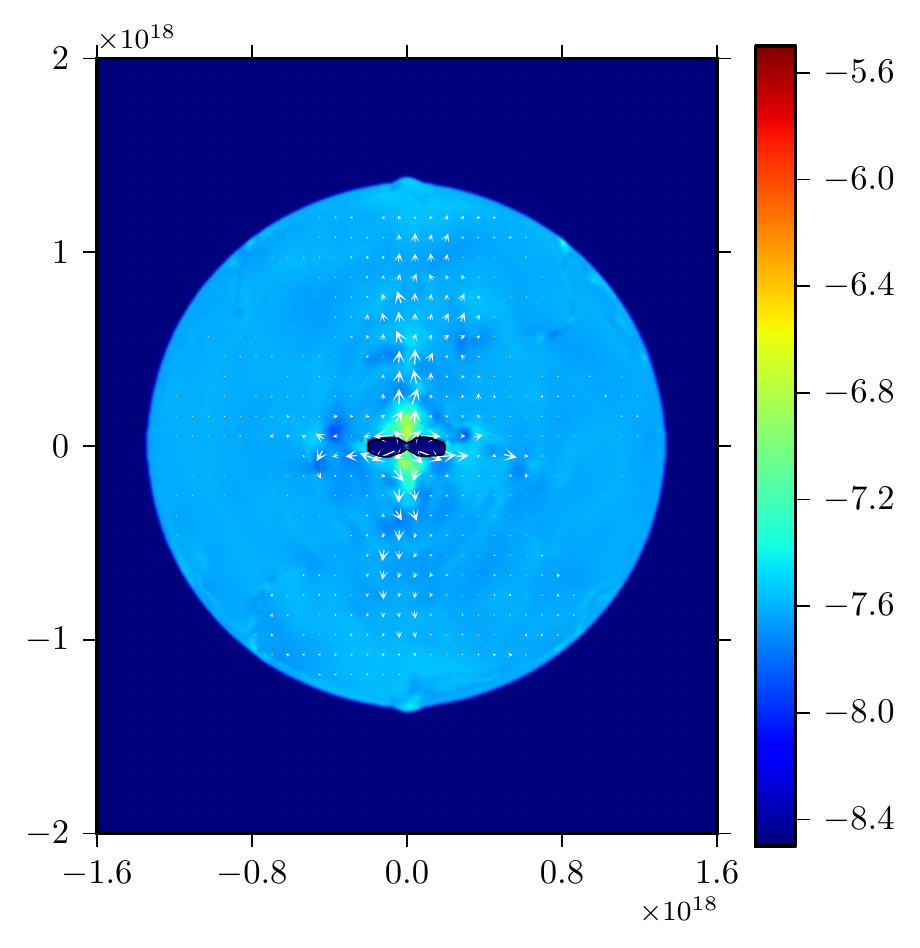}
\includegraphics[width=85mm]{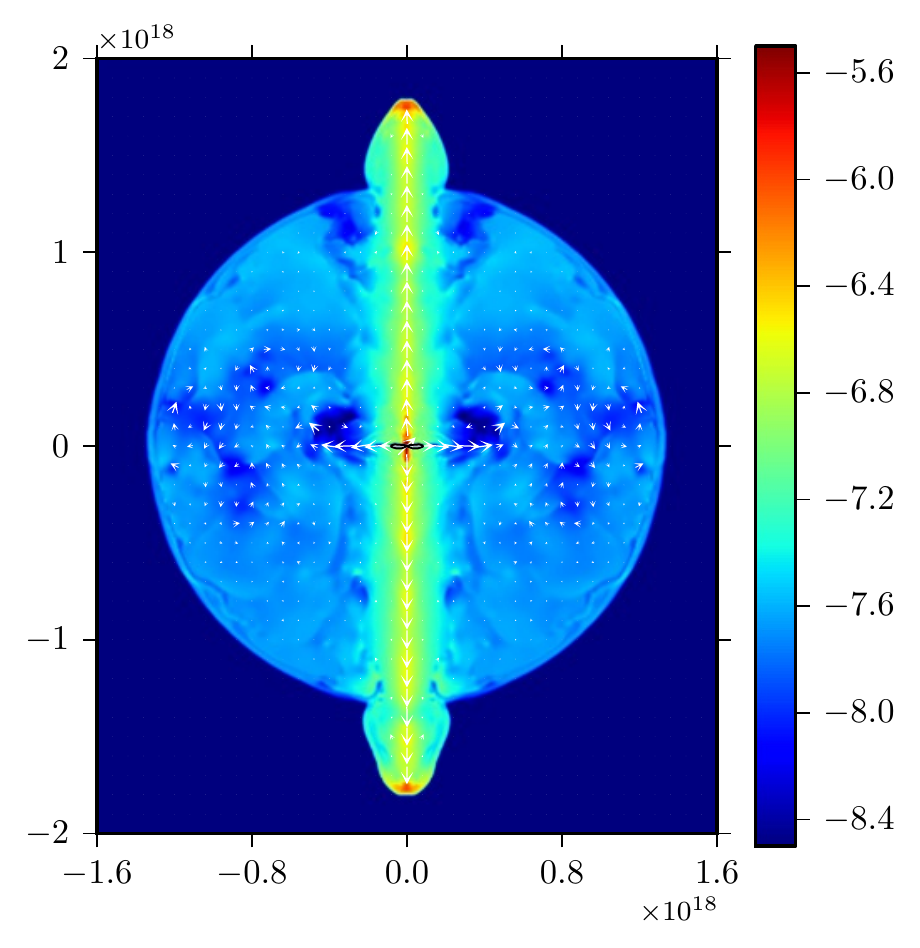}
\caption{Dependence of the total pressure distribution, $\log_{10}p_{\rm tot}$, 
on the imposed symmetry.  
The left panel shows the pressure distribution in the $xz$-plane of the 
3D simulation run B3Dhr and the right panel in the corresponding 2D run B2Dhr, 
both at the time of 51 years from the start of the simulations.  
The white arrows show the in-plane velocity vectors.  The strong axial compression 
observed in this and previous 2D simulations is an artefact of the imposed 
symmetry.  }
\label{fig:ptotSlice}
\end{center}
\end{figure*}

One of the most striking features of the previous 2D axisymmetric simulations 
of PWN is their strong axial compression due to the hoop stress of their purely 
azimuthal magnetic field.  These results agree very well with the theoretical 
models of the z-pinch configurations of PWN by \citet{begelman1992}. With the 
increased magnetisation of the pulsar wind, the pressure distribution over the 
interface with the supernova shell becomes so much enhanced near the polar 
region that the nebula expansion in the polar direction becomes significantly 
faster compared to the one in the equatorial direction. The polar jets begin to 
drill holes through the supernova shell and produce escape routes for the 
PWN plasma (see the right panel of figure~\ref{fig:ptotSlice}).  
Such breakouts have been observed in similar earlier simulations, aimed to explain 
the origin of GRB jets in the magnetar model \citep{bucc-07,bucc-08}.

In contrast, the additional degree of freedom in our 3D simulations allows a 
fairly uniform distribution of the total pressure in the nebula (see the 
left panel of figure~\ref{fig:ptotSlice} ).  Thus, the assumption that the 
PWN plasma adopts a magneto hydrostatic z-pinch configuration as proposed in
\citet{begelman1992} is an over-simplification.  The symmetry of the pulsar 
wind is not preserved inside the nebula due to kink instability of this 
configuration (see section~\ref{sec:jet-morphology}). 
Only in the very vicinity of the termination shock, which is dominated by 
the plasma injected into the nebula recently,  we still observe 
a noticeable axial compression. This is the formation region of the 
polar jets ( plumes ).

\subsection{Evolution of the termination shock}
\label{sec:evol-term-shock}

\begin{figure}
\begin{center}

\includegraphics[]{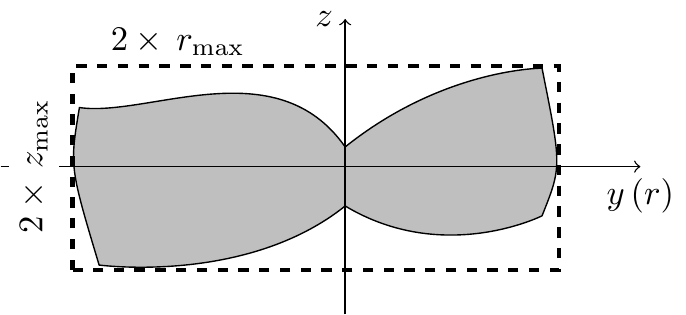}
\caption{Sketch of the termination shock.  The shock size is measured
  in the $yz$-plane by fitting the wind region into a rectangle with
  base lines $2 r_{\rm max}$ and $2 z_{\rm max}$.  }
\label{fig:TS}
\end{center}
\end{figure}

To investigate whether the observational constraint on the 
termination shock size can
be satisfied, we now focus on the temporal evolution of the shock.
Given the fact that in the simulations the shock shape can be quite 
complex compared to that of our simple analytical model, 
we need to come up with a robust procedure for measuring its sizes.  
The sizes given below are measured as follows. First a rectangle is 
fitted  around the unshocked wind region as shown in figure~\ref{fig:TS}.  
Then the horizontal and vertical lengths of this rectangle are identified 
with doubled $r_{\rm max}$ and $z_{\rm max}$ respectively. 

\begin{figure*}
\begin{center}
\includegraphics[width=80mm]{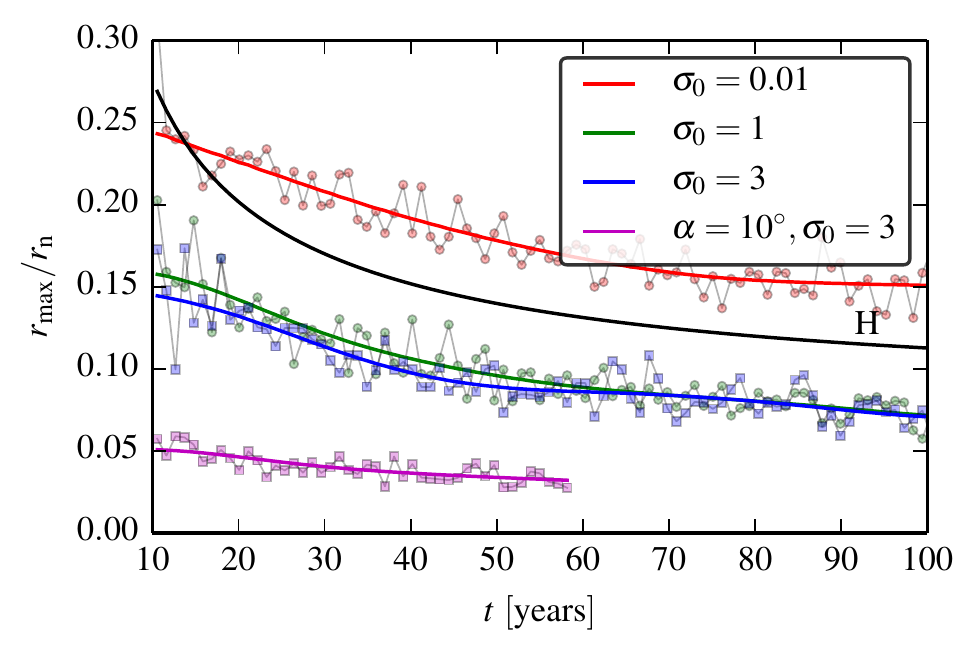}
\includegraphics[width=80mm]{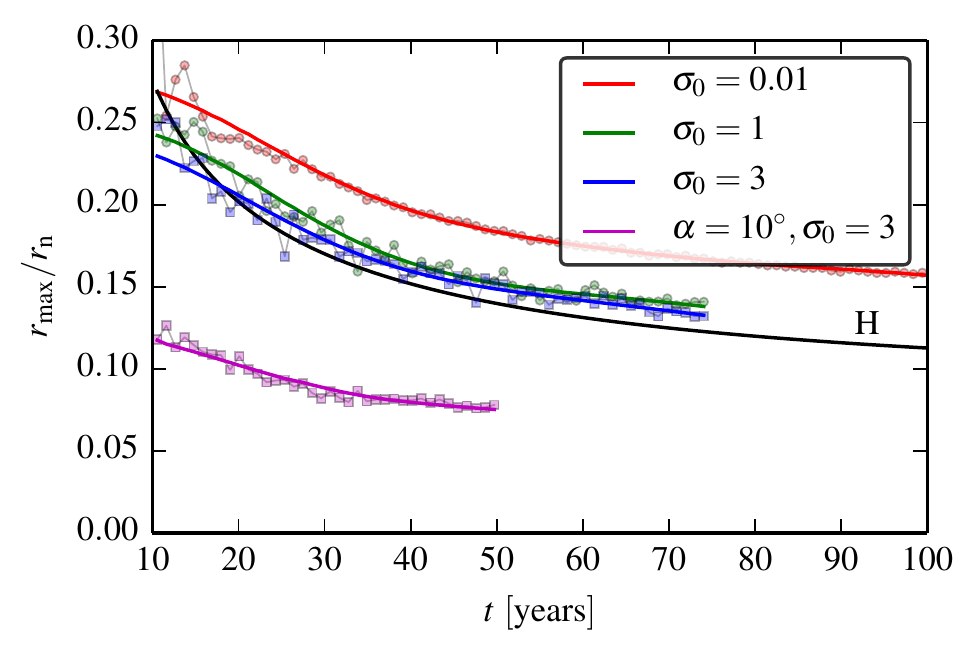}
\includegraphics[width=80mm]{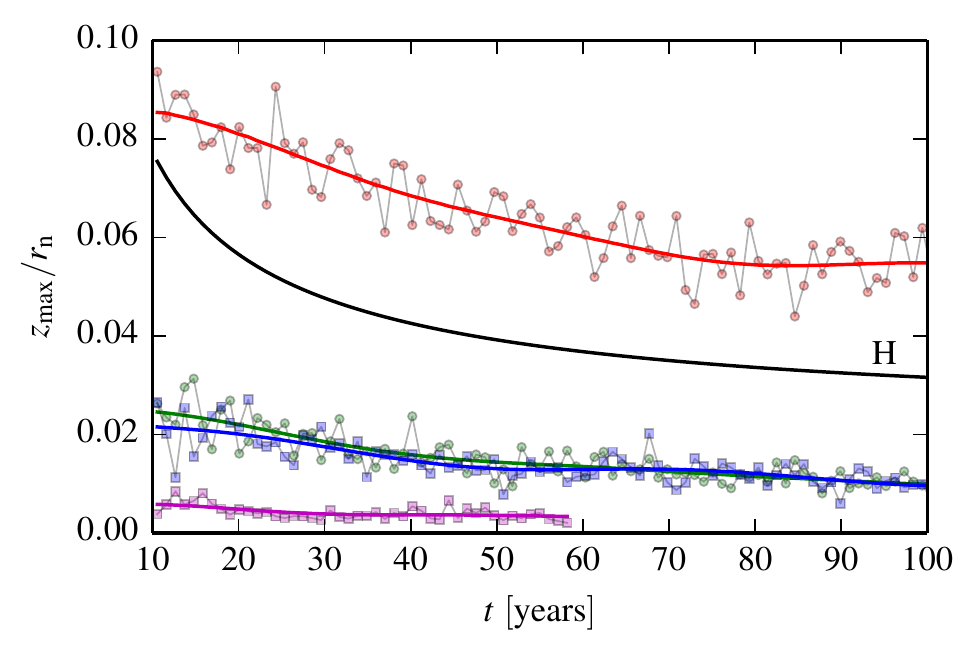}
\includegraphics[width=80mm]{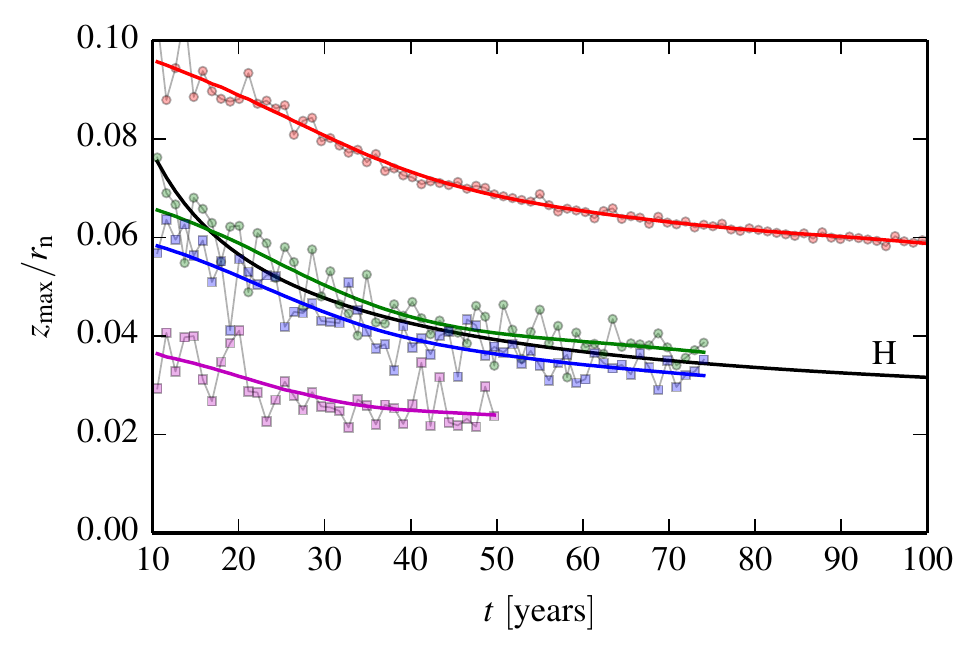}
\caption{Maximum horizontal and vertical extent of the TS in units of
  the volume averaged nebula radius.
  The left panels show the 2D runs for models \{A2D, B2D, C2D, D2D\} 
  while the 3D results with the equivalent setups
  \{A3D, B3D, C3D, D3D\} are shown in the right panels.  The smooth lines show the 
  low-pass filtered data and hence illustrate the overall long-term  trend. 
  The markers shows the actual shock size and hence illustrate the short-term 
  shock variability. The hydrodynamic expectation based on equation~(\ref{eq:r0hydro}) 
  is given by the  black solid line labeled with letter ``H''.  }
\label{fig:rmax}
\end{center}
\end{figure*}

The results are shown in figure \ref{fig:rmax}, together with the 
prediction of our reference analytic model (\ref{eq:r0hydro}). 
The most striking and 
important result is that the shock sizes in 3D runs with high wind 
magnetization and $\alpha=45^{\circ}$ follow very closely the 
analytical curves, even better that the model with $\sigma_0=0.01$. 
This is in great contrast with the \citet{kc84a} model, where 
for such high magnetizations the shock size collapses to zero.   
Moreover, as the initial magnetization $\sigma_{0}$ increases from $1$ to $3$, the
shock size remains virtually unchanged. This is not overly
surprising, given that the energetically  most important parameter 
$\bar{\sigma}_{\rm s}$ (see table \ref{tab:simulations}) changes only
by $\sim40\%$.  
In our 2D simulations, the agreement with the reference model 
is not as good but can still be considered as satisfactory, 
with deviations within a factor of less than two. 

As expected, the difference between the mean shock sizes in 2D and 3D 
simulations is particularly minor in the models with $\sigma_{0}=0.01$. 
However, the short term variability of the shock is much more pronounced 
in the 2D case, indicating that the symmetry condition promotes this 
variability. On the other hand, the fact that in 3D simulations the 
shock variability is strong for models with high $\sigma_{0}$ and 
almost absent in the model with $\sigma_{0}=0.01$ shows that the 
magnetic field plays a key role in exciting and sustaining this 
variability.

For magnetizations $\sigma_0\ge 1$, the difference between 2D and 3D 
results becomes stronger, resulting in shock sizes which 
are $\sim 1.5$ times larger in radial direction and $\sim 2$ times 
larger in vertical direction in the 3D models compared to the 
corresponding 2D models. This is consistent with the lower axial 
compression of the nebula in the 3D solutions as described in Sec.\ref{sec:gs}.  
As a further consistency check, we run simulations with the 
obliqueness $\alpha=10^{\circ}$, which gives higher magnetization 
of the plasma supplied into the nebula and hence promotes stronger 
axial compression. As expected, in these runs we obtained smaller 
termination shocks.  

\begin{figure*}
\begin{center}
\includegraphics[width=85mm]{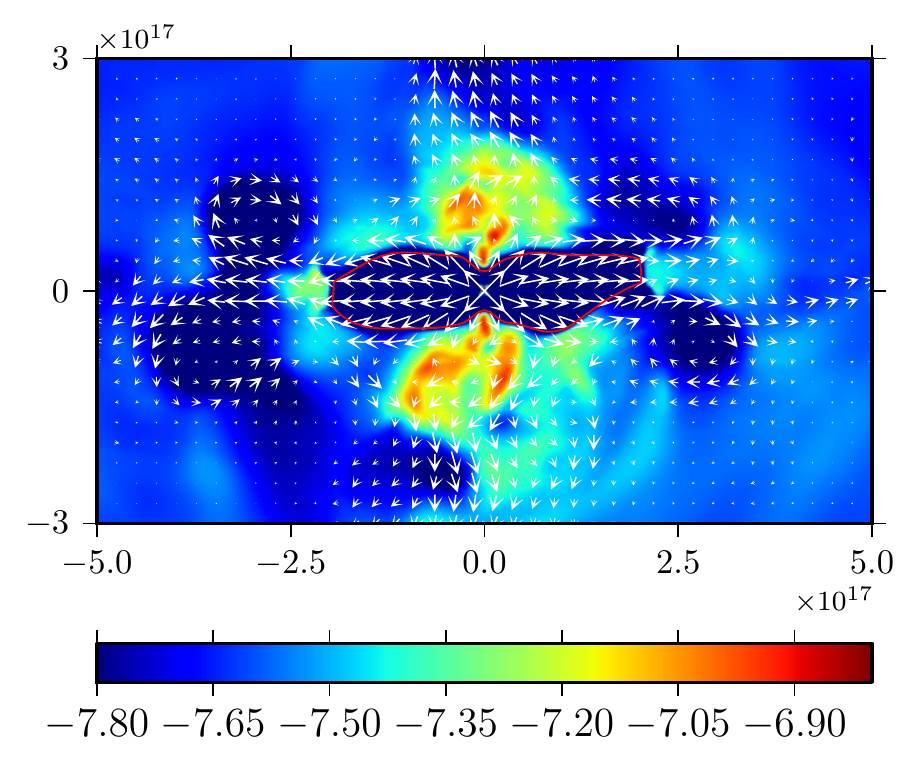}
\includegraphics[width=85mm]{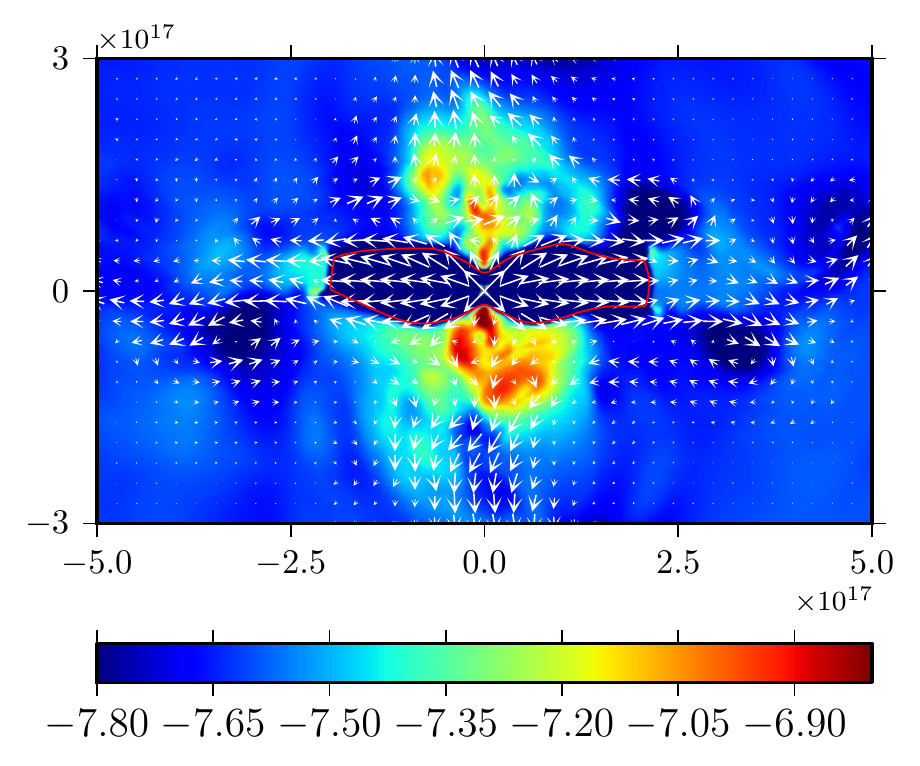}
\includegraphics[width=85mm]{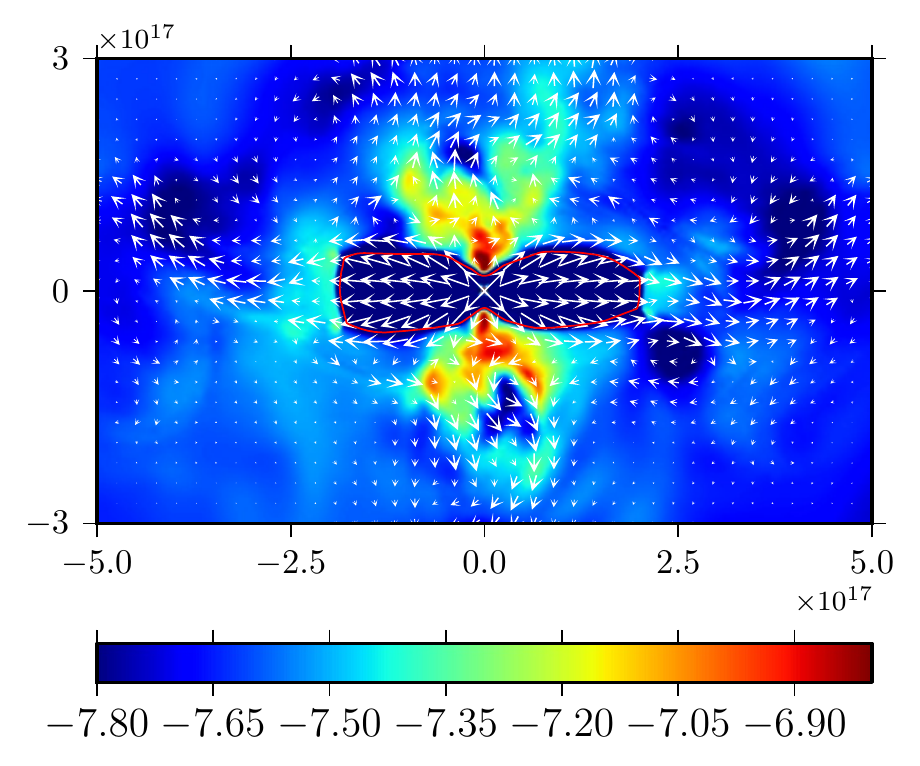}
\includegraphics[width=85mm]{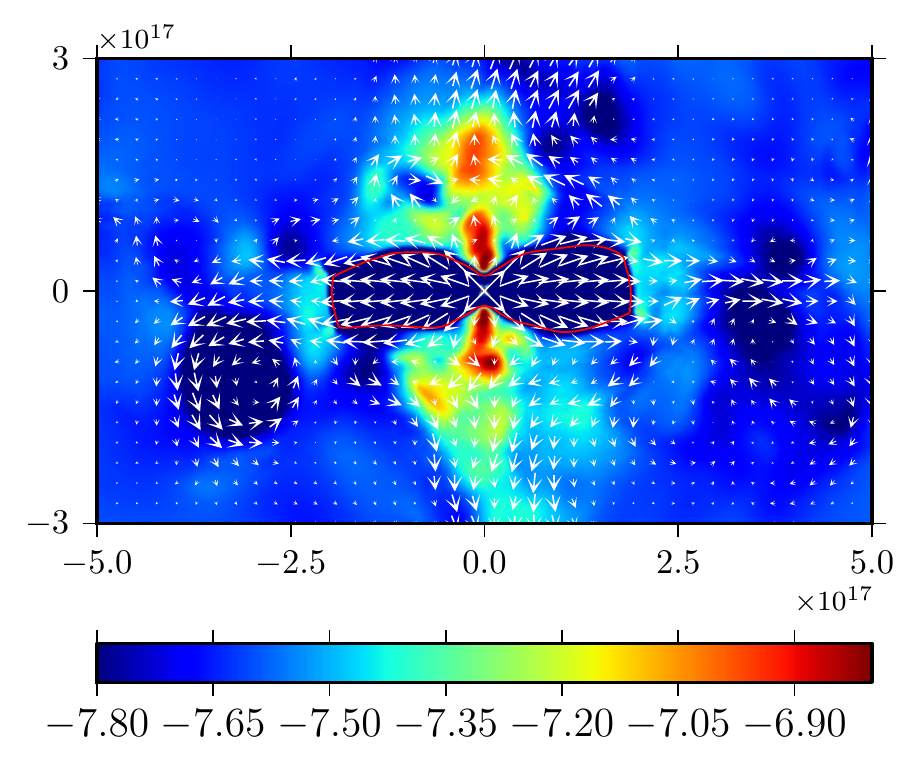}
\caption{Variability of the termination shock in the 3D simulation
  B3Dhr. The color images show the logarithm of total pressure 
 $\log_{\rm 10}p_{\rm tot}$ in the $xz$-plane and the in-plane velocity 
 vectors are shown as white arrows. The red line shows the location of the 
 termination shock. The snapshots are taken at $t=44,45,46,47$ years 
 (left to right, top to bottom).  
  One can the quasi-periodic vortex-shedding from the termination shock 
  and its distortion by the highly variable PWN flow. 
  }
\label{fig:shockTime}
\end{center}
\end{figure*}

Figure~\ref{fig:shockTime} illustrates the typical flow behavior in the 
vicinity of the termination shock in the 3D simulation B3Dhr. One can see
that this 3D solution reproduces the typical morphology seen 
in the earlier 2D simulations \citep{komissarov2004},  with its 
Mach belt, polar compression region, rim shock and the fast 
``$\Upsilon$-flow'' just outside of the oblique section of the 
termination shock.  
This figure also shows the vortex-shedding by the termination shock. 
The vortices are formed near the Mach belt and then move outwards.   
The newly formed vortices seem to interact with the termination shock, 
modulating the width of its Mach belt.  For example, in the first panel, 
which features strong vortices near the Mach belt, its width is much smaller 
than in the fourth panel, which shows no strong vortices close to the 
termination shock. The figure also shows very irregular and highly 
variable total pressure and velocity fields in the immediate surrounding 
of the shock, thus exposing the highly dynamic nature of the PWN 
flow in this region. It is not surprising that the termination shock 
responds to this flow by changing its shape.       

On the other hand, the unsteady termination shock itself is a source 
of variable downstream outflow, seeding waves, shocks and vortices into 
the nebula.  For example, as the shock
obliqueness increases due to its vertical compression, the downstream flow
velocity also increases (obeying the jump conditions) and a feedback
cycle is established.  This quasi periodic nature of the TS was
discussed by \cite{camus2009}, who commented on the difficulty in 
distinguishing the cause and effect of the shock perturbations.

The most significant difference between the 2D and 3D simulations 
is in the nature of the axial compression and its impact on the 
termination shock. In the 2D simulations, waves converge on the 
symmetry axis in a highly coordinated way, creating very strong 
variability of the total pressure right at the funnel of the termination 
shock and having very strong impact on the shock size along the axis. 
In the 3D simulations, the non-axisymmetric instabilities disrupt  
the highly  coordinated structure of the 2D flow near the polar axis. 
The total pressure does not display such a strong maximum at the 
axis and the amplitude of its variability is reduced. A much more 
extended region of enhanced pressure around the axis is created instead. 
This explains the somewhat lower amplitude of the shock radius variations 
in the 3D simulations (cf. figure \ref{fig:rmax}).  With these
differences in mind, we find that the mechanism of coupling the 
shock perturbations to the variations in the nebula flow still 
operates in the full 3D case and results in substantial variations 
on less than one year timescale.

\subsection{Magnetic dissipation in the nebula}

\begin{figure*}
\begin{center}
\includegraphics[width=80mm]{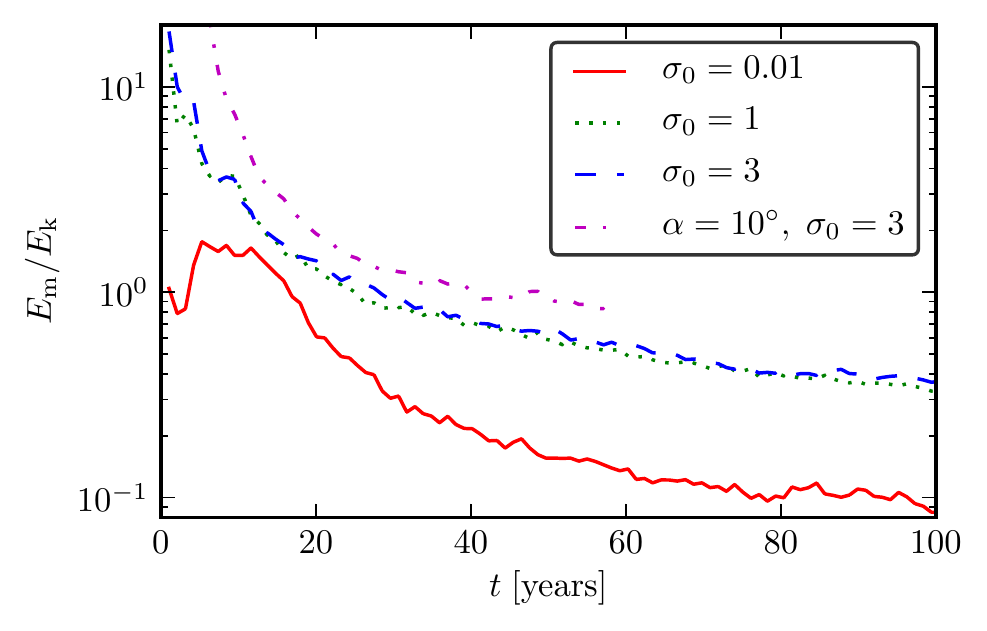}
\includegraphics[width=80mm]{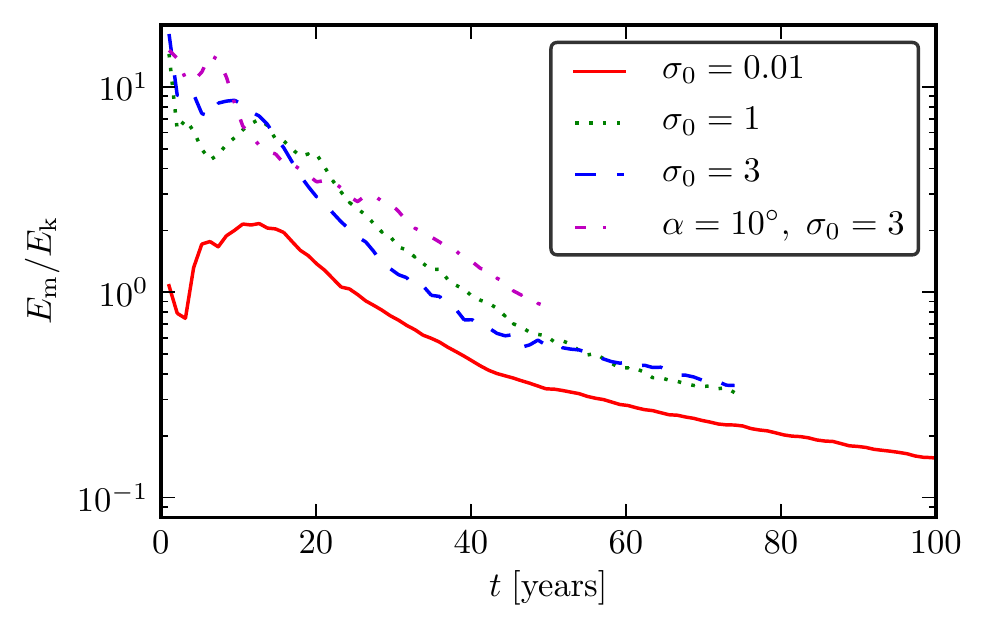}
\includegraphics[width=80mm]{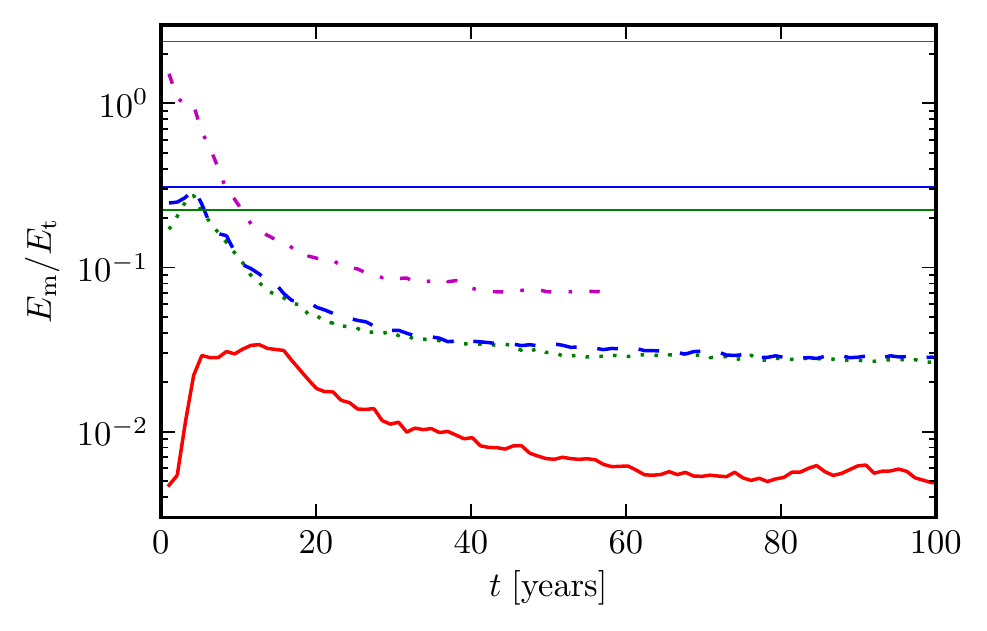}
\includegraphics[width=80mm]{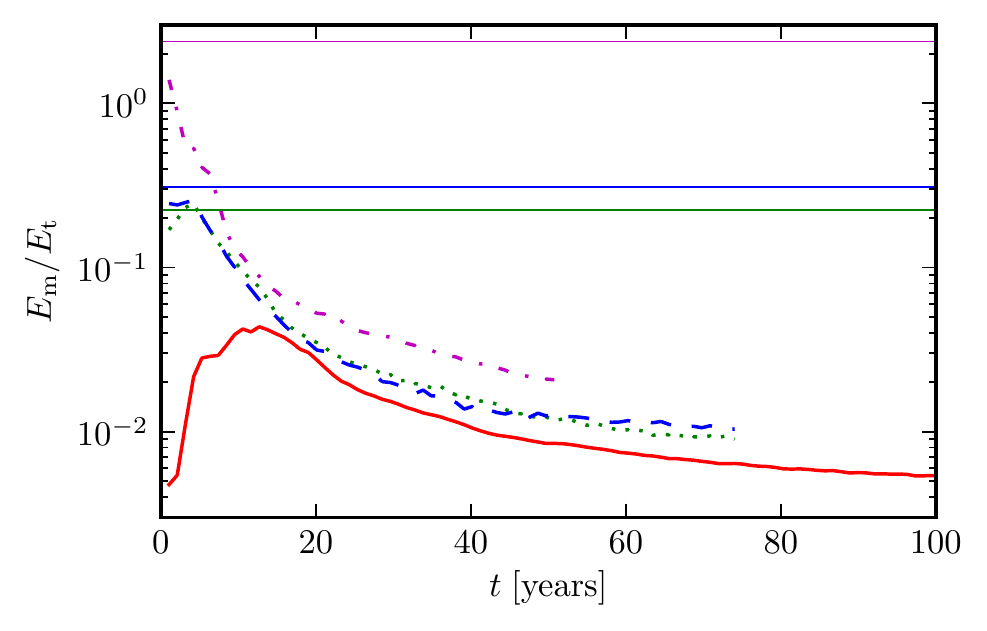}
\caption{Magnetic dissipation in 2D and 3D simulations.  The top panels show the 
  ratio of the total magnetic and kinetic energies, whereas the bottom ones the ratio 
  of total magnetic and thermal energies. The data from the 2D runs 
\{A2D, B2D, C2D, D2D\} are shown in the left panels 
  while the 3D results for the runs with the equivalent setups  
  \{A3D, B3D, C3D, D3D\} are in the right panels.  The solid horizontal
  lines of the lower panels show the values expected in the case when  
  the randomization of magnetic field is not accompanied by its dissipation.  }
\label{fig:dissipation}
\end{center}
\end{figure*}

Surprisingly, even the 2D cases show an acceptable shock size when
compared to the Crab Nebula observations via our 
analytic reference model.  This is due to the dissipation of the 
nebula magnetic field, which occurs when loops of opposite
polarity come together in the turbulent nebula flow.  Its overall effect on 
the nebula energetics is demonstrated by figure \ref{fig:dissipation} which
shows the ratios of the magnetic to the kinetic and thermal energies
in the nebula.

One can see that all simulations follow rather similar evolution 
of $E_{\rm m}/E_{\rm k}$, the ratio of magnetic and kinetic energies.
At the start of the simulations,
$E_{\rm m}\simeq10\, E_{\rm k}$ and after $100~\rm yr$,   
$E_{\rm m}\lesssim\,E_{\rm k}$.

The results for  $E_{\rm m}/E_{\rm t}$ are  even more conclusive.  
If the magnetic field was only randomized, as
suggested by \cite{begelman1998}, but not subject to significant 
dissipation then one would expect this ratio to remain unchanged during 
the nebula lifetime.  Because the magnetic hoop stress is no longer
significant, the randomized magnetic field behaves like an 
ultrarelativistic ideal gas \citep{heinz2000} and hence evolves in the same 
way as the real gas of particles. Hence, $E_{\rm m}/E_{\rm t}$ stays the 
same.  From figure
\ref{fig:dissipation} it is evident that our numerical solutions do not 
fully support the Begelman's hypothesis as in addition to the randomization 
a significant amount of magnetic energy is converted 
into the thermal energy of plasma. Obviously, this energy exchange between 
the two ``gases'' does not have a direct impact on the global 
flow dynamics -- it remains governed by the equations of relativistic hydrodynamics.   
Importantly, the magnetic dissipation does not require
complete randomization of the field which would be difficult to
reconcile with the observations.  At least in the central parts,
the optical polarization measurements
\citep{1979ApJ...227..106S,hickson1990} clearly indicate dominance
of the azimuthal component.

\begin{figure*}
\begin{center}
\includegraphics[]{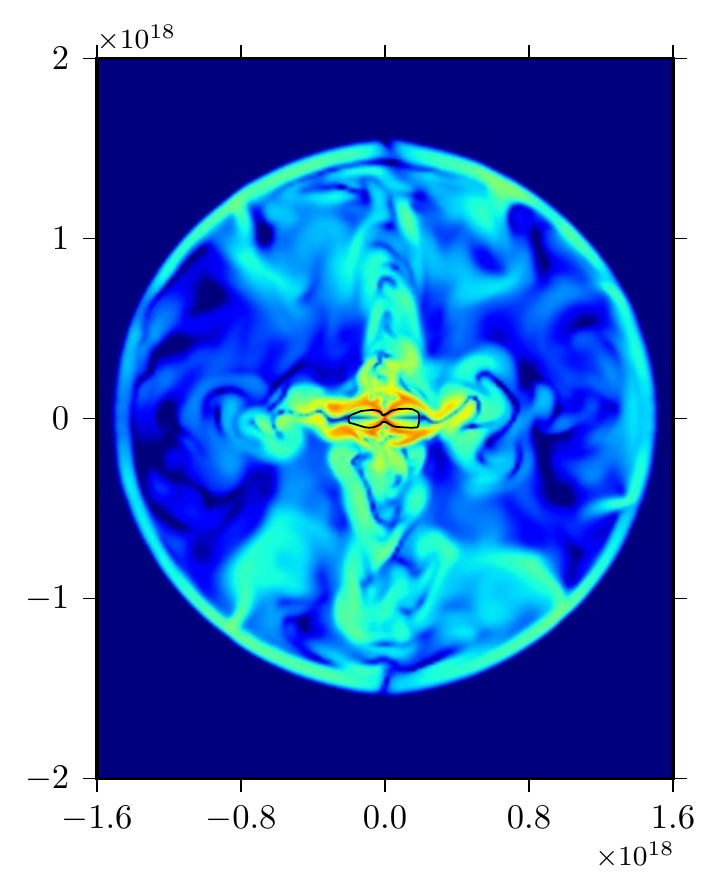}
\includegraphics[]{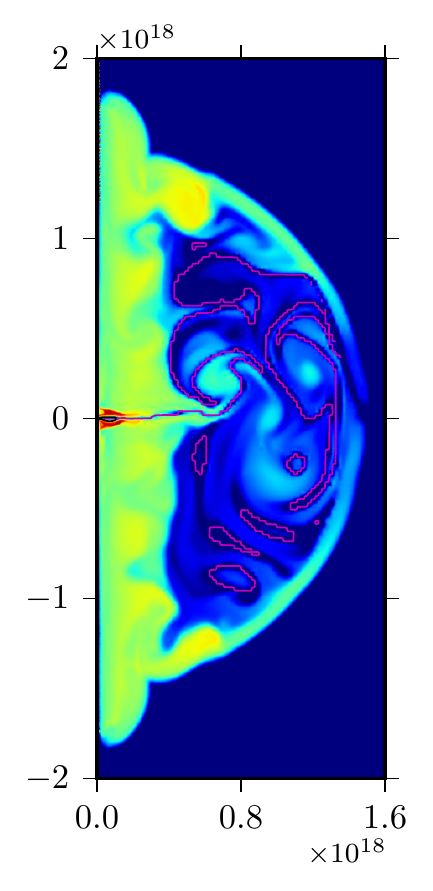}
\includegraphics[]{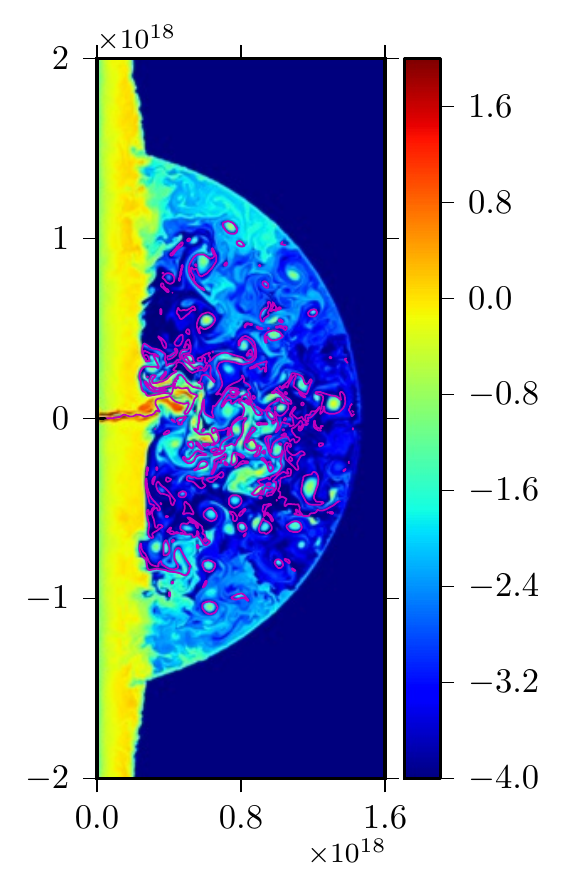}
\caption{Magnetic dissipation regions in 3D and 2D simulations. The images 
  show  the distribution of $\log_{10}e_{\rm m}/e_{\rm t}$ for the models with
  $\alpha=45^{\circ}$ and $\sigma_{0}=1$  at $t = 70$ yr.  From
  left to right: the $yz$-slice for the run B3D, the data for B2D and B2Duhr 
  (the B2Duhr run has eight times higher resolution inside the nebula compared to B2D).  
  The magenta line in plots for the 2D runs shows the locus of points where the 
  magnetic flux changes its sign and the black contour shows the termination shock.  
  In the 2D cases, a strong polar jet develops, which is shielded from the rest of nebula
  by its backflow.  }
\label{fig:dissSlice}
\end{center}
\end{figure*}

\cite{lyutikov2010c} argued that the Crab Nebula flow remains largely 
axisymmetric and the opposing magnetic loops meet at the equator and the 
polar axis, which become places of strong magnetic dissipation. 
However, even in our 2D simulations the situation is rather different as
the so-called equatorial current sheet quickly leaves the equator and 
densely fills most of the nebula volume, being tangled by its turbulent eddies 
(see figure \ref{fig:dissSlice}). Thus, the magnetic dissipation occurs 
over most of the nebula volume.

As to the polar region, in the 2D simulations it is occupied by the 
polar jet which is shielded from the rest of the  nebula by
its backflow and is almost free from the magnetic dissipation 
(see figure \ref{fig:dissSlice}).    
In the 3D simulations, the situation is qualitatively different as 
no strong jet and backflow develop (we discuss the ``plume'' further in section
\ref{sec:jet-morphology}). The non-axisymmetric motions of the plasma 
in the polar region promote displacement of magnetic loops and create 
additional sites of magnetic dissipations there.     
This is why the magnetic dissipation is stronger in 3D. For example, 
the 3D models with $\alpha=45^{\circ}$ and $\sigma_{0}\ge1$ reach 
$E_{\rm m}/E_{\rm t}  \simeq 0.01$ by the time of $\sim 80$ years, 
whereas the corresponding 2D models seem to saturate at 
$E_{\rm m}/E_{\rm t}\simeq 0.03$ (see figure~\ref{fig:dissipation}).
Incidentally, the combined data on the synchrotron and inverse Compton emission 
of the Crab Nebula also lead to $E_{\rm m}/E_{\rm t}\simeq 0.03$ within 
its ``one zone model''. This is in conflict with the much higher 
mean value expected on theoretical grounds for the plasma injected into 
the nebula by the pulsar wind and hence suggestive of efficient  
magnetic dissipation inside the nebula \citep{komissarov2013}.

{\bf Since we integrate equations of ideal RMHD, the magnetic dissipation 
occurs at the grid-scale via numerical resistivity. However, to become 
efficient it requires creation of ever smaller scales structures in the 
magnetic field distribution. The processes which drive the development 
of such structures, and occur on scales above the grid-scale,  must 
ultimatelety determine the dissipation rate and it is important that 
their dynamics is captured sufficiently accurately.      
The most stringent test we can perform to this end is to check the resolution 
dependence of the observables inferred from the simulations.  
This is described in detail in appendix \ref{sec:convergence} and for the 3D 
simulations the results are encouraging: Upon doubling the resolution for the 
run B3D, we obtain very close results both for the termination shock size 
and the magnetic dissipation rate inferred from the nebula energetics.  
While the influence of the numerics can not be ruled out completely yet 
( one has to wait until even higher resolution simulations become available 
and other numerical schemes tried) 
this lets us to believe that the high degree of dissipation observed in our 
simulations is not far from being realistic.

The combined observational data on the synchrotron and inverse Compton emission 
of the Crab Nebula lead to $E_{\rm m}/E_{\rm t}\simeq 0.03$, within 
its simplified ``one zone model'' \citep{MH10,komissarov2013}. 
Our 3D data show saturation of this parameter around  
$E_{\rm m}/E_{\rm t}  \simeq 0.01$ by the time of $\sim 80$ years, 
(see figure~\ref{fig:dissipation}). This suggests that the magnetic dissipation 
in our simulations may be a bit excessive but not off the scale. On the other hand, 
the strength of magnetic field in our simulations is far from being uniform. 
In fact, in contrast to the 1D model of Kennel-Coroniti, it is significantly lower 
in the outskirts of the nebula compared to the vicinity of the termination shock. 
This rises questions about the reliability of estimates based on the one zone model.}

\subsection{Azimuthal magnetic flux}

The energy content of PWN is not solely determined by the 
magnetic dissipation and other dynamical factors may become 
important too. One may argue that a complimentary, and perhaps  
more direct, approach to quantification of the magnetic 
reconnection-driven dissipation should involve nebula's azimuthal 
magnetic flux.

In axisymmetry, the total azimuthal flux $\Phi$ contained within PWN can 
be defined via the integral 

\begin{align}
\Phi \equiv \int\limits_{S} |B_{\phi}(r,\theta)| dS\,   
\label{eq:flux}
\end{align}
over any cross section of the nebula containing the symmetry axis. 
In the absence of magnetic reconnection, this flux increases at the 
rate given by the flux injection rate of the pulsar wind. 
For a stationary wind this rate is constant. 
However, in the absence of axisymmetry, one has to specify a particular 
cross section or to average over the azimuthal angle.  
Here, we simply measure the toroidal flux of the 3D runs in the $yz$-plane.
In any case, the resultant
quantity is no longer conserved even in the absence of magnetic 
reconnection, which reduces the usefulness of this parameter.    

\begin{figure*}
\begin{center}
\includegraphics[width=80mm]{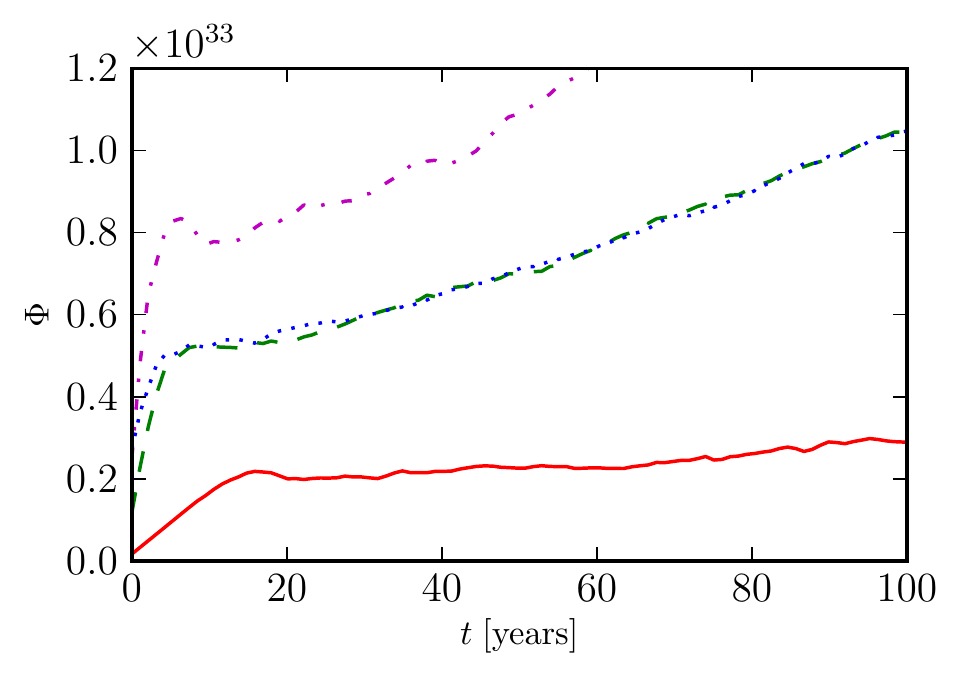}
\includegraphics[width=80mm]{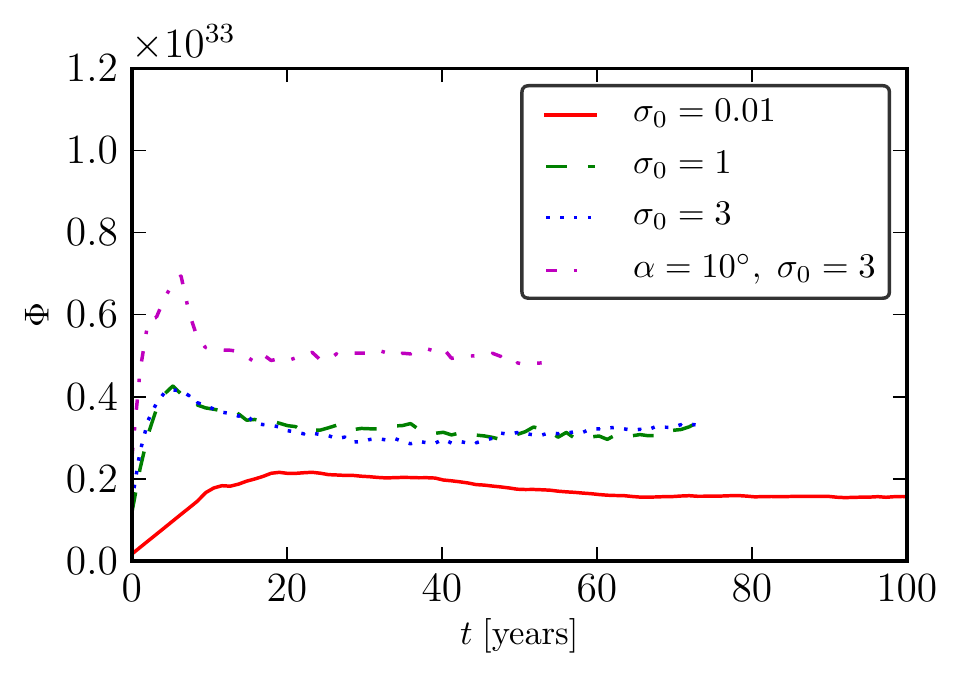}
\caption{Evolution of the total azimuthal magnetic flux $\Phi$ as defined 
  by equation~(\ref{eq:flux}). The left panel shows the results for the 
  2D runs  \{A2D, B2D, C2D, D2D\}  while the results for the 3D runs with 
  equivalent setups \{A3D, B3D, C3D, D3D\} are shown in the right panel.  
  The evolution departs from the flux conservation law 
  after few light crossing times. In the 3D runs the flux saturates, whereas 
  in the 2D runs it keeps increasing at approximately linear rate. } 
\label{fig:fluxes}
\end{center}
\end{figure*}

The left panel of figure \ref{fig:fluxes} shows the temporal 
evolution of $\Phi$ in our 2D simulations. One can see that the flux does 
not increase at a constant rate but evolves through at least two phases. 
During the first phase the growth of $\Phi$ is linear and at the rate 
determined by pulsar wind. In the models with $\sigma_{0}\ge1$ this 
phase lasts for only few years, whereas in the low sigma  model 
it continues for  $\sim15$ years. At the end of this phase, the 
nebula flow becomes fully turbulent, speeding up dramatically the 
annihilation rate of the magnetic flux. As the result, $\dot{\Phi}$ is 
significantly reduced. These two phases can also be seen in the plots 
of the magnetic energy evolution (see figure~\ref{fig:energetics} ). 
In the 3D models, $\Phi$ remains approximately constant during the 
second phase (see the right panel of figure~\ref{fig:fluxes}). 
However, because the nebula expansion has not reached the self-similar phase, 
we can not say that this is a characteristic of the long term nebula
evolution.       

As we have commented above, the azimuthal flux is a less meaningful
measure of magnetic reconnection in 3D models. 
Further progress in this direction could be
made via procedures locating individual current sheets in the 3D data 
and measuring local energy conversion rate,
for example by using an algorithm as proposed by
\cite{2010PhRvE..82e6326U,zhdankin2013}.  We leave such an effort to
future work employing higher spatial resolution as needed for the
detection of the current sheets.

\subsection{Magnetic field structure}\label{sec:magn-field-struct}

Observations of optical polarized light from the Crab Nebula indicate a 
rather complex magnetic structure.  The early images of polarized intensity due
to \cite{michel1991a, fesen1992} revealed a ``hourglass'' structure 
aligned in the north-south direction, which was interpreted as synchrotron
emission originating in a predominantly azimuthal magnetic field.  Near 
the nebula boundary the morphology is complex and indicative of a
``layered'' configuration with the magnetic field wrapping around 
individual filaments (see the detailed account of \cite{hester2008}).
Thus, the observations suggest that the inner 
Crab Nebula  is dominated by the azimuthal component
freshly supplied by the pulsar wind.  
In a 3D flow,  the non-axisymmetric effects such as the
kink instability and 3D turbulence tend to randomize 
the magnetic field. Thus, the question is how quickly 
these processes destroy the ordered field in our simulations  
and whether their results agree with the observations in this respect.

\begin{figure}
\begin{center}
\includegraphics[width=80mm]{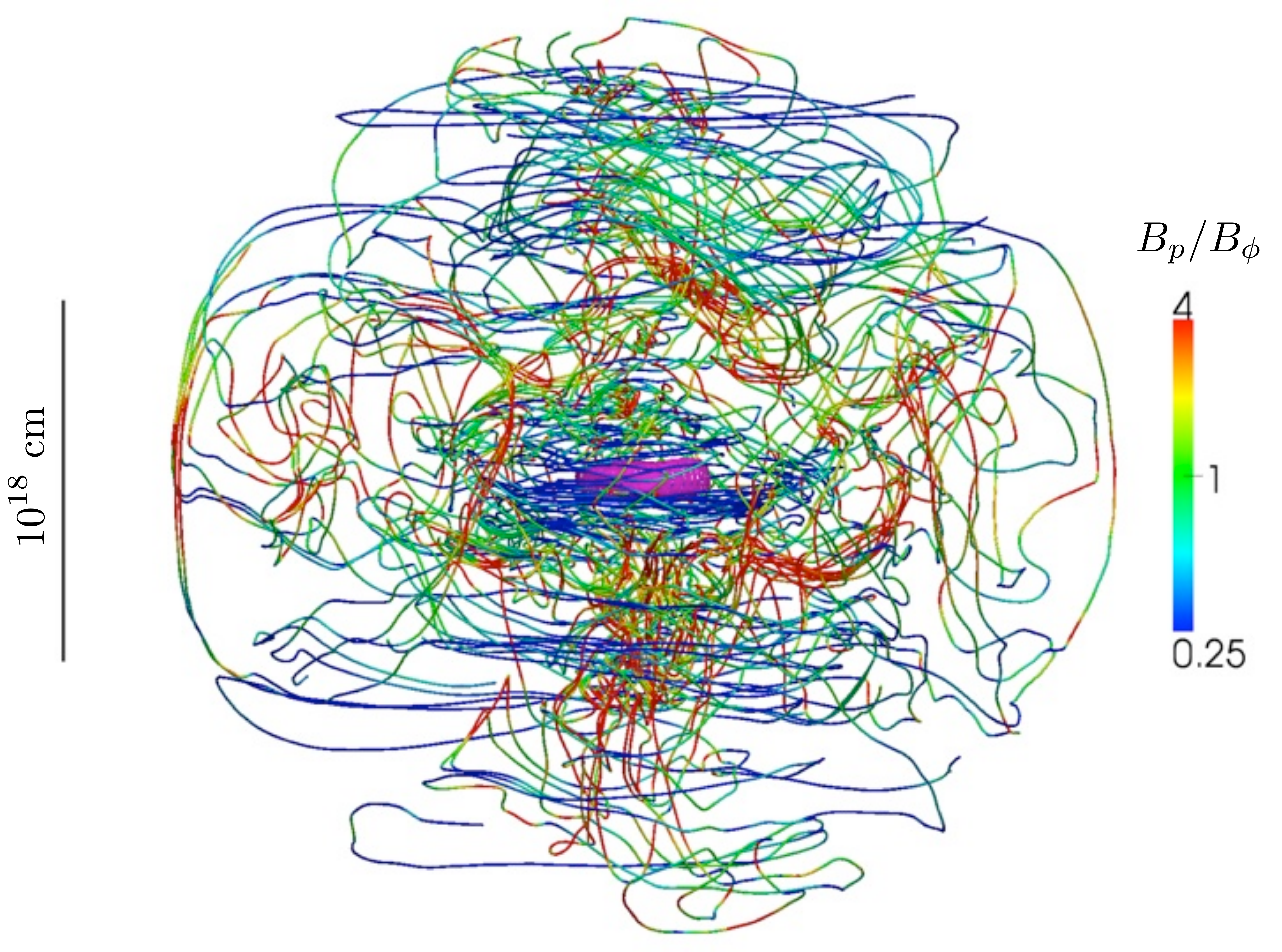}
\caption{Field lines in the high resolution run B3Dhr at
  $t=50~\rm years$.  The lines are coloured according to their 
  orientation, sections with dominating azimuthal component being blue
  and those with dominating poloidal component red. In order to trace both 
  the inner and the outer structure,   
  the seed points of the field line integration are randomly placed on two
  spheres with radii of $4\times10^{17}~{\rm cm}$ and $1.2\times10^{18}~{\rm cm}$.  
  The surface of the termination shock is also shown, using the magenta contour.}
\label{fig:fieldlines}
\end{center}
\end{figure}

\begin{figure*}
\begin{center}
\includegraphics[width=80mm]{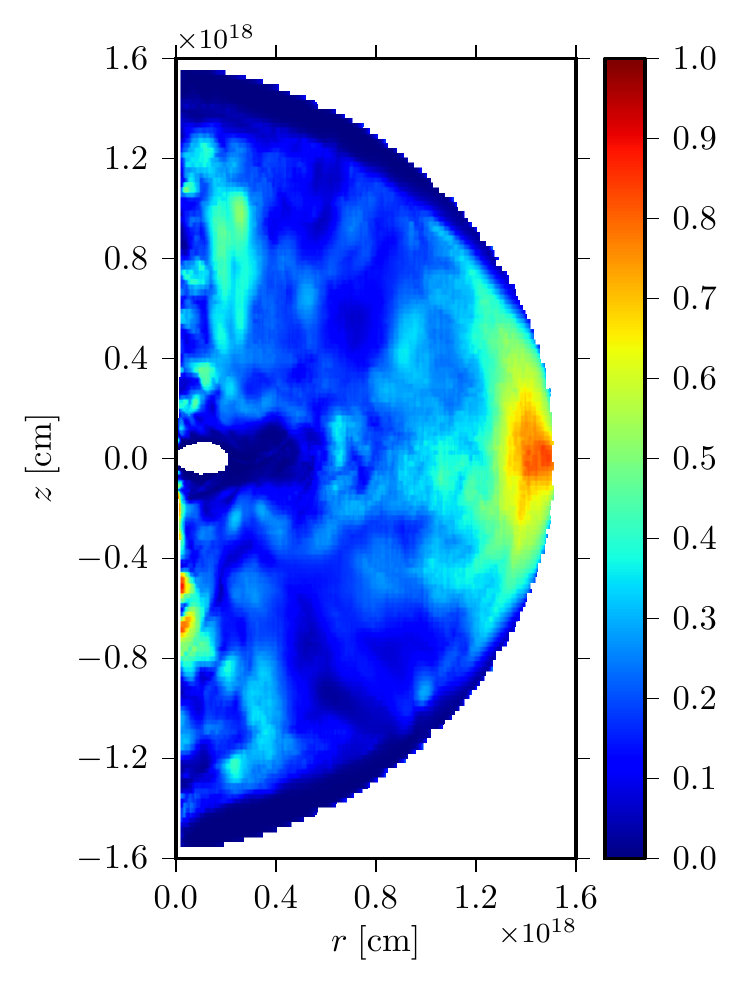}
\includegraphics[width=82mm]{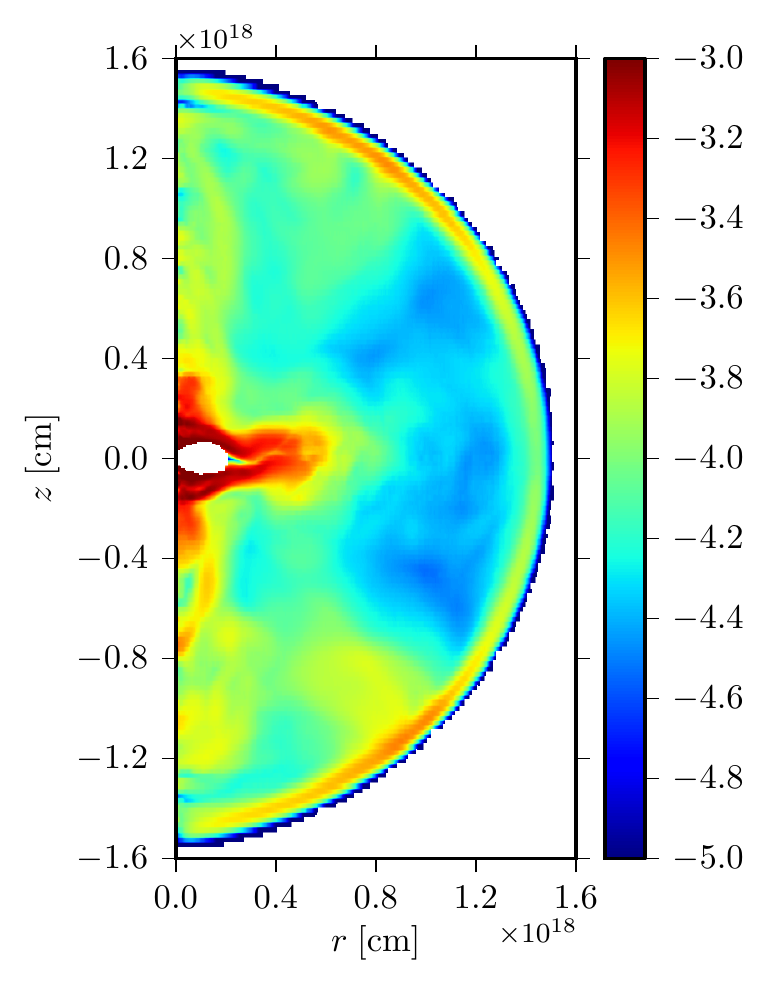}
\caption{Anisotropy of the magnetic field and its strength in the simulation run B3D. 
The left panel shows the parameter $\bar{\alpha}=\langle B_{p}^{2}/B^{2} 
\rangle_{\phi}$ at the time $t\simeq70~$ years. The right panel shows 
the angular averaged field strength, $\log\langle |\mathbf{B}/{\rm 1 Gauss}|\rangle_{\phi}$, 
at the same time.  The supernova shell and pulsar wind 
regions are not represented in the plots. The increase of the magnetic field 
strength near the outer radius of the PWN is probably an artifact. }
\label{fig:aniso}
\end{center}
\end{figure*}

Figure \ref{fig:fieldlines} illustrates the typical complex magnetic 
field structure in the 3D simulated PWN.
It is evident that with the loss of axisymmetry, the highly ordered 
structure of the magnetic field in the pulsar wind 
does not survive in the nebula where the field becomes fairly random.  
However, we can still identify regions of predominant field direction.  
For this purpose, we introduce the anisotropy parameter
\begin{align}
\bar{\alpha}=\langle B_{p}^{2}/B^{2} \rangle_{\phi}
\end{align}
where the average is taken over the azimuthal direction.  
This quantity for the simulation run B3D is shown in figure \ref{fig:aniso}.  
One can see that the azimuthal component still  dominates 
near the termination shock and that regions of predominantly 
\emph{poloidal} field arise close to the jet and in the equatorial
region next to the nebula boundary, where its magnitude is rather weak
(see the right panel of figure \ref{fig:aniso}).

The magnetic field magnitude varies substantially throughout the 
simulated nebulae. The strongest field is found just outside of 
the termination shock, where it is roughly ten times
stronger compared to the mean field in the rest of the nebula
volume (see the right panel of figure \ref{fig:aniso}).
Comparing the left and right panels of figure \ref{fig:aniso}, we find
that regions of strong field are dominated by the azimuthal component 
(with a Pearson correlation coefficient between the anisotropy
$\alpha$ and $|\mathbf{B}|$ of $-0.17$).  This explains why  
the observed degree of polarisation of the Crab nebula is so high 
near its center and why the polarization vectors suggest azimuthal 
field.  

\subsection{Jet morphology}
\label{sec:jet-morphology}

Perhaps the most interesting region in our simulations is the polar
flow that is produced due to the hoop stress of the azimuthal field 
via the so-called toothpaste effect.  This polar flow is strikingly
different in 2D and 3D models.

\begin{figure*}
\begin{center}
\includegraphics[width=0.54\textwidth]{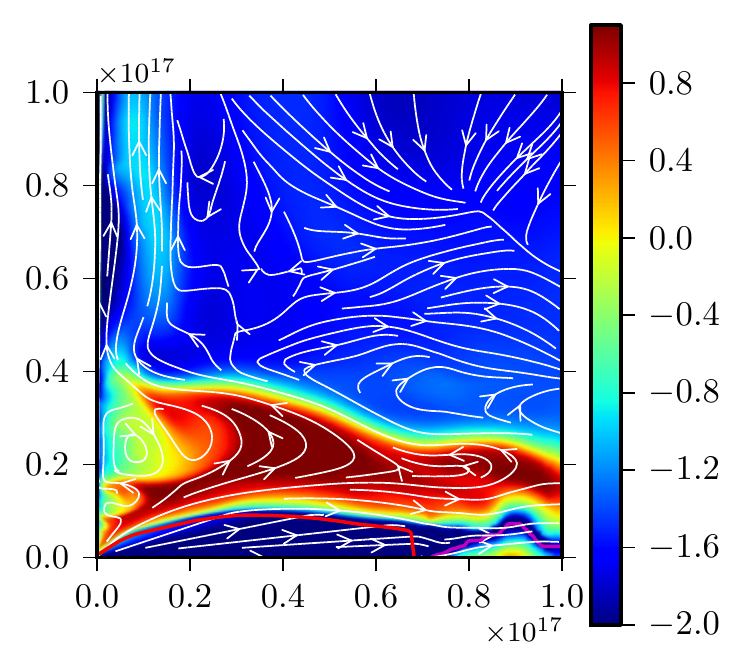}
\includegraphics[width=0.45\textwidth]{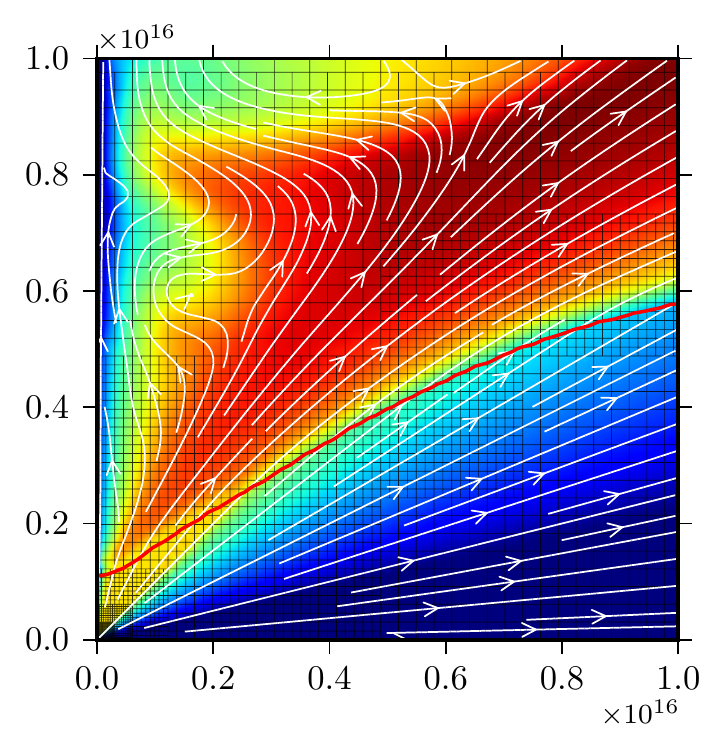}
\caption{The flow structure near the termination shock in the axisymmetric model 
B2Dvhr at the time t=100 years from the start of the simulations. 
{\it Left panel:} The colour image shows
the generalised magnetisation, $\log_{10}\sigma_{\rm s}$.  
The ``momentary'' stream lines show how the flow originating in the magnetized polar 
region becomes focussed back towards the axis to form the jet.  
One can also see the jet backflow which provides additional compression of the 
termination shock.  The red line shows the position of the termination shock 
whereas the magenta line shows the equatorial current sheet. 
{\it Right panel:} As in the left panel but for the very inner region. This 
plot also shows the computational grid.  This illustrates the grid refinement 
at the termination shock and towards the origin.  Note that the scale of this 
image is 100 times smaller than the radius of the initial nebula bubble. 
}
\label{fig:sigmaZoom}
\end{center}
\end{figure*}
 
The formation region of the polar flow in the 2D simulation run 
B2Dvhr is illustrated in figure \ref{fig:sigmaZoom}.   
To better identify flow structure we show a close-up image 
of the size of the termination shock and also zoom on the 
flow near the very origin\footnote{The latter image also illustrates the
grid refinement near the termination shock and at the origin.}. 
We find it helpful to distinguish two jet-like features: the ``polar beam'' 
and the ``plume''.  By the polar beam we understand the post termination 
shock flow that is fed by the highly magnetised unstriped polar section 
of the pulsar wind.  By the ``plume'' we understand the polar outflow 
which formed by the axial collimation of the flow originating from the 
striped section of the pulsar wind. Surely, there is no discontinuity 
separating these wind sections, and hence the division between the beam and the plume 
is not sharp. This is only a matter of scales. Hence, 
the right panel of figure~\ref{fig:sigmaZoom} shows the dynamics of the 
polar beam, whereas the left panel illustrates the formation of the 
plume. In particular, one can see the increase of the magnetisation 
in the flow away from the symmetry axis until the magnetic hoop stress
becomes strong enough to stop the expansion.         
  
The linear stability analysis of axisymmetric z-pinch configurations of 
relativistic plasma, first performed by \cite{begelman1998},  gives 
the timescale for the fastest growing mode relevant to our application  

\begin{align}
  t_g \sim \sqrt{\frac{4\pi \rho c^2}{B^2}+\frac{16\pi p}{B^2}}~
  \frac{r}{c} \, , 
\end{align}
where all quantities are measured in the co-moving frame and
$r$ denotes the cylindrical radius.
For a strong shock, the shocked flow is relativistically hot 
with $p \gg\rho c^2$ and the adiabatic index $\gamma= 4/3$.  
Hence, 

\begin{align}
t_g \sim (2 \beta)^{1/2} \frac{r}{c} \label{eq:tgbeta} \, .
\end{align}
Since, $t_g$ cannot be shorter than the light crossing time, 
this result only applies to plasma with $\beta > 1/2$.
With the low-sigma model of \cite{rees-gunn-74} in mind,
\cite{begelman1998} thus predicted that the azimuthal magnetic field 
will be disrupted at the distance  $\sim 3 r_{\rm s}$ from the 
termination shock.

\cite{lyubarsky2012} argued that the stability of 
\emph{collimating} flows is not adequately described by the
\cite{begelman1998} model.  Instead, he suggested that the polar beam 
becomes unstable at the re-collimation point 
\begin{align}
  z_{\rm c}=\frac{\pi}{\sqrt{6}}\theta_0^2 r_{\rm max} \, .
\label{eq:recollimation}
\end{align}
In the derivation, it was assumed that the initial opening angle of the beam is 
$\theta_0\ll1$.

\begin{figure*}
\begin{center}
\includegraphics[width=80mm]{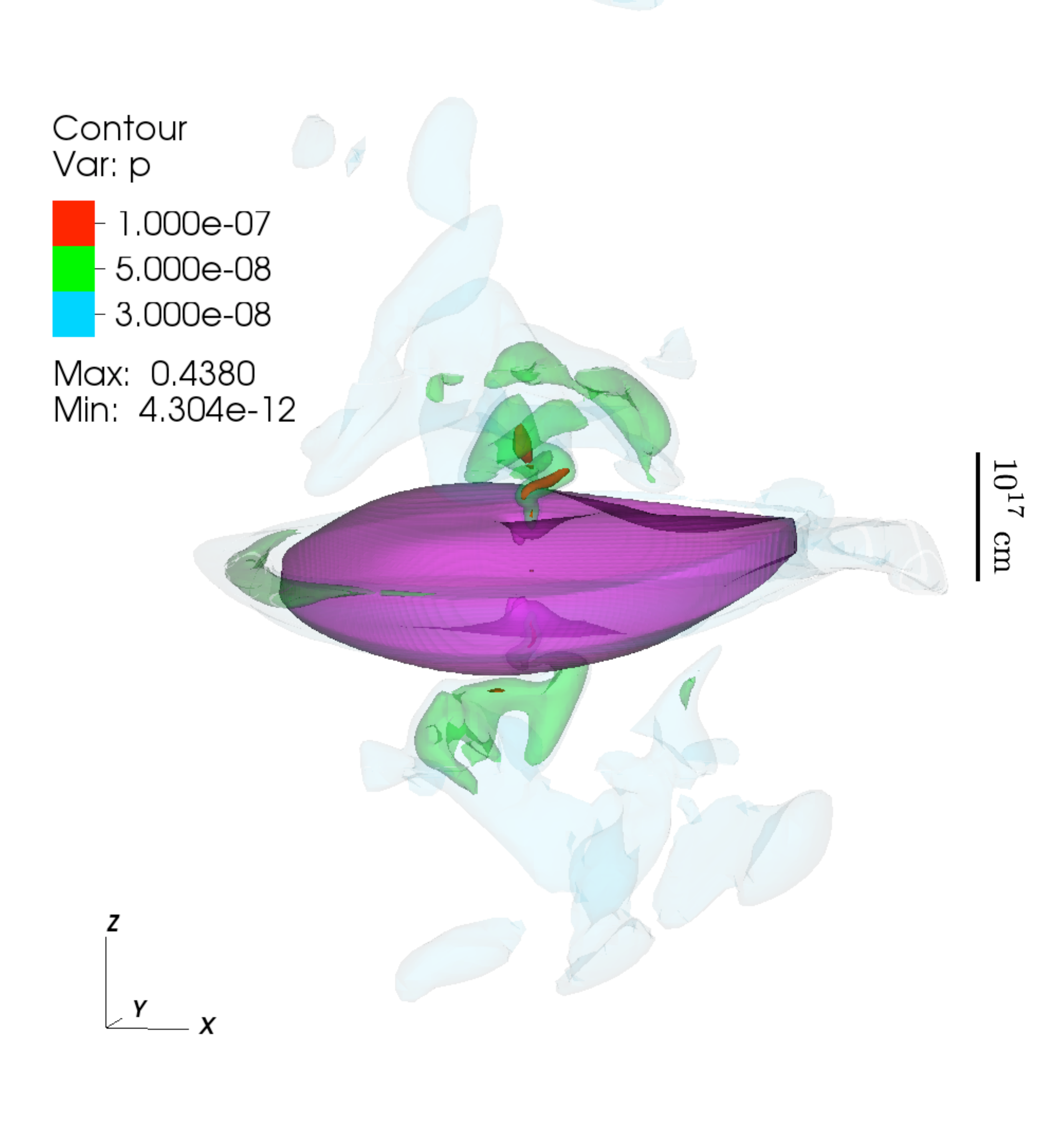}
\includegraphics[width=80mm]{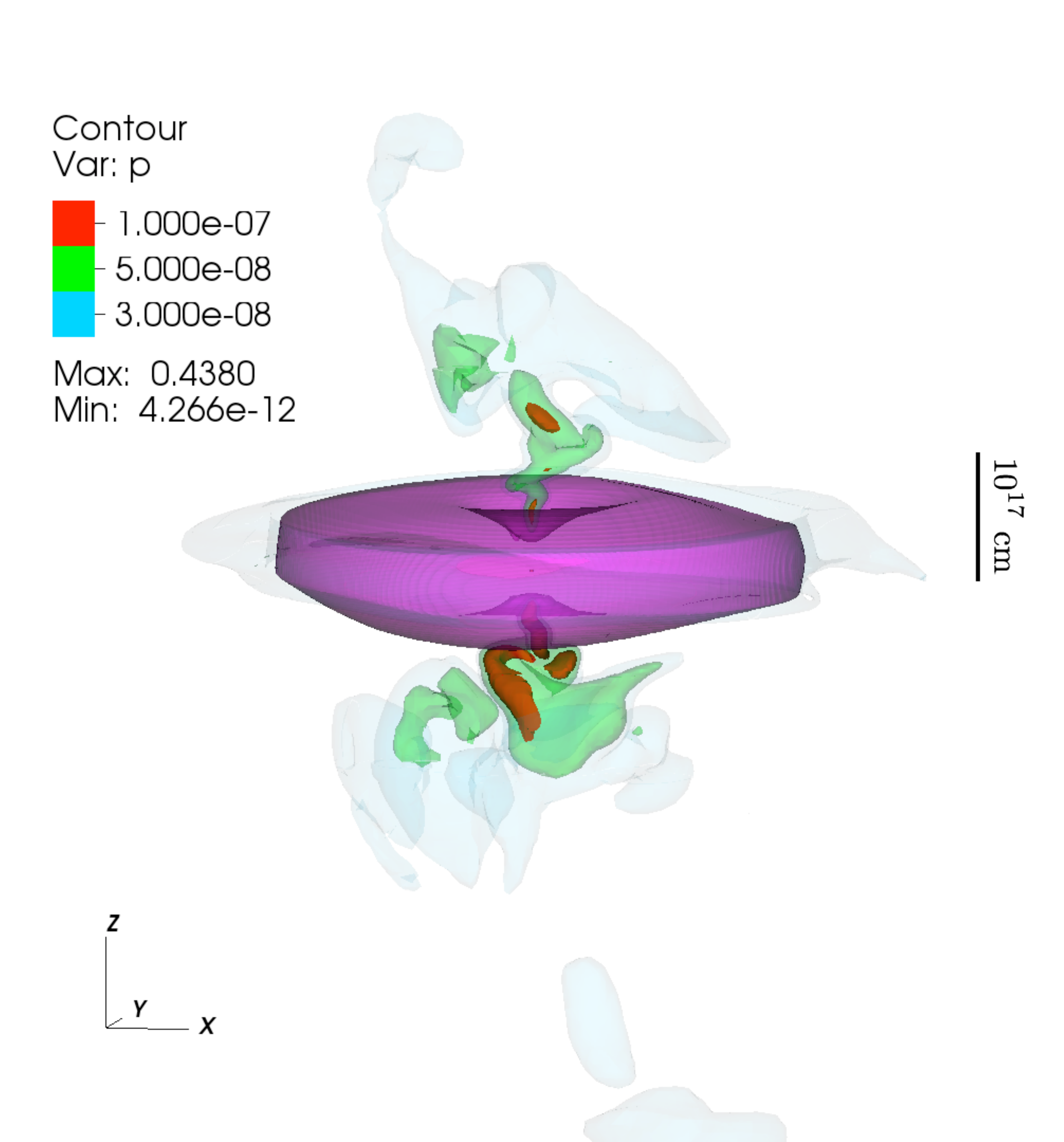}
\includegraphics[width=80mm]{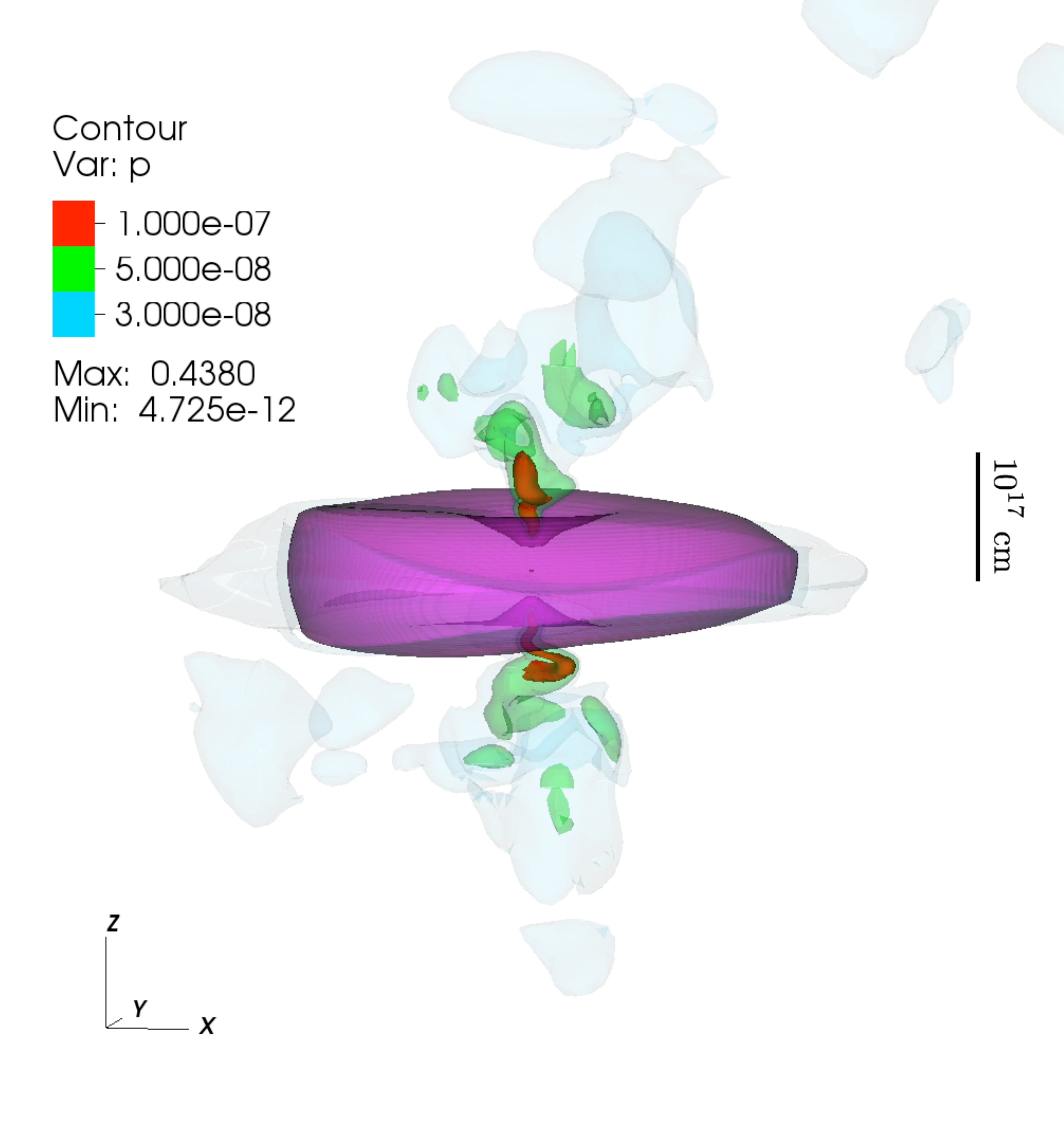}
\includegraphics[width=80mm]{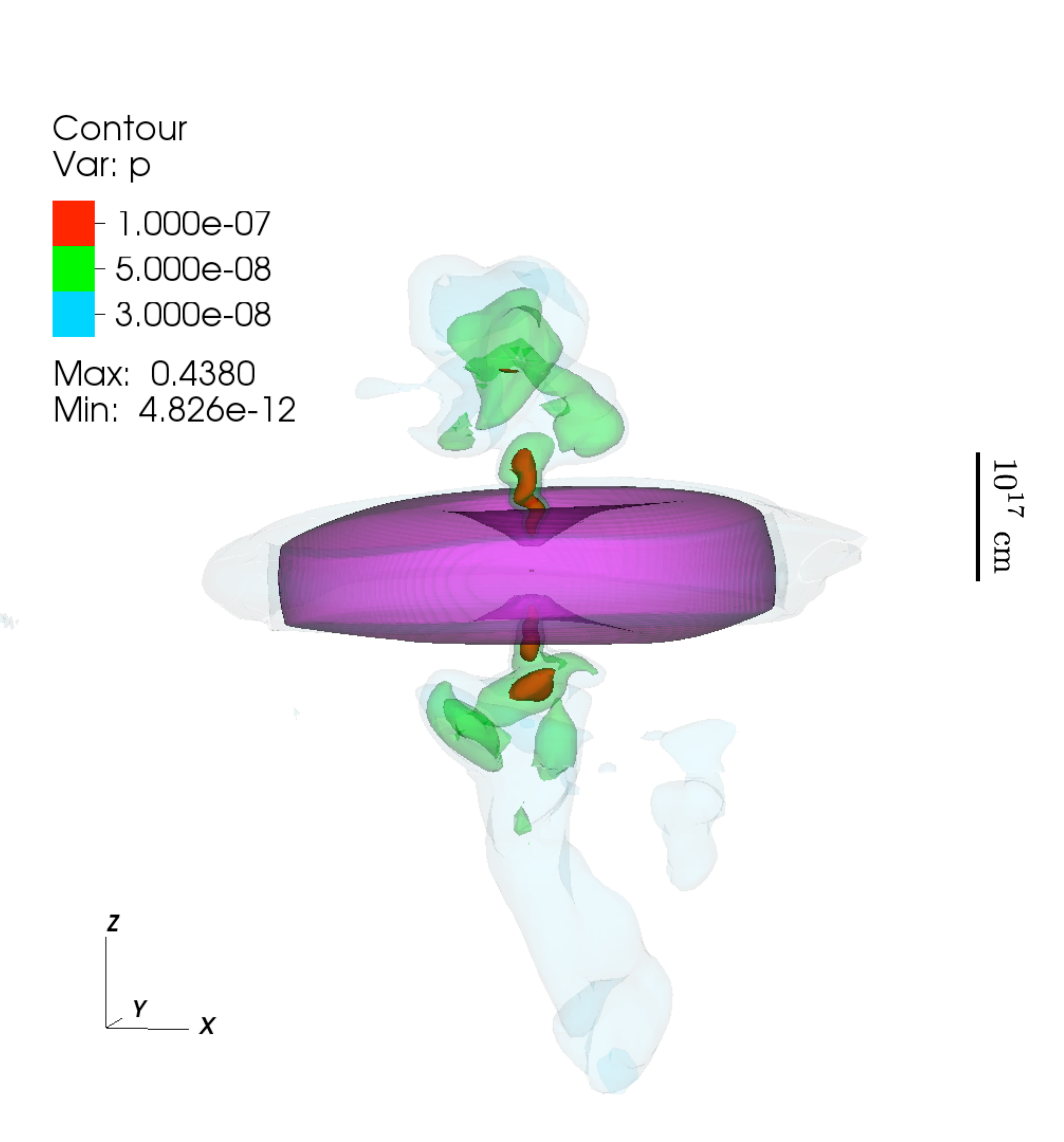}
\caption{ Termination shock and ``inner'' jet. 
  The images illustrate the simulation run B3Dhr 
  at $t=44,45,46,47$ years (left to right, top to bottom). 
  The termination shock surface is shown in magenta. The inner jet region is 
  ``illuminated'' using the level surfaces of the thermal pressure. Their 
  colour scheme is given in the plots. One can see that the polar outflow 
  is collimated into a jet with high pressure spine already inside the ``funnel'' 
  of the termination shock.  The jet exhibits corkscrew-type distortions and
  becomes disrupted not far from its origin.  
  A different representation of these snapshots is shown in figure \ref{fig:shockTime}. 
}
\label{fig:renderjet}
\end{center}
\end{figure*}

The morphology of the polar beam in our 3D simulations is shown in
figure \ref{fig:renderjet} for the high resolution run B3Dhr.
Applying equation (\ref{eq:tgbeta}) to this case, we obtain the linear
growth time scale of $\sim10^{-2}~\rm years$, much shorter than the
light crossing time of the nebula.  Given the grid resolution of 
$2.4\times10^{15}~\rm cm$ at the termination shock, the flow is thus
expected to become kink unstable after traversing only few grid
cells.  Indeed, the pressure isocontours of figure
\ref{fig:renderjet} appear heavily perturbed already on the scale of
$\sim10^{16}~\rm cm$.  The ``ridge'' of the beam resembles a corkscrew,
typical for the kink instability \citep[e.g.][]{2009ApJ...700..684M}.
The polar beam is thus highly variable and shifts orientation on less
than monthly timescales. This may be relevant to the rapid variability 
of the ``sprite'' feature, which is located at the base of the Crab's X-ray 
jet \citep{hester2002}.

\begin{figure*}
\begin{center}
\includegraphics[width=80mm]{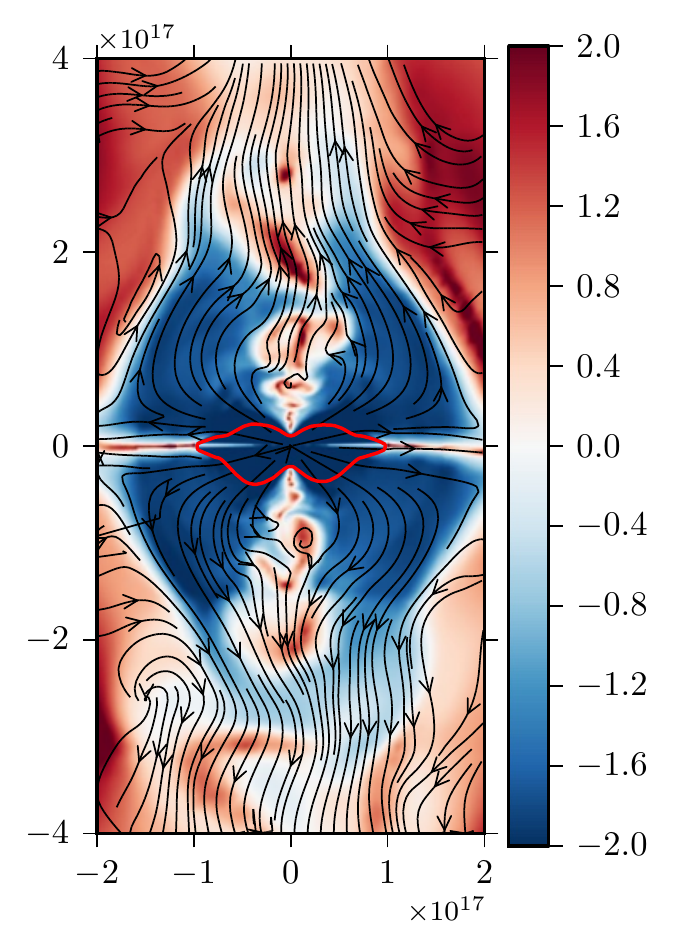}
\includegraphics[width=80mm]{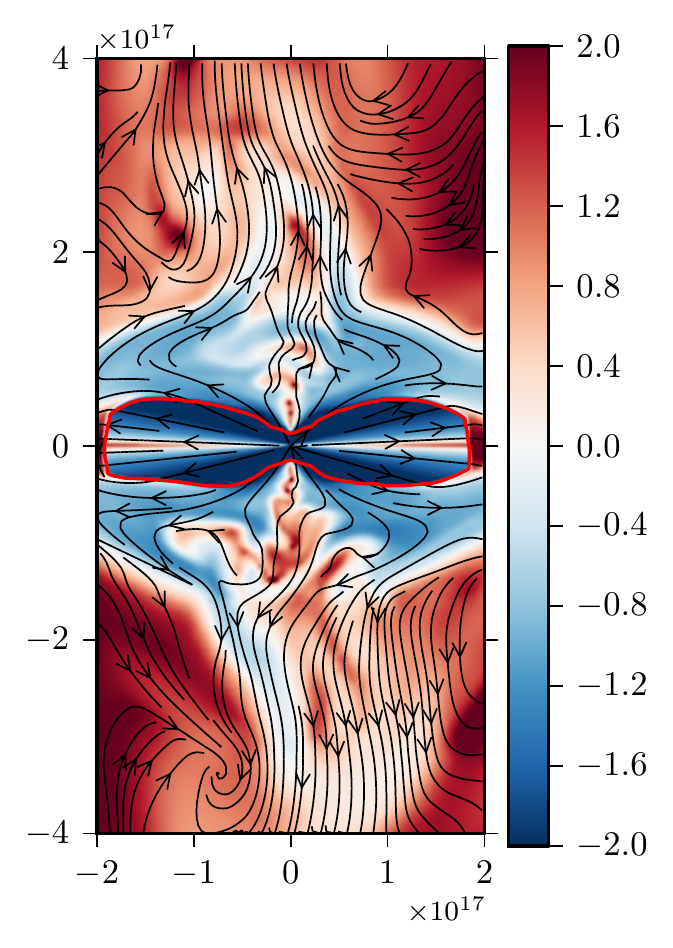}
\caption{Formation of the polar plume. 
  The left panel shows the data for the simulation run D3D ($\alpha=10^\circ$) 
and the right panel for C3D ($\alpha=45^\circ$). In both cases $\sigma_0=3$ 
and $t=50~$years. The colour images show the distribution of 
$\log_{10}\beta$ in the $yz$-plane, the black lines the momentary streamlines, 
and the red line the termination shock. The streamlines clearly indicate 
gradual collimation of the flow around the highly dissipative axial region.  
A coherent plume, roughly in the axial direction, is obtained after approximately 
one or two shock radii, depending on the obliqueness of the pulsar wind.  }
\label{fig:betaSlice}
\end{center}
\end{figure*}

If indeed the disruption of the beam follows its re-collimation then 
its maximal propagation length should be influenced by the opening 
angle of the unstriped region of the pulsar wind (see equation 
(\ref{eq:recollimation}).  We investigate
this by comparing the simulation C3D featuring $\alpha=45^\circ$
with the simulation D3D which has $\alpha=10^\circ$.  Both these 
runs are highly magnetised in the polar region ($\sigma_0=3$).  
Figure~\ref{fig:betaSlice} shows $\log_{10}\beta$ for these two cases, 
in the $x=0$ plane, as well as the momentary stream lines 
of the velocity vector field (black lines) and the
termination shock (red contour).  From these images, it
becomes clear that there is no precise ``recollimation point'' -- 
each flow line tends to hit the axis at a different point, 
more distant for higher initial opening angle. 
In both cases, the polar beam becomes rapidly unstable leading to a 
cork-screw with an opening angle of roughly $10^\circ$.  
As to the differences between these solutions, one can see that 
the D3D model has a somewhat larger magnetically dominated central 
region, especially when compared to the equatorial radius of the shock. 
The magnetisation of this region is higher too. Finally, the shape of 
its termination shock resembles a bird's beak, as its radius 
increases rapidly after entering the narrow striped wind zone 
(figure \ref{fig:betaSlice}).

\begin{figure*}
\begin{center}
\includegraphics[width=85mm]{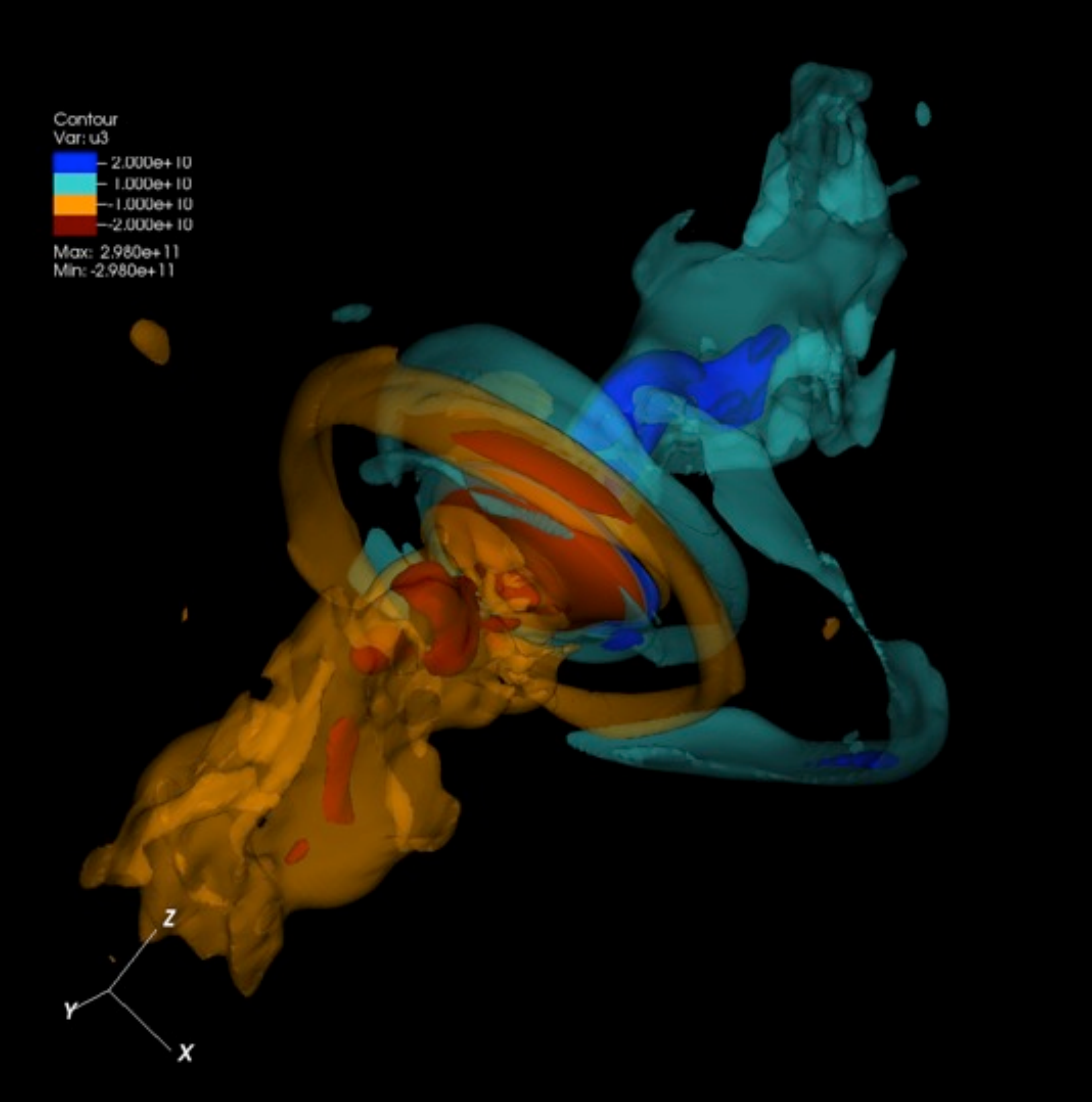}
\includegraphics[width=85mm]{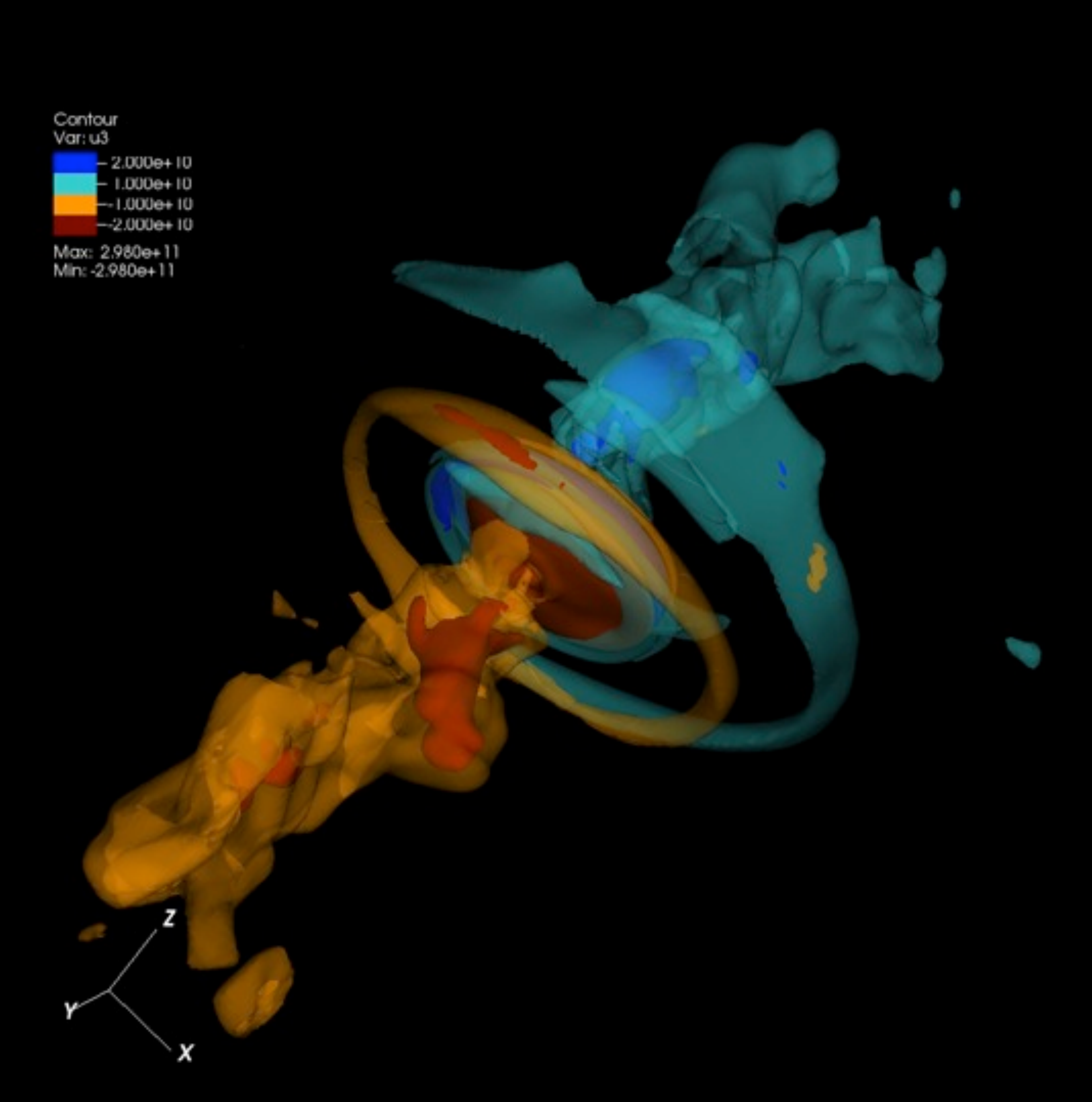}
\includegraphics[width=85mm]{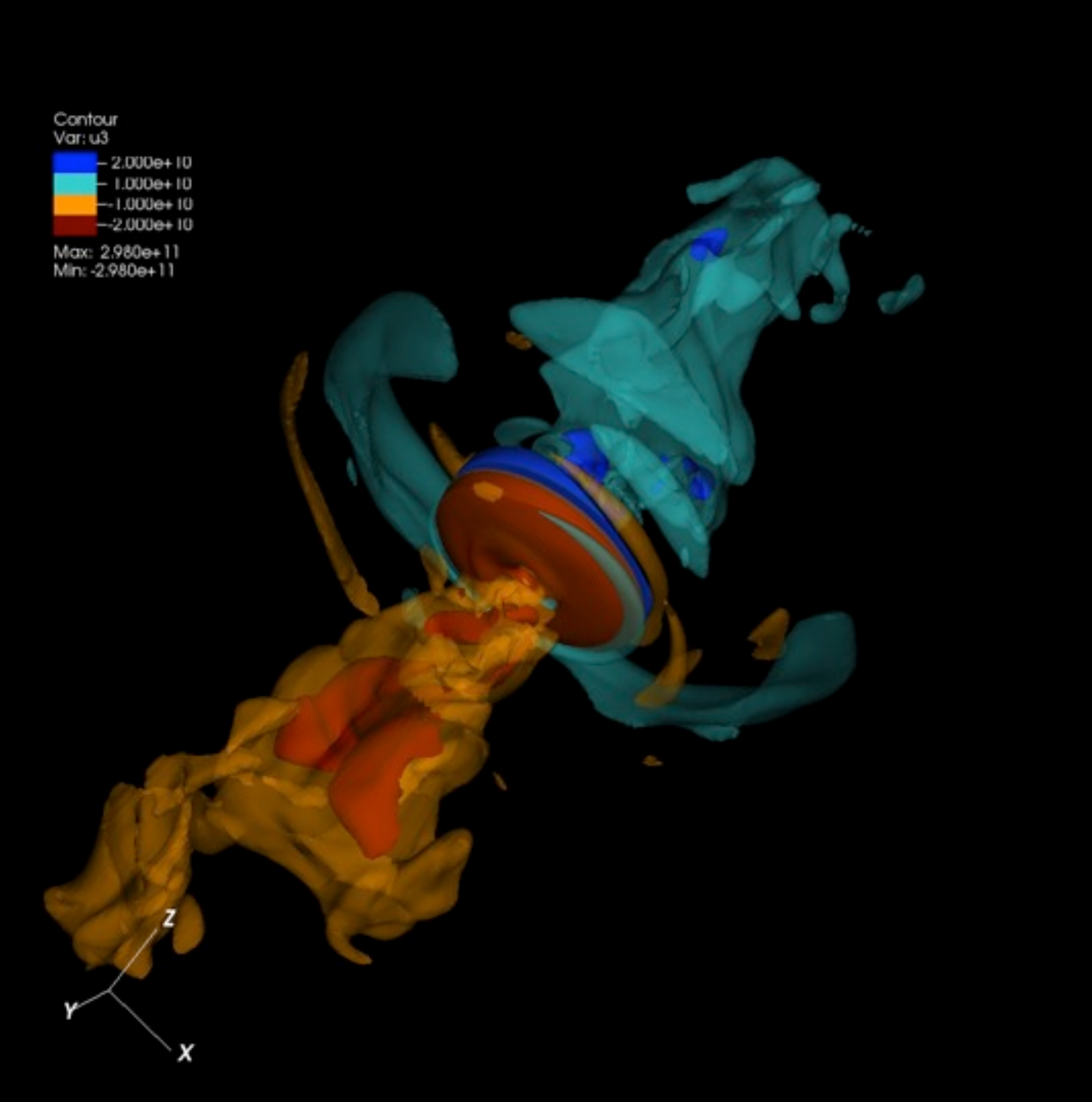}
\includegraphics[width=85mm]{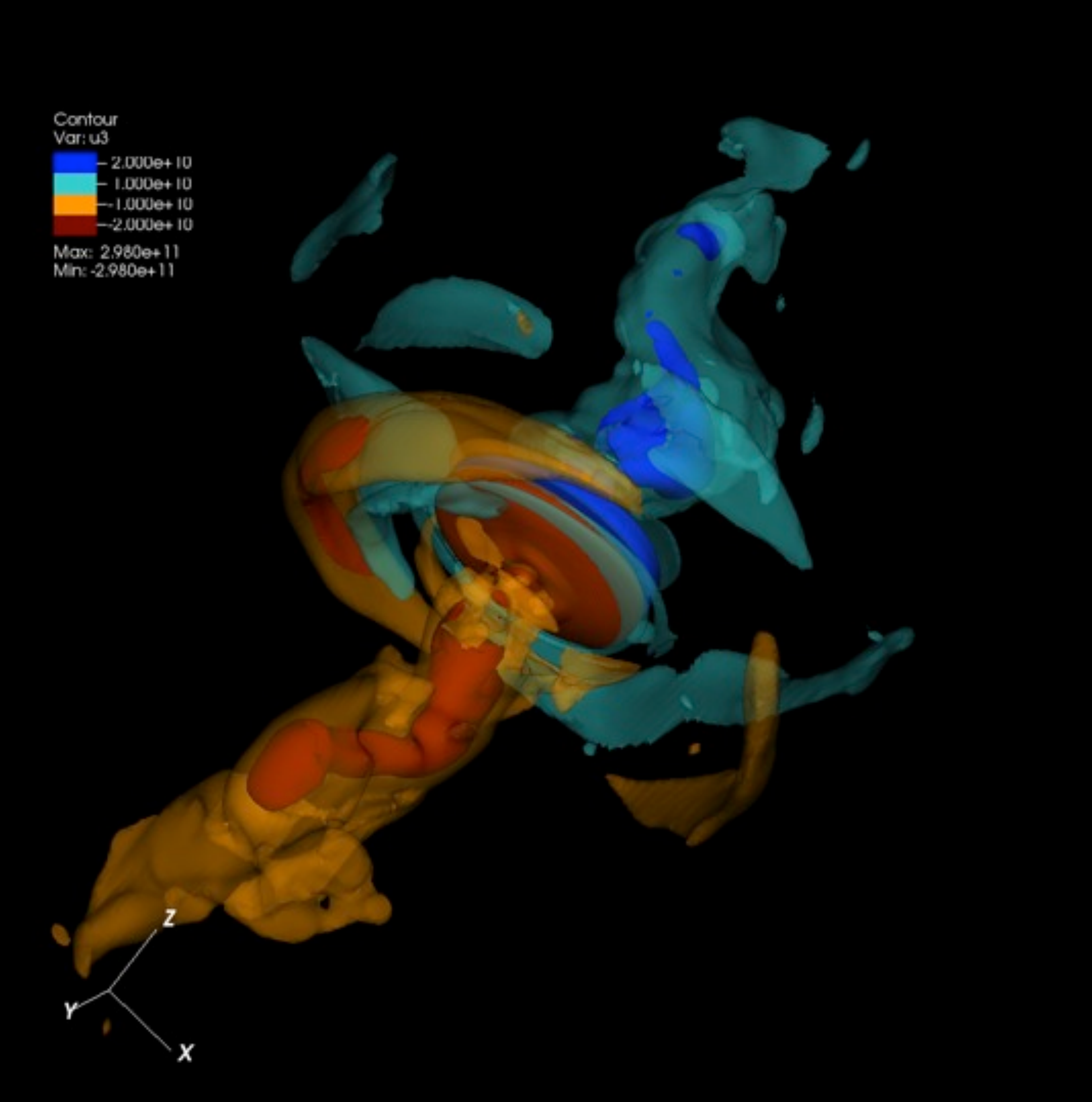}
\caption{High velocity flow of inner PWN. Once again we show a 3D rendering 
of the data for the run B3Dhr at $t=44,45,46,47$ years (left to right, top to
bottom). Now the level surfaces of $u_z$ are used to ``illuminate'' the regions 
of high speed flow inside the simulated PWN. The level values are  
$u_z=(-2/3c,-1/3c,1/3c,2/3c)$. Twin jets (plumes) form downstream of the
  termination shock due to the ``toothpaste effect''.  These jets remain
  coherent for several termination shock radii until they fragment and
  dissolve into the nebula.  The right-hand panels indicate helical
  distortions, characteristic of the kink instability.  }
\label{fig:plume}
\end{center}
\end{figure*}

The plume is formed via axial collimation of the flow lines 
originating in the striped wind zone. 
In the otherwise turbulent nebula body, this ``plume'' sustains a
directed flow for at least $\sim 6$ shock radii until it eventually
fragments and dissolves (see figure~\ref{fig:plume}). 
{\bf
It can visually be identified in iso-contours of velocity with $|u_z|=1/3c$ surrounding a faster spine that occasionally exceeds $0.7c$.
This is consistent with observations of Vela and Crab, supporting a pattern speed of $0.3-0.7 ~c$ for Vela \citep{PavlovTeter2003} and $\sim0.4c$ in the case of Crab \citep{hester2002}.  
}
In the simulations, we observe signs of kinking
and fragmentation of the plume with an over-all good resemblance to
the Crab jet as well.  

Recently, \cite{DurantKargaltsev2013} showed that the Vela pulsar jet
can be modelled as a helix, steadily turning on the timescale of 400 days.
To this we comment that while the helical distortions due to the kink
instability can be identified in individual snapshots of our
simulations, we find that this structure is typically disrupted in
less than a year.  The helix eventually re-grows to be disrupted again 
shortly thereafter.  The complex dynamics of the PWN makes it
hard to identify a comprehensive mechanism or timescale.  This process
can be observed in the second panel of figure \ref{fig:plume}, which shows
the approaching helically deformed jet just before its disruption; two
years later (last panel of figure \ref{fig:plume}) a new helix is
formed.  In the snapshot taken at $t=48$ years, not presented in 
the figure, it is dissolved again.

\section{Synthetic Synchrotron Maps}\label{sec:synchrotron}

\subsection{Particle injection prescriptions}\label{sec:recipes}

\begin{figure*}
\begin{center}
\includegraphics[width=80mm]{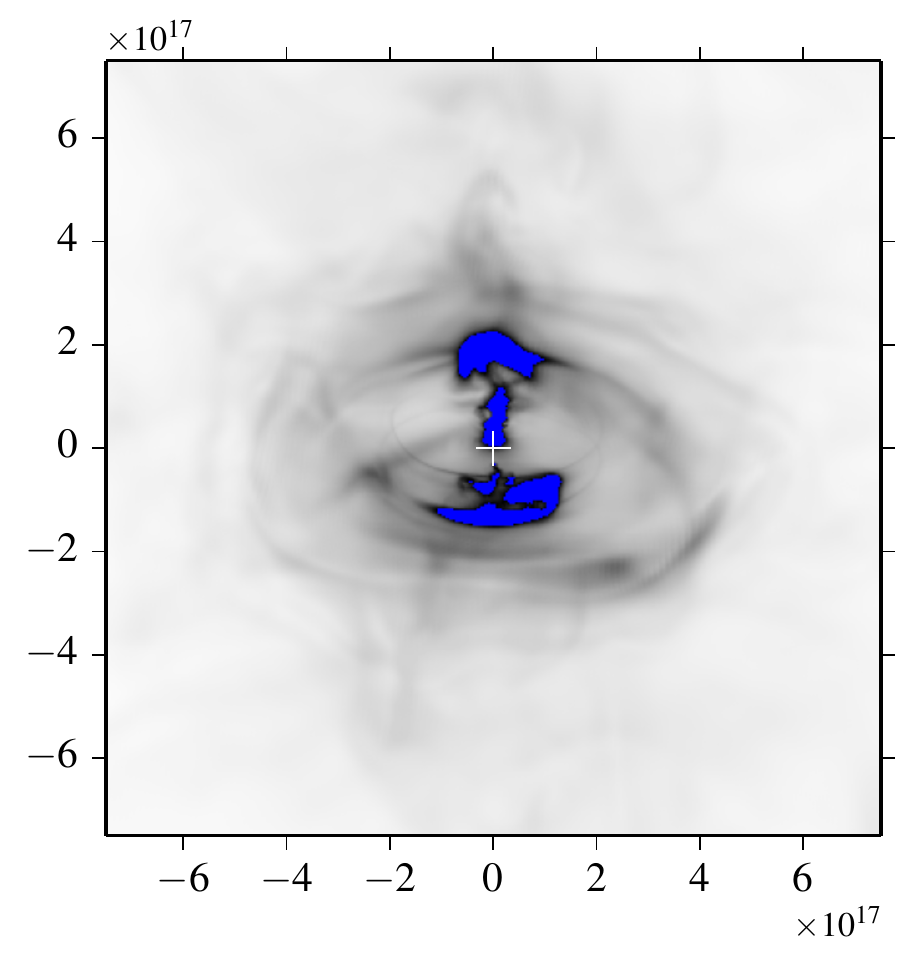}
\includegraphics[width=80mm]{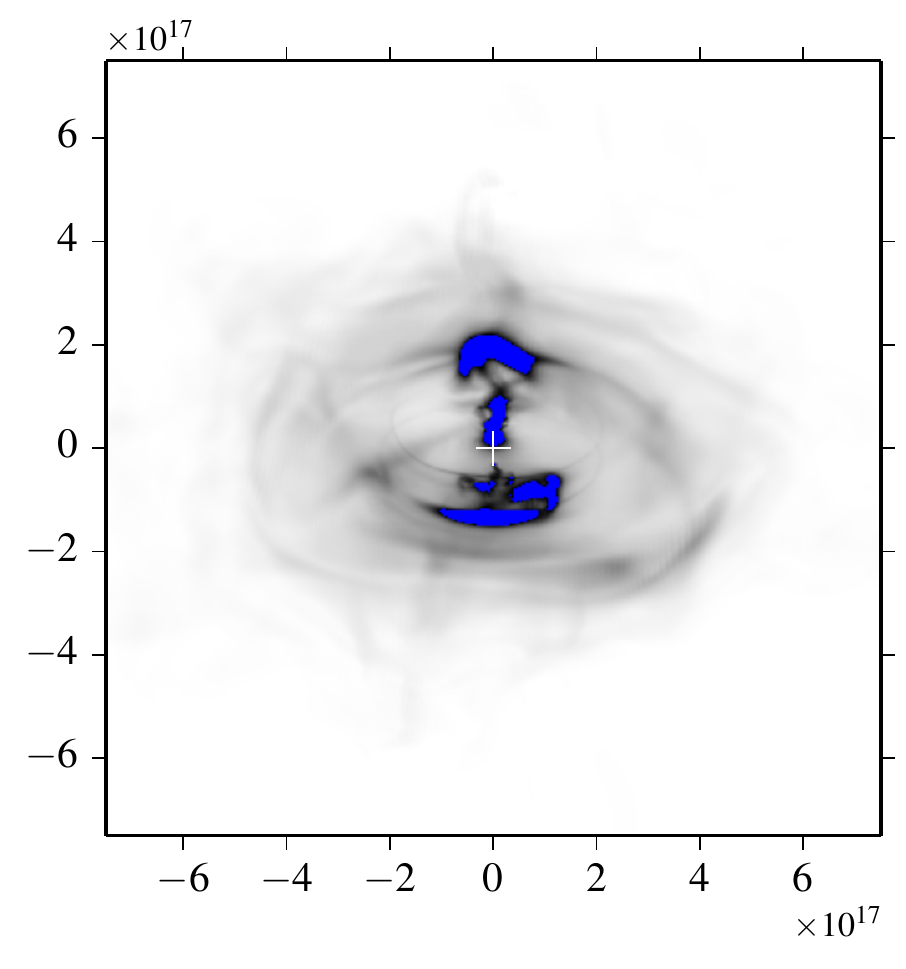}
\includegraphics[width=80mm]{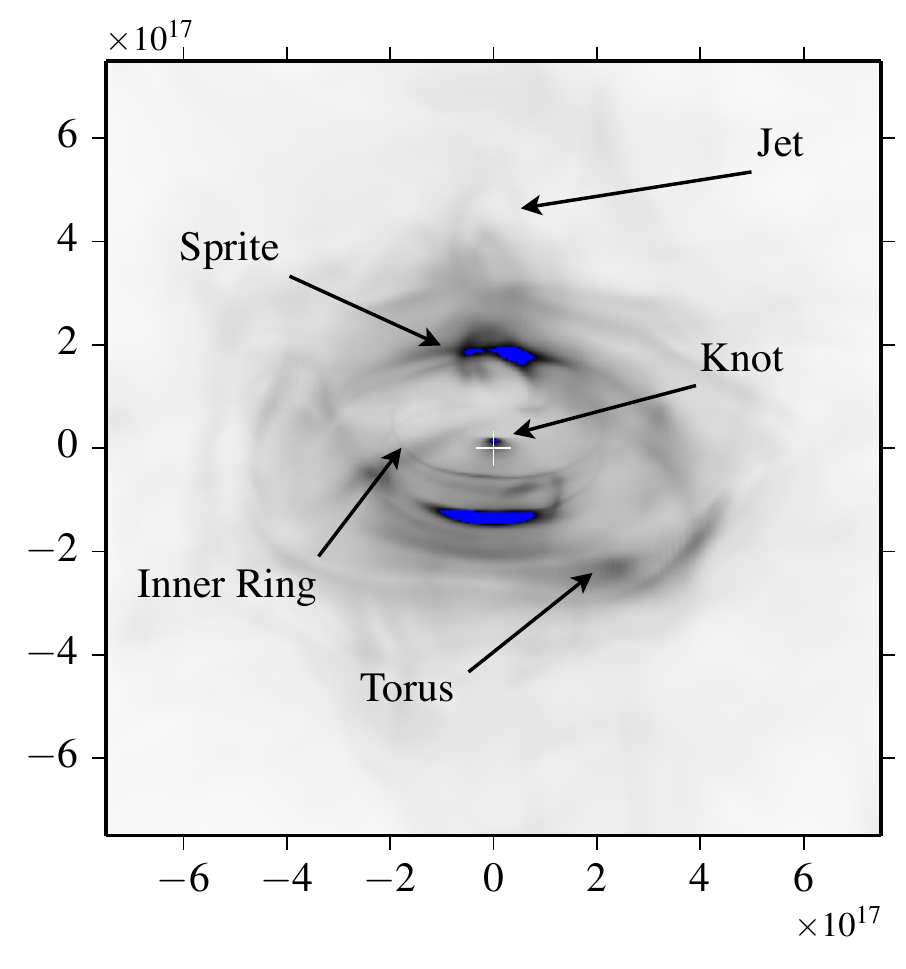}
\includegraphics[width=80mm]{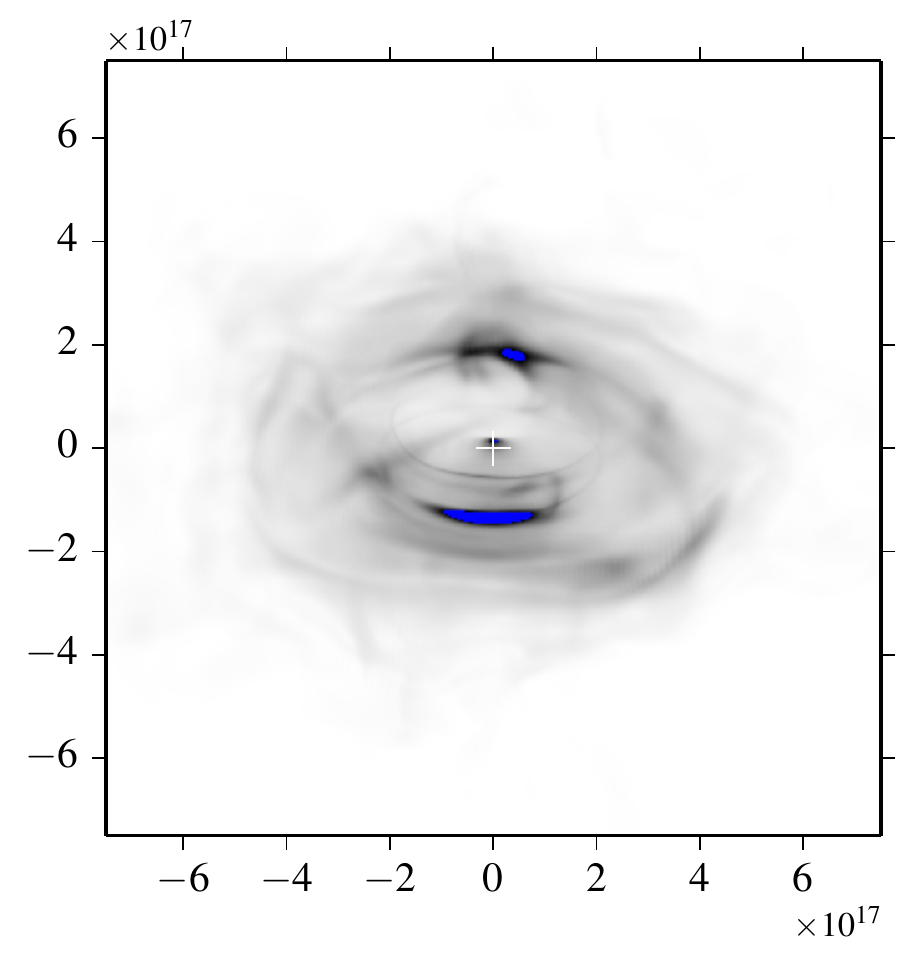}
\includegraphics[width=80mm]{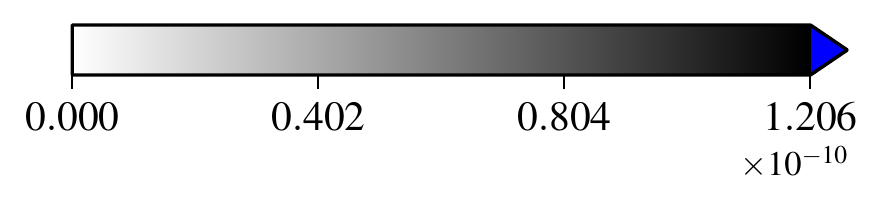}
\includegraphics[width=80mm]{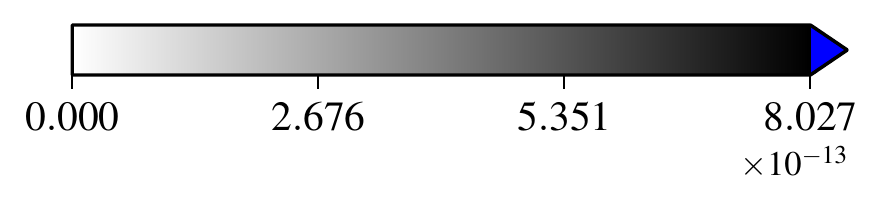}
\caption{Optical and X-ray synthetic synchrotron images of inner PWN. The images 
are produced for the simulation run B3Dhr. 
Left panels show the optical images ($\nu=10^{15}$Hz) and the right panels 
show the X-ray images ($h\nu=1$ keV). The top panels show the results based on 
the recipe A for the injection of relativistic particles and the bottom panels show 
the results based on the recipe B. {\bf In all these plots the brightness is shown using a 
linear scale up to $1/10$ of the maximum intensity of both recipes}.  
The location of the pulsar is marked with white ``+'' and saturated pixels are indicated blue. The 
recipe A gives rise to excessively bright sprite  and inner jet. 
No clear polar plume is seen in any of these maps. 
}
\label{fig:recipes}
\end{center}
\end{figure*}

We now compare the images obtained with different considerations for
the particle injection (section \ref{sec:initialization}).  To adopt a
view similar to the Crab Nebula, we choose the viewing angle measured
from the pole as $60^\circ${\bf, the rotation axis thus tilts by $30^\circ$ out of the plane of the sky}.  The assumption that the relativistic electrons and
positrons are injected isotropically at the nebula termination shock
(recipe A) results in the top optical image of figure~\ref{fig:recipes}.  
It exhibits a very bright jet-like feature, which is not present in the 
maps of the Crab nebula,  between the pulsar and the \emph{sprite}. 
The sprite itself and the opposing part of the torus are also excessively 
bright.
 
When the relativistic particles are injected only in the
striped wind section of the termination shock (the recipe B), the synthetic
images are much more similar to that of the Crab Nebula.  The bright inner 
jet-like feature is absent and the emission from the sprite is significantly 
reduced. The sprite is produced by those streamlines of the striped-wind that 
first  converge back to the axis and thus transport emitting particles 
to the base of the polar plume. As to the plume itself, 
while this structure is certainly present in the velocity plots,  
it does not clearly come out in the synthetic
images.  ``Lighting up'' the jet might require additional cycles of
particle acceleration, perhaps in the unstable inner region at the
base of the plume as ventured by \cite{lyubarsky2012}.

We indicate the locations of the various features mentioned previously
in the lower left panel of figure \ref{fig:recipes}.  {\bf The right panels of 
figure~\ref{fig:recipes} show the synthetic X-ray images in the Chandra band.  
Due to the synchrotron burn-off, X-ray photons originate only from the torus region.}
Both prescriptions for particle
injection give rise to the \emph{torus} morphology, but have
difficulties in reproducing the X-ray \emph{jet} of the Crab Nebula.   
{\bf Overall we find the prescription B more suitable and all the 
emission data in the rest of the paper are obtained using this prescription.
}

\subsection{Wisps}

\begin{figure}
\begin{center}
\includegraphics[width=80mm]{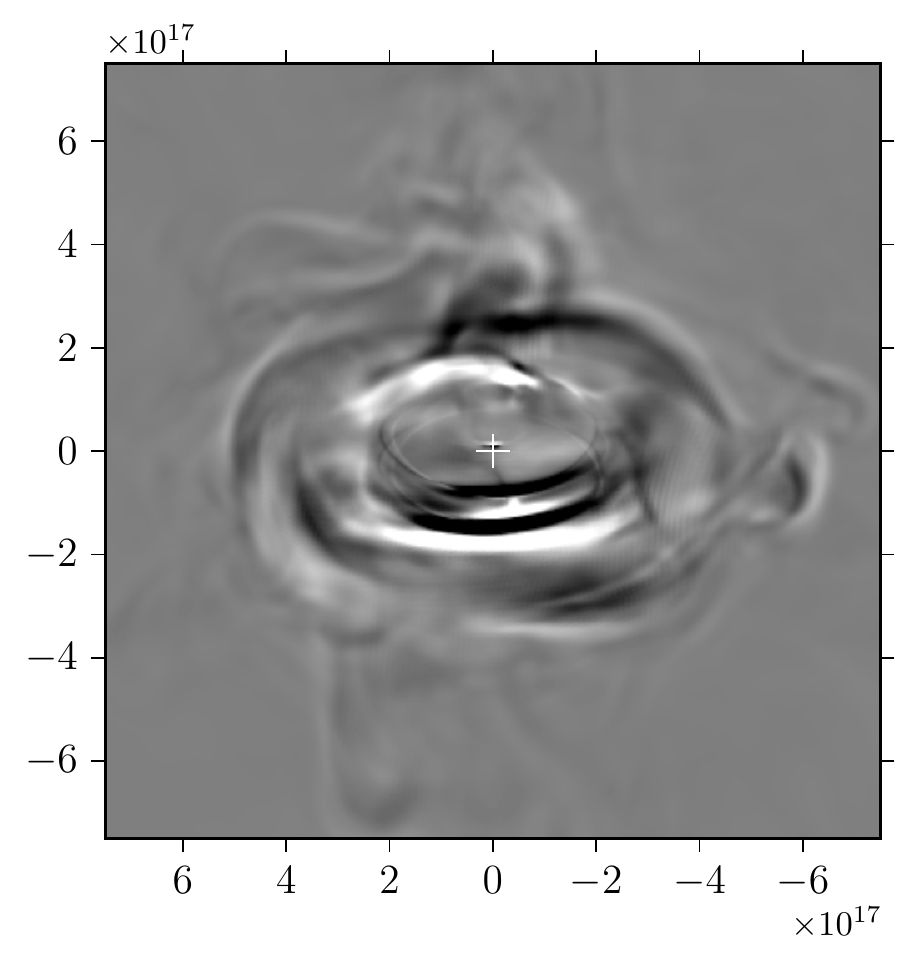}
\caption{ Moving wisps. The plot show the difference between two 
optical images separated by $\sim110$ days, thus  
highlighting the variability of the emission of the inner PWN. 
In particular, the plot shows that the wisps move outwards (with an apparent 
velocity of $\sim1/3 c$).  One can also see that the inner knot has shifted 
slightly away from the origin.  A fair amount of variability is also seen in the
sprite.  
}
\label{fig:difference}
\end{center}
\end{figure}

Using the prescription B, as more suitable,  
we now focus on the production of concentric wisps
near the termination shock in the high resolution run B3Dhr.  The
motion of the azimuthal wisps can be visualised by subtracting two
subsequent images, as done for example by \cite{hester2002} using
Hubble and Chandra imaging of the Crab nebula.  Outward moving wisps
are thus indicated as light arches followed by the corresponding
darker features on the inside (figure ~\ref{fig:difference}).  In
accordance with \cite{hester2002} and \cite{camus2009}, we space our
synthetic observations by $110$ days.  The observed features move out
with the speed of $\sim 1/3\rm c$, roughly in agreement with
the observations \citep{hester2002}.  However, likely due to the lower spatial
resolution in our 3D runs compared to that of \cite{camus2009}, we can
identify only two distinct wisps.  
Figure~\ref{fig:difference} also indicates a  
fair amount of variability is also present in the sprite, and shows a change
in the position of the inner knot.

In the \emph{online material}, we show a synthetic ``Hubble movie'' produced 
from ten snapshots taken around $t=50$ years and separated in time 
by $\Delta t = 0.1 ~\rm years$ for the simulation run B3Dhr.  
It clearly demonstrates the outward motion of the wisps and the variability 
of the knot.

\subsection{Inner knot}\label{sec:knot}

Figure~\ref{fig:recipes} shows a knot-like feature, which is located 
only slightly off the pulsar position in 
the vertical direction.  In the Crab nebula, a similar 
knot was first discovered in the high resolution Hubble observations
by \cite{hester1995}.  Its dynamical origin was explained by
\cite{ssk-lyub-03} as Doppler-boosted emission of the plasma just outside of 
the oblique section of the termination shock.  
The knot is an especially interesting feature of
the PWN as it might be the origin of most of the gamma-rays
observed from the Crab Nebula \citep{komissarov2011}.  \cite{MoranShearer2013}
showed that while the intensity of the knot may vary significantly
with time, the optical polarisation signal remains remarkably stable
and high (with a degree of $0.59\pm0.019$).

\begin{figure*}
\begin{center}
\includegraphics[width=40mm]{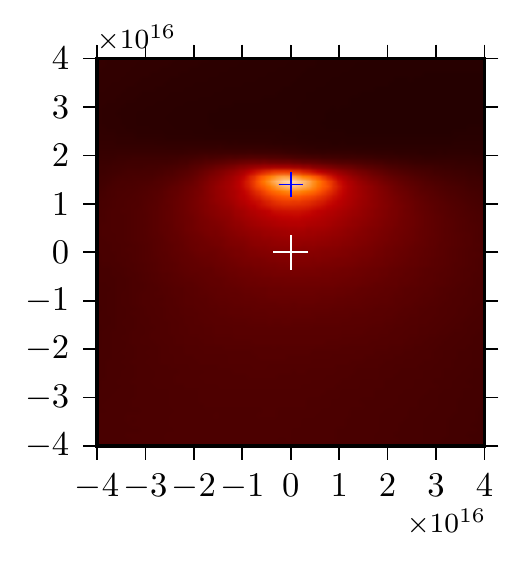}
\includegraphics[width=40mm]{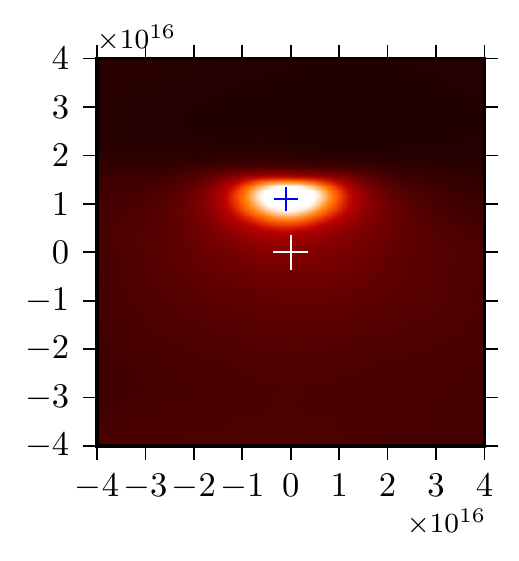}
\includegraphics[width=40mm]{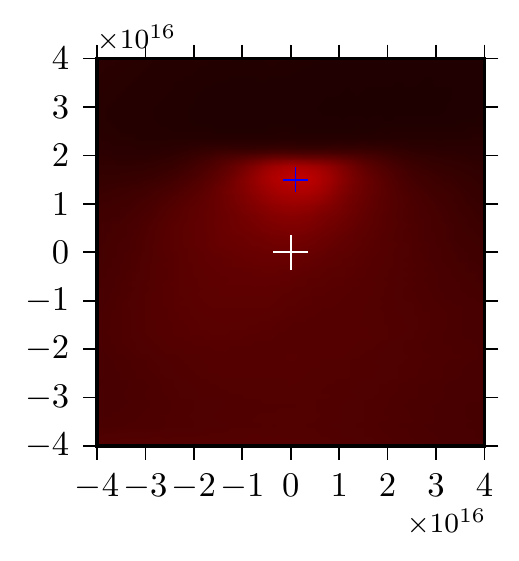}
\includegraphics[width=40mm]{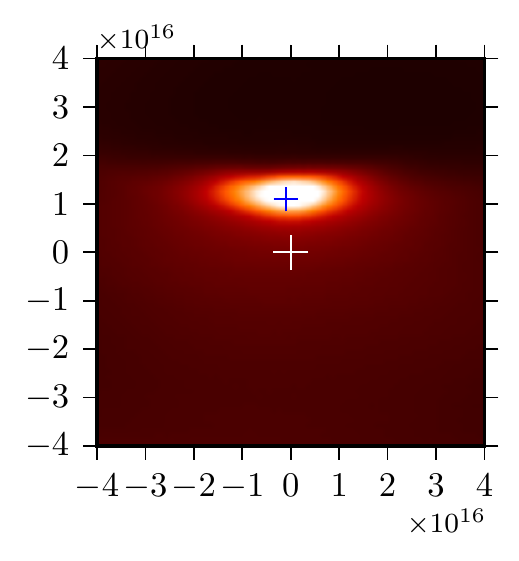}
\includegraphics[width=80mm]{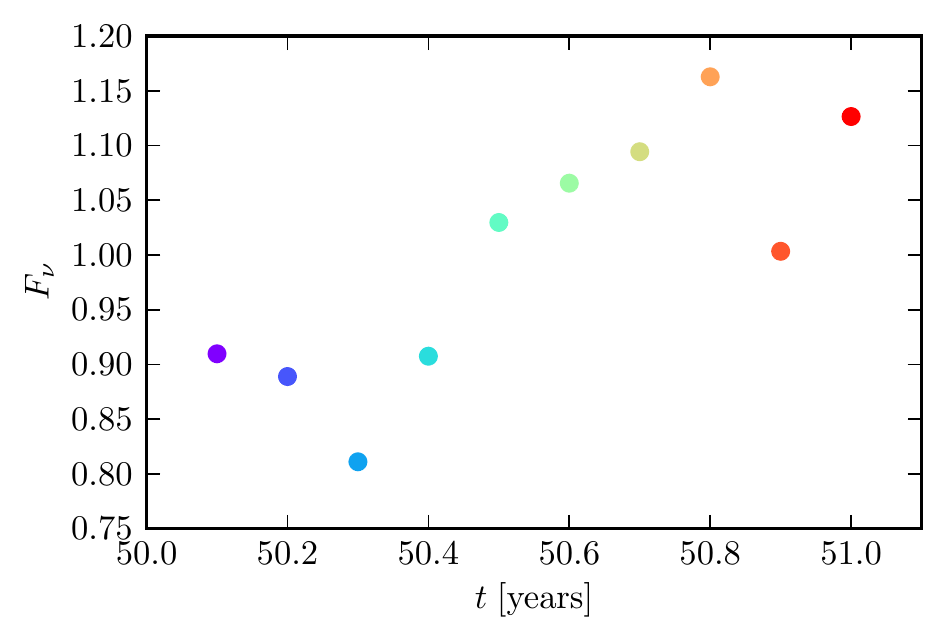}
\includegraphics[width=80mm]{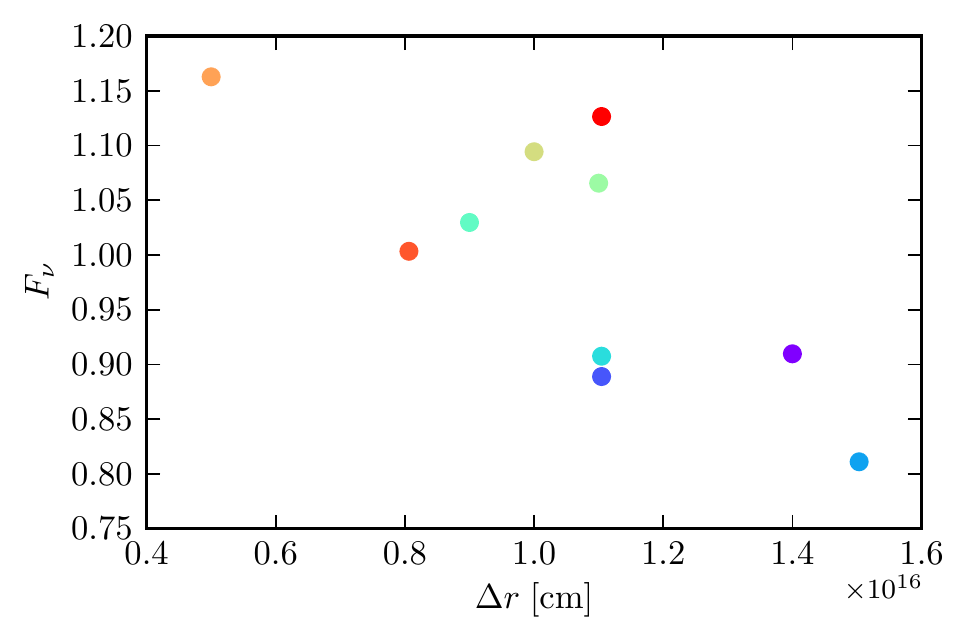}
\includegraphics[width=80mm]{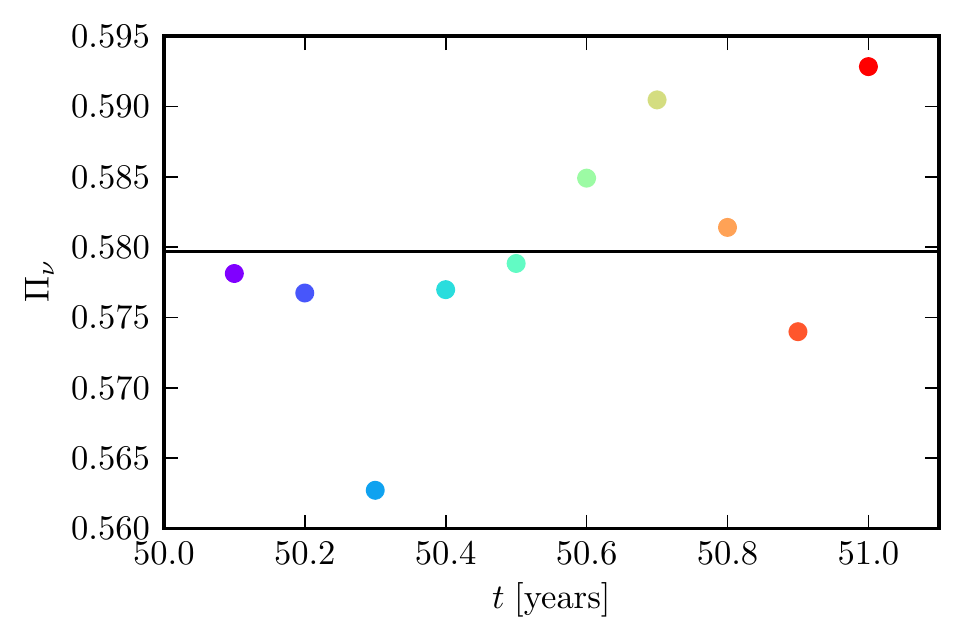}
\includegraphics[width=80mm]{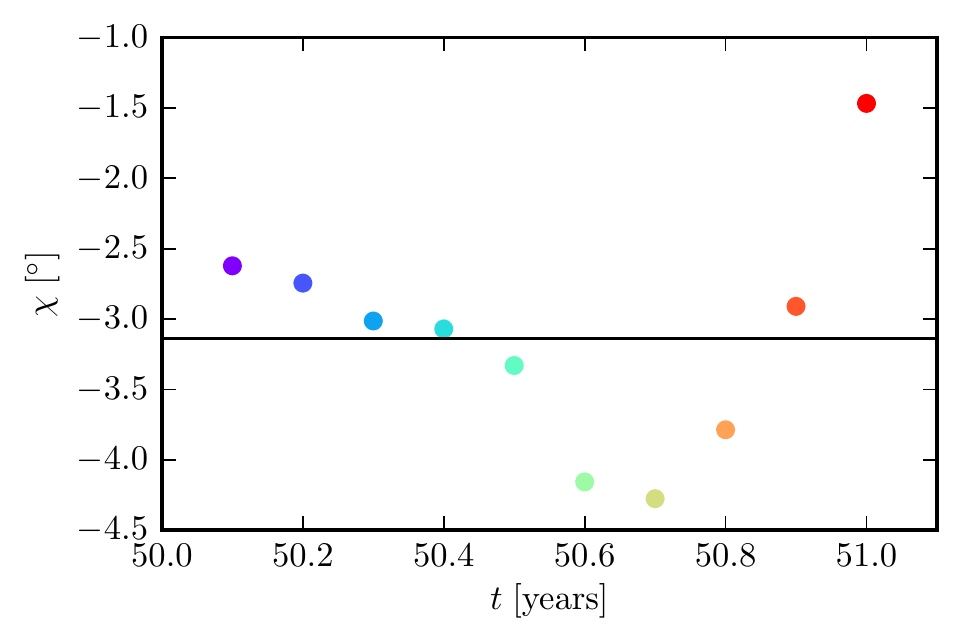}
\caption{Variability of the inner knot.  The top row
  shows the optical brightness distribution (linear scale) of the knot 
  in the simulation
  B3Dhr for the consecutive times $t=\{50.1,50.2,50.3,50.4\}~\rm
  years$.  In this run, the numerical resolution at the shock is
  $1.1\times10^{15}~\rm cm$, allowing close zooming-in.  As before,
  the position of the pulsar is marked by a white cross and we
  indicate the location of the intensity peak with a smaller blue cross.
  In the middle left panel, we show the corresponding total flux of the 
  knot emission (normalised to the temporal mean value) against the time 
  and  in the middle right it is shown against the separation $\Delta r$ 
  between the pulsar and the intensity peak.
  The data-points on the panels are identified by their colour. Over the 
  considered time-span, the knot flux $F_\nu$ exhibits variability 
  with the standard deviation of $11\%$.
  The flux is anti-correlated with the
  separation, with the Pearson correlation coefficient of $-0.81$.  The lower
  panel shows the unresolved polarisation degree and the electric vector
  position angle of the knot emission.  These display only very weak variability.  
  The (temporal) mean polarisation degree is $0.58\pm0.01$ and the mean 
  polarisation angle is $-3.1^\circ\pm0.8^\circ$. Thus, the electric vector is 
  in close alignment with the symmetry axis of the wind (the reported errors give
  the standard deviation).  The horizontal lines just show the mean values.  
  }
\label{fig:knotVariability}
\end{center}
\end{figure*}

We now focus on the variability of the knot in the synthetic ``Hubble
movie'' data set.  Thus, we zoom into the central region where the
flux is dominated by the emission from this bright feature.  
Figure \ref{fig:knotVariability} illustrates the variability of 
knot's shape, position and flux over the period of one year.   
The time scale of the variability is around one month. For the knot 
flux, the standard deviation is about $10\%$ but between
$50.3$ and $50.8$ years it increases systematically by $35\%$.  
This is comparable to the $\sim 40\%$ brightening over a period of 
two months of the Crab's inner knot reported recently  
by \cite{MoranShearer2013}.
 
The latter authors also report a relation between the knot flux and
its position.  To investigate the knot position in detail, we
calculate the distance $\Delta r$ between the intensity peak and the
pulsar location in the plane of the sky.  The middle panel of figure
\ref{fig:knotVariability} shows the knot flux as a function of $\Delta r$.
The figure indicates an anti-correlation for which we obtain the 
Pearson correlation coefficient between $\Delta r$ and $F_\nu$ of $-0.81$.
Thus, as the knot approaches the pulsar, its optical flux tends to
increase -- in good agreement to the findings of
\cite{MoranShearer2013}.

The bottom panels of figure \ref{fig:knotVariability} show the
degree of polarisation and the electric vector position angle (EVPA) of the
synthetic  knot emission 
{\bf as measured from the projected jet axis in the anti-clockwise 
direction.}
One can see that the knot polarisation 
remains nearly constant.  Within the error, the polarisation degree, 
$0.58\pm0.01$, agrees with the early off-pulse polarisation
observations by \cite{JonesSmith1981,SmithJones1988} {\bf and particularly 
the recent Hubble data curated by \cite{MoranShearer2013}, who separated the pulsar
and knot contributions.} 
The EVPA angle varies between $-4.5^\circ$ and $-1.5^\circ$ and thus it is 
closely aligned with the rotational axis of the pulsar. Such a weak variability 
agrees with the observations of the Crab's inner knot \citep{MoranShearer2013}.  

\subsection{Nebula polarisation}

{\bf 
We have already discussed the polarisation of the knot 1 in Sec.\ref{sec:knot}.
Here we describe the polarisation properties of the rest of the simulated nebulae. 
The polarisation of the total flux of the Crab nebula in radio band is 
$\Pi= 8.8\pm0.2\%$ with EVPA $\chi=149\fdg9 \pm 0\fdg2$ \citep[90~GHz,][]{Aumont10}, 
in optics $\Pi= 9.3\pm0.3\%$ with $\chi=160\fdg0 \pm 0\fdg8$ \citep{Oort56}, and in
X-rays $\Pi= 19.5\pm2.8\%$ with $\chi=152\fdg6 \pm 4\fdg0$ \citep[5.2~keV,][]{WSK78}. 
These measured EVPAs are very close to the PA of the Crab torus symmetry axis, 
$\chi\simeq126^o$ \citep{NgRomani2004} and presumably reflect the orientation of 
the magnetic loops injected into the nebula by the pulsar wind.  
The higher degree of polarisation in X-ray is naturally explained as a result of 
the synchrotron burn-off. Indeed, because of the burn-off the X-ray emission originates 
from the inner region of the Crab Nebula where the randomization of the magnetic field 
is not as strong as in the rest of the nebula volume. For optical- and radio-emitting 
electrons the burn-off time is much longer and as the result all parts of the PWN 
contribute to the integrated flux. The Faraday depolarisation at 90~GHz is very 
small \citep{bietenh-91b} and this explains the similarity of the radio and optical 
data.  

From our 3D simulations we find $\Pi=45\pm1\%$ and 
$\chi=-0.79^\circ\pm1.07^\circ$ (where $\chi=0$ corresponds to the symmetry axis)  
at 1~keV and $\Pi=34\pm1\%$ and $\chi=-2.60^\circ\pm0.83^\circ$ 
in the optical band. While we also find the X-ray polarisation stronger than 
the optical one, the polarisation degrees based on the simulation are significantly 
above the values found for the Crab Nebula. The difference could be due to the shortness 
of the 3D runs. For example, this does not leave sufficient time for  
developing dense filaments via the Rayleigh-Taylor instability (see Sec.\ref{sec:RT}).
The dynamic interaction of the PWN flow with the filaments is likely to enhance 
randomisation of the magnetic field. Another possibility is an excessive magnetic 
dissipation in the simulations, resulting in smaller contributions to the total 
flux from the outer regions of the nebula where the magnetic field is less ordered.  
    
We now discuss in details the polarisation features of our synthetic synchrotron maps
for the jet-torus region.  Figure \ref{fig:polarisation} shows the optical
intensity maps with overlying photon b-field vectors, which are scaled according
to the polarisation degree. For comparison, we include maps based on our 
2D simulations as well. In 2D, the predominant polarisation e-field
direction is that of the projected symmetry axis (see the top panels of
figure ~\ref{fig:polarisation}), the symmetry dictating exact alignment 
on the polar axis.  Away from the axis, the b-field vectors tend to curve 
around, tracing the azimuthal magnetic field. However along the bright inner ring, 
the polarisation vectors do not trace the magnetic field.  This is due to the
relativistic effect of polarisation swing ( see Eq. \ref{eq:swing}), which is
strong because the flow velocity just after the termination shock is still quite 
close to the speed of light. The polarisation degree is highest in the
jet and in the ring. \citet{del-zanna2006} obtained similar results based 
on their 2D simulations.  

In 3D, the correspondence between individual features in the 
intensity and polarisation degree maps is even more striking 
(the bottom panels of figure~\ref{fig:polarisation}). In particular,   
the inner and individual wisps are clearly identified in both of them.  
Along the wisps, the polarisation b-vectors are generally aligned with 
them. The polarisation degree is close to the theoretical maximum in 
the forward wisps and in the knot.  Not only fine features but also quite 
extended regions can have high polarisation. For example, large 
patches of high polarisation are seen  near all four corners of the 
polarisation degree map of Fig.~\ref{fig:polarisation}.

\begin{figure*}
\begin{center}
\includegraphics[width=70mm]{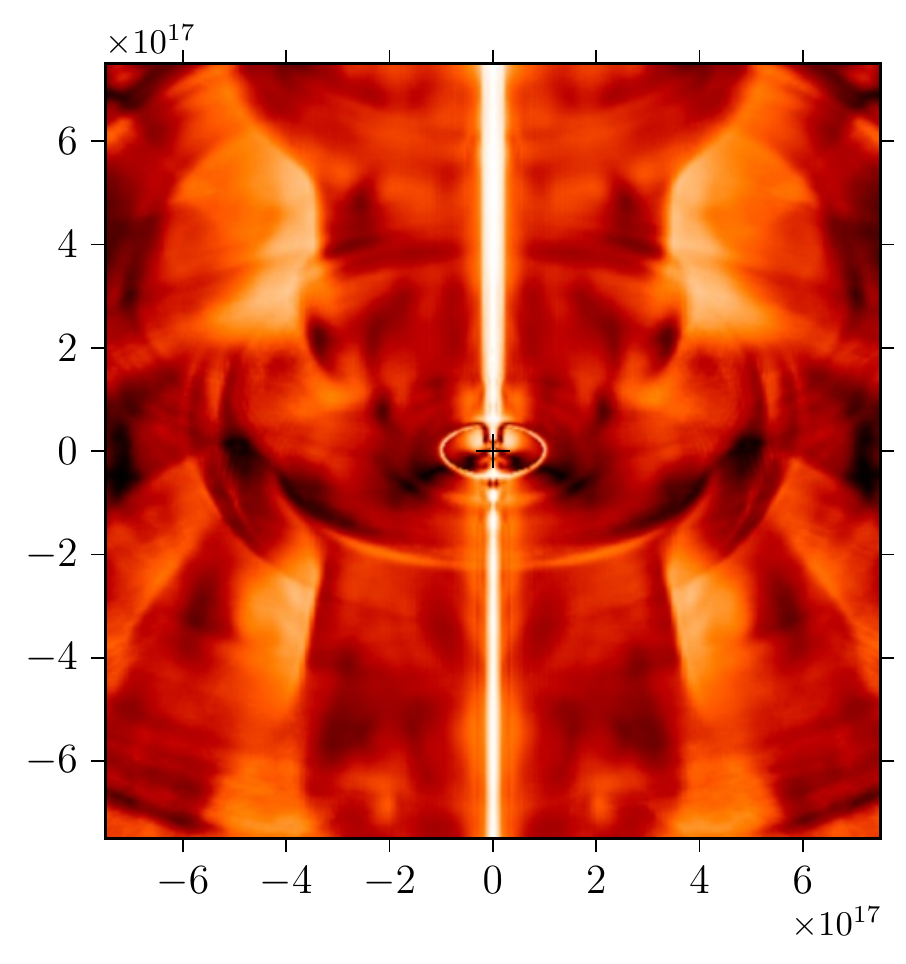}
\includegraphics[width=70mm]{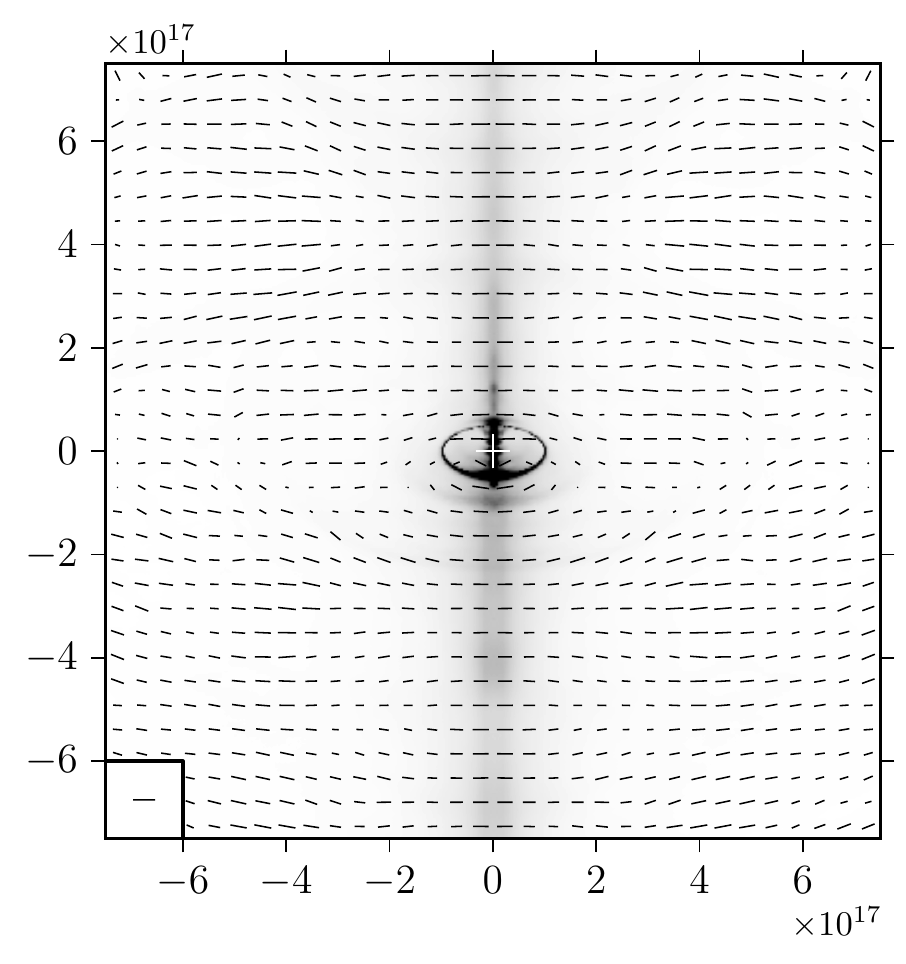}
\includegraphics[width=70mm]{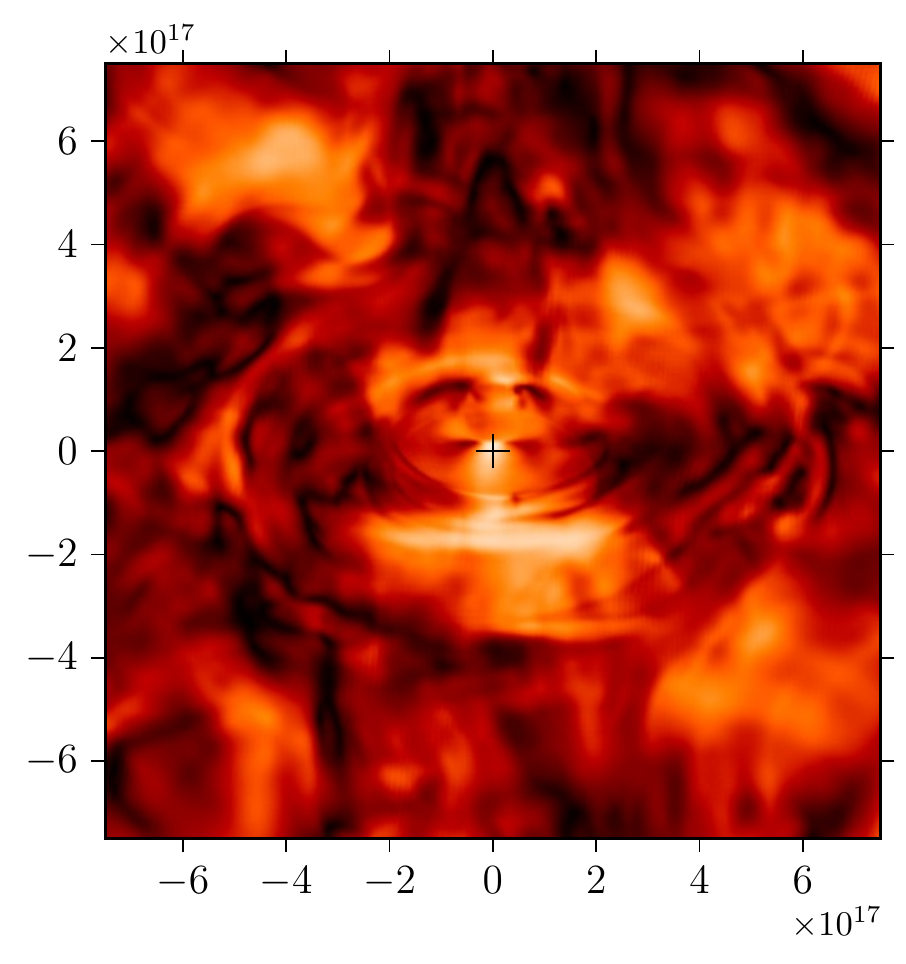}
\includegraphics[width=70mm]{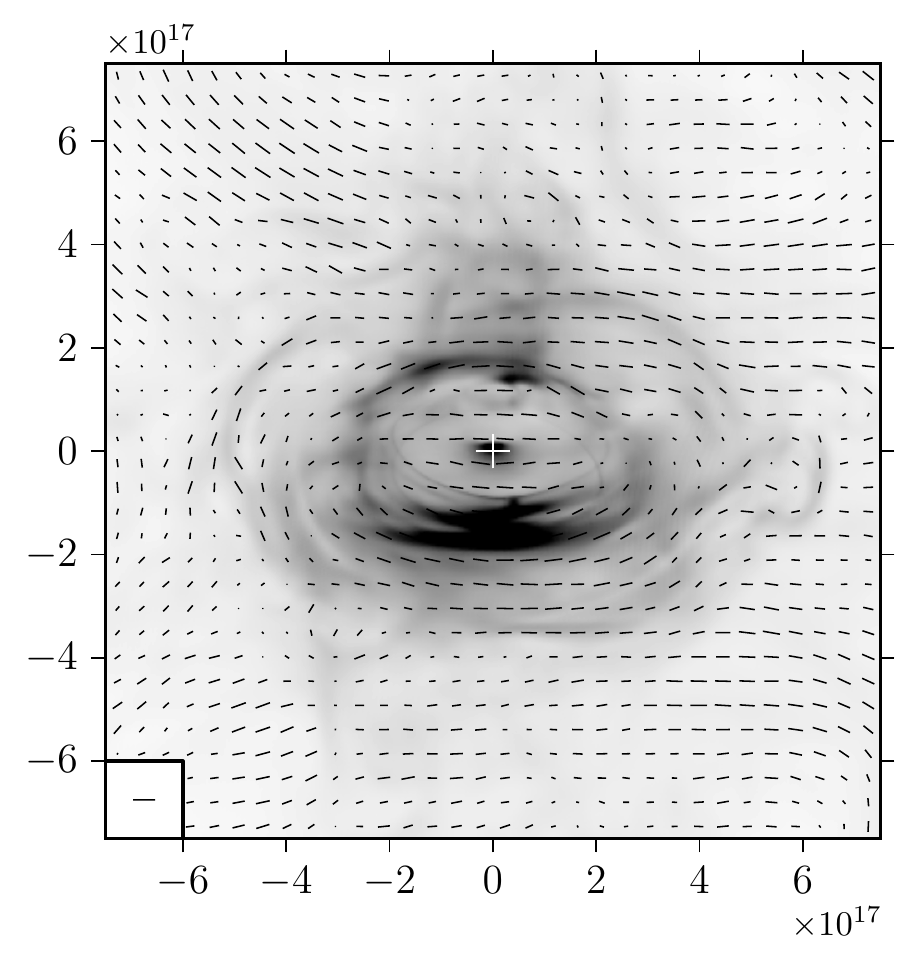}
\includegraphics[width=70mm]{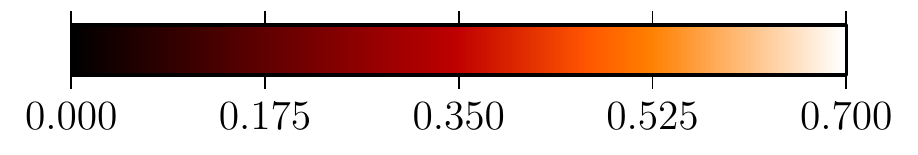}
\caption{Synthetic {\bf optical } polarization maps of inner PWN. 
The top panels show the maps based on the 2D simulation run B2Dhr 
and the bottom panels the maps based on the 3D run B3Dhr, both at $t=51~$years.
In the left panels, the colour images show the degree of polarization. 
In the right panels, the grey scale images show the optical intensity ( linear scale )
and the bars show the polarisation vectors. The vectors are
normalised to the maximum polarisation degree of $(p+1)/(p+7/3)\simeq0.7$ 
(indicated in the inset).    
Both the inner knot and the wisps are highly polarised with polarisation 
degree close to its theoretical maximum.  
}
\label{fig:polarisation}
\end{center}
\end{figure*}

According to \citet{Hester-08} the derived magnetic field in the torus region is aligned 
with wisps, like in our maps, but only after subtraction of the background 
contribution to the polarization maps (see his figure~6), what we did not do.  
This difference may again indicate somewhat suppressed emission 
from the outside of the torus in our results. 
\citet{Hester-08} also claims very high polarization of the wisp emission, close to 
$\Pi=70\%$ in some cases, which is in agreement of the very high polarisation of 
the bright wisps in our synthetic maps. In more recent paper, \citet{MoranShearer2013} 
give more moderate, but still high estimates $\Pi\sim 40\%$ for the wisps. The outcome is 
apparently sensitive to the background subtraction algorithm. They also find that EVPA 
of the wisp emission is approximately aligned with the torus axis.      

The reader is once more directed to the \emph{online material} for an animation of the time-series of the nebula polarisation. 
}

\section{Development of Rayleigh-Taylor filaments}\label{sec:RT}

\begin{figure*}
\begin{center}
\includegraphics[width=80mm]{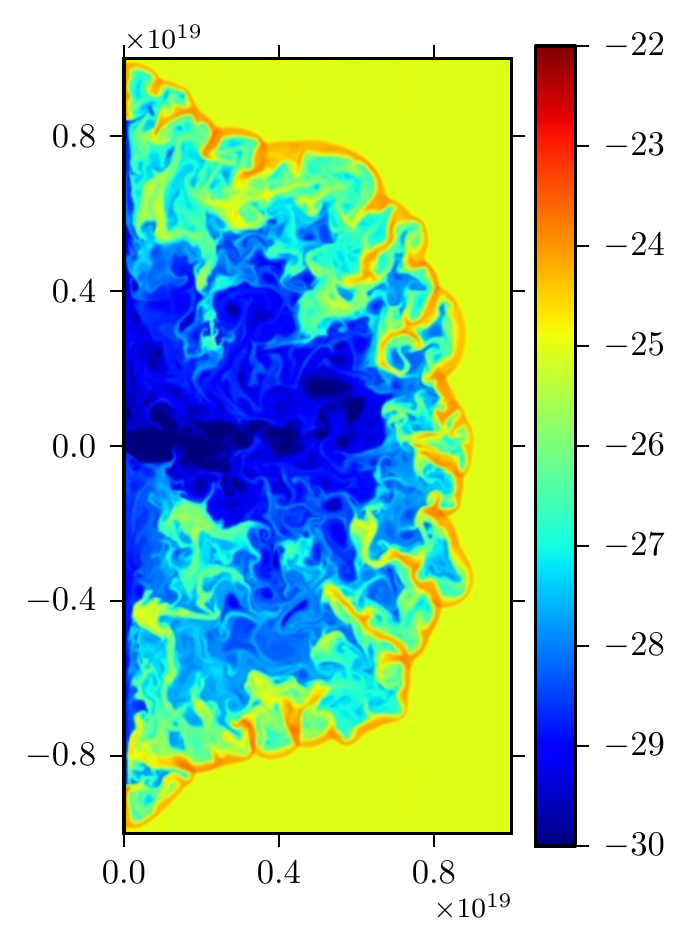}
\includegraphics[width=61mm]{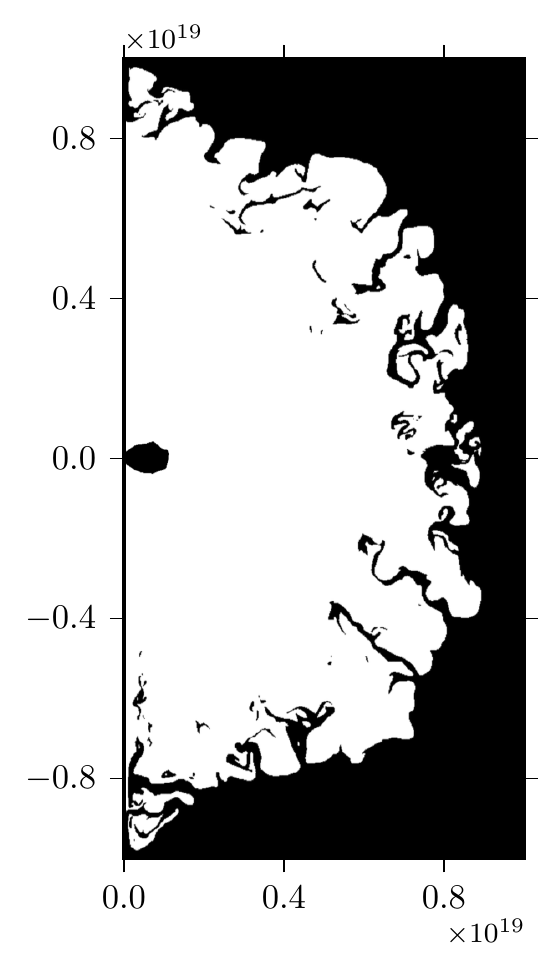}
\caption{Rayleigh-Taylor ``fingers'' in 2D simulations.  
The left panel shows $\log\rho$ at the end of the simulation run
A2D ( $t\simeq 1060$ years ).
Note that in this low magnetisation run ($\sigma_0=0.01$) the jet is not able
to penetrate far into the SNR. 
The right panel illustrates the way of determining the nebula radius $r_{\rm n}$.  
It shows the mask used according to Eq. (\ref{eq:mask}) in order to 
separate PWN from the supernova shell.  The nebula radius is then obtained as the 
volume average $r_{\rm n}=\left(3V/4\pi\right)^{1/3}$.  
The conspicuous ``fingers'' form via the Rayleigh-Taylor (RT) instability of
the contact discontinuity between the PWN and the supernova shell. 
The typical ``mushroom'' morphology of the RT-fingers, characteristic of the 
non-linear stages of this instability, is not seen here,  most likely 
because of the interaction with the turbulent flow of PWN.
}
\label{fig:filaments}
\end{center}
\end{figure*}

We run some of our 2D simulations for very long time, 
comparable to the age of the Crab Nebula. Their results show the development of 
a filamentary structure (see figure~\ref{fig:filaments}),  
which is very similar to the famous thermal filaments 
of this nebula \citep[see reviews by][]{davidson1985,hester2008}.  
\citet{ChevalierGull1975} were first to suggest that the filaments are formed 
via the Rayleigh-Taylor (RT) instability of the accelerating 
interface separating the pulsar wind nebula from the supernova shell. 
A further confirmation was given by \cite{HesterStone1996}, who analysed
the linear stability of such an interface.  Hydrodynamic
numerical simulations of the RT instability in plerions were first
performed by \cite{jun1998} and they corroborated the general RT scenario.
For the incompressible and non-relativistic regime, and only in the case of 
plane symmetry, the growth rate of the \emph{magnetic} RT instability for modes of
wavenumber $k$ is
\begin{align}
\omega^2 = \frac{g k
  (\rho_2-\rho_1)-B_{||}^2k^2/2\pi}{\rho_2+\rho_1} \, ,
\label{eq:RTclassic}
\end{align}
where $B_{||}$ denotes the magnetic field component tangential 
to the interface and parallel to the wave vector and
$g$ is the interface acceleration \citep{Chandrasekhar1961}.  
One can see that the magnetic field plays a stabilizing role, reducing the 
growth rate. 
\citet{bucciantini2004} studied the development of RT instability under the 
conditions relevant to PWN  by means of 2D relativistic MHD simulations. 
They found that the interface becomes stabilised when the magnetic 
pressure due to $B_{||}$ approaches the thermal pressure. 
In our axisymmetric simulations $B_{||}=0$ and thus the magnetic field plays 
no role during the linear phase.

In the application to PWN, the RT-problem can be further simplified
with the observation that $\rho_2\gg\rho_1$ and $\rho_2\gg p/c^2$,
hence the growth rate reduces to $\omega^2\simeq gk$.  For example,
Bucciantinti et al. (2004) derived the growth rate for a self-similar
expansion with density profile of the SNR $\rho_2\propto
r^{-\alpha}t^{\alpha-3}$ leading to an expansion law of $r_{n}\propto
t^\beta$ with $\beta=(6-\alpha)/(5-\alpha)$ \citep{chevalier1992}.
Due to the nebula expansion, the wave-length of a given perturbation,
$\lambda_{\rm f}$, increases in time.  Specifically,

\begin{align}
\lambda_{\rm f} = \theta_{\rm f} r_{\rm n}
\end{align}
or $k=2\pi/\theta_{\rm f} r_{n}$, where $\theta_{\rm f}$ is the angular
``size'' of the perturbation.  Hence, 

\begin{align}
\omega(t)^2 = \frac{2\pi \beta (\beta-1)}{\theta_{\rm f} t^2}  \, .
\end{align}
For our model of a supernova shell with constant density profile ($\beta=6/5$), 
this yields 
 
\begin{align}
\omega(t)\simeq1.2 \theta_{\rm f}^{-1/2} t^{-1} \, .
\end{align}
Thus during the phase of self-similar expansion, the growth rate of 
large wavelength perturbations with $\theta_{\rm f}>1$ is reduced compared to 
that of small wavelength perturbations with $\theta_{\rm f}<1$.  

We now analyse the development of filaments in our simulations. 
Figure \ref{fig:filaments} shows the distribution of the rest-mass density 
for the low sigma model A2D at the age $\sim 1060~$years\footnote{The 
nebula age is given by the simulation time $t$ plus the initial 
time $t_0=r_{\rm i}/v_{\rm i}\simeq210~$years, assuming initial expansion 
with constant $v_{\rm i}$.}. 
The filaments originate from the shocked supernova shell 
and extend into the nebula for a distance of up to $1/4\,r_\ind{n}$.  
The initially spherical shock front is now heavily perturbed and bulges 
out between the filaments.  Note that some of the filaments become 
detached from the shell.  This was also observed by \cite{jun1998} 
and attributed to the secondary Kelvin-Helmholtz (KH) instability at 
the shear layer between the ``in-falling'' filament and the PWN.  
We observe that the filaments do not exhibit the characteristic 
``mushroom caps'', seen in that study, most likely due to the highly turbulent 
state of the nebula. Although the linear theory predicts fastest growth 
for small scale filaments, these are also the most easily disrupted 
due to secondary instabilities and merging.  

A naked eye inspection of the results presented in figure~\ref{fig:filaments} 
suggests a predominating length scale for the separation
between the filaments, which is about the size of the termination shock. 
It is presently unclear which process sets this scale.  Since the filament
separation is much larger than the grid scale, we can rule out 
numerical reasons related to grid-based viscosity and diffusion.  
It seems possible that the scale is set by the largest eddies in 
the turbulent nebula flow, whose size is  also of the order of the 
termination shock radius.  We plan to address this issue in a forthcoming study 
on the outer structure of the Crab nebula.

\begin{figure}
\begin{center}
\includegraphics[width=80mm]{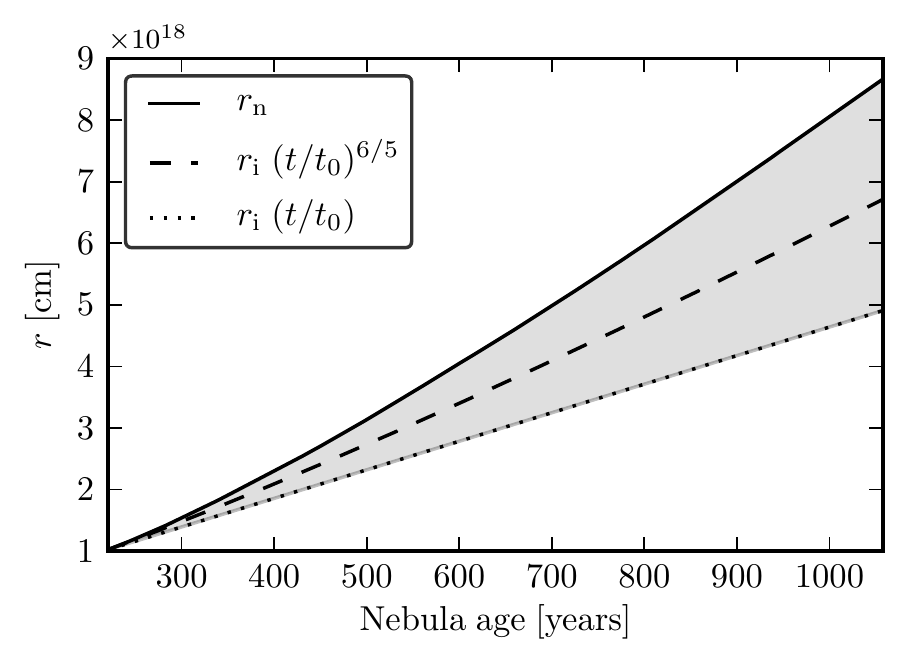}
\caption{Accelerated expansion of PWN.  
  The dotted straight line shows the PWN radius $r_{\rm n}$ based entirely on the 
initial expansion rate of the supernova shell in the simulation A2D.  
The dashed line shows the radius expected in the self-similar model, a power 
law with the index $1.2$. The solid line shows the actual evolution of the 
PWN radius in this run, which can be fitted with a power law of the index $1.37$.  
  The maximal length of the RT-fingers is given by the difference between
  the solid and the dotted curves.
}
\label{fig:rn}
\end{center}
\end{figure}

The length of the RT-filaments in our simulations cannot exceed 

\begin{align}
l_{\rm max}(t) = r_{\rm n}(t) - v_{\rm i} t \, ,
\end{align}
where $v_{\rm i}$ is the initial speed of the interface. This provides us with 
a simple test for the origin of the filaments.   
The nebula radius can be obtained empirically from the 
nebula volume measured according to Eq. (\ref{eq:mask}). For reference, the
corresponding mask is shown in the right panel of figure~\ref{fig:filaments}. 
The results of the test are shown in figure~\ref{fig:rn} and they 
give $l_{\rm max}/r_{\rm n}\simeq 0.5$ at the end of simulation A2D, which is twice 
the typical filament length. Thus, they are not in conflict with the RT scenario.  
The expansion law of figure~\ref{fig:rn}
is well described by the power-law $r_{\rm n}(t)\propto t^{1.37}$, which 
is only slightly steeper than the self-similar index of $6/5$.  
This discrepancy may be caused by the mixing within the RT unstable 
layer that transports the shell material into the nebula and thereby 
increases its effective radius.

None of the previous 2D simulations of 
PWN \citep{komissarov2004,del-zanna2004,bogovalov2005,camus2009},
captured the development of the RT instability. 
We believe that this is due to the insufficient resolution in these 
studies at the interface between the PWN and the supernova shell.  
They were mainly concerned with the inner regions around the 
termination shock and used spherical coordinates, thus quickly 
loosing the resolution with the distance from the origin.   
In contrast, we used cylindrical coordinates and obtain uniform
resolution throughout PWN.  In combination with AMR this has allowed 
us to obtain much higher resolution at the interface and hence 
capture the growth of RT fingers.

\section{Discussion}
\label{sec:summary-discussion}

{\bf  
Our simulations show that a number of flow
features discovered in the earlier 2D relativistic MHD simulations of
PWN are preserved in 3D.}
First of all, in the vicinity of the termination shock the flow of freshly 
injected plasma still separates into the equatorial outflow, which gives 
rise to the Crab torus, and the polar outflow, which is identified with the 
Crab jet. This was expected as the separation is due to the anisotropy of the 
pulsar wind power distribution and the hoop stress of the azimuthal magnetic 
supplied by this wind. {\bf Both these factors are inherent to the PWN models 
and they are introduced in the 2D and 3D simulations via identical inner boundary 
conditions, describing the pulsar wind.}   
Although in 3D the azimuthal field is disrupted, this takes time and in the 
vicinity of the termination shock, where the separation occurs, the disruption 
is minimal. 

In the equatorial region close to the nebula boundary, we find that
the poloidal field direction can dominate over the azimuthal one.  In
this region however, the average field strength is ten times less than
the field close to the termination shock.  This large range of
magnetic field strength calls into question the validity of
``one-zone'' spectral modelling of the Crab.

Secondly, the termination shock remains unsteady in 3D, and its
unsteady dynamics still yields highly inhomogeneous outflow from the
shock. In the synthetic synchrotron maps, this outflow appears as a
collection of wisps emitted from the shock location, in a qualitative
agreement with the observations of the Crab Nebula.  This unsteady
dynamics is less violent compared to that reported in the 2D
simulations of \citet{camus2009}, and the wisps are produced less
frequently. This may originate in the significantly lower resolution
of our simulations in the vicinity of the termination shock, 
leading to higher numerical viscosity and
diffusion, which normally result in erasing small scale features
and may damp the shock oscillations. On the other hand, the enhanced
coherency of the flow in 2D may lead to artificial strengthening of
the impact which the motions in the simulated PWN make on the shape of
the termination shock.  In addition, we find that perturbations of the
termination shock may also be triggered by non-axisymmetric
instability of the polar jet.

However, our simulations have also demonstrated that the additional
degree of freedom introduced in 3D models of PWN has a strong effect
on their global dynamics. The disruption of the z-pinch configuration
of the axisymmetric models of PWN in our 3D simulations results in the
almost complete disappearance of the strong axial compression observed
in the previous 2D numerical simulations. {\bf It is preserved only above 
the polar section of the termination shock, the region where the 
polar jets (plumes) are produced. This is in agreement with \cite{mizuno2011b} 
who found a fast transition to the non-linear phase of the kink instability. 
In the plane of the sky, the Crab Nebula is noticably more extended in 
the direction of its jet compared to the direction of the main axis of 
its ``torus''. 
\citet{begelman1992} explained this as a consequence of the magnetic 
z-pinch. In our 3D simulations, the nebula does not show such an 
elongation but remains very much spherical. One possible explanation for 
this could be an excessive magnetic dissipation -- stronger magnetic field would  
produce stronger z-pinch. However, the calculations of \citet{begelman1992} 
show that in order to achive the observed elongation of Crab Nebula in this 
model the magnetic pressure should be of the order of the thermal pressure, 
which seems to be in disagreement with the sub-equipartition magnetic field 
derived from the observations \citep{MH10,HA04}.  
Alternatively, the observed elongation of the Crab Nebula may have nothing to 
do with the z-pinch but rather with the non-spherical geometry of the supernova 
ejecta. This could be a result of non-spherical supernova explosion 
\citep[e.g.][]{MBA06,Burr07}. }

{\bf In the high-$\sigma$ models, we find a truly dramatic difference 
between the global evolutions of PWN in 2D and 3D simulations. Thanks to 
the artificially preserved z-pinch configuration of the 2D models, they develop 
extremely strong polar jets, which burst through the supernova shell. 
In contrast, in the 3D models the polar outflows are less powerful and 
eventually lose collimation and coherency via the kink instability.   
This difference shows that one has to be very careful and not to give 
too much credit to the results of 2D relativistic MHD simulations of 
the jet production.  In particular, our 2D jets are very similar, as well as 
the whole setup with purely azimuthal field, to the jets developed in the
2D simulations of the magnetar jets by \citet{bucc-07,bucc-08}, which
are most likely artefacts too. }

{\bf \citet{MignoneStriani2013} studied the 
3D dynamics of relativistic jets injected into an expanding cavity, with 
application to the Crab jet. They concluded that the observations are best 
fitted by the models with comparatively low jet Lorentz factors ($\Gamma \approx 2$), 
which is not far from we find  in our simulations.  
However, their setup is somewhat less suitable to the Crab Nebula, with the 
termination shock, equatorial outflow and self-consistent jet production 
not being incorporated. The 
jet instability is triggered as a result of its interaction with the strong backflow, 
which develops when this jet hits and ``erodes'' the supernova shell. 
Such a strong backflow does not develop in our simulations. 
}

The dynamics of our 3D numerical solutions shows that the
magnetization of pulsar winds does not have to be as low as in the
Kennel-Coroniti 1D model in order to agree with the observations. In
fact, we find that even with the wind magnetisation as high as
$\sigma_\ind{0} = 3$, the size of the termination shock is still very
close to the one predicted by an unmagnetised wind model.\footnote{We
could not consider higher values of $\sigma_\ind{0}$ because of the
limitations of our numerical code.}  Overall, our results provide
strong support to the anticipation of \citet{begelman1998} that the
dynamics of 3D MHD models of PWN will differ from that of 1D models in
a way that can solve the $\sigma$-problem. The main reason behind this
anticipation was the randomisation of magnetic field at the non-linear
stages of the magnetic kink mode instabilities. Although the
magnetic field does indeed become highly randomised, another related
factor also comes into play and makes a strong impact -- the magnetic
dissipation inside the nebula.

Due to the magnetic dissipation in the striped wind and in the nebula,
the energetics of the simulated PWN is dominated by thermal energy.
Inside the nebula, the magnetic energy dissipation proceeds
via two routes. First, it is the annihilation of opposing magnetic flux 
tubes that are blown into the nebula from the two hemispheres of the 
pulsar wind. The flux annihilation is amplified by the turbulence in the 
nebula which causes efficient mixing of the opposing polarities and 
transports magnetic energy to smaller scales via the turbulent cascade.
Our high resolution run indicate the emergence of an inertial
range in the magnetic energy spectrum with the power-law slope of
$-5/3$. Second, in the polar region, current sheets are produced via the
kink instability of the beam and plume. 

The rate of this dissipation in our simulations was high enough to
result in the magnetic energy making only a small contribution to the
total energy budget of the nebula. This made the composition of PWN
similar to what is expected in the 1D Kennel-Coroniti model, which has
no magnetic dissipation, in the case of particle-dominated pulsar
wind. Unexpectedly, even our 2D numerical models had enough magnetic
dissipation to not differ too much from the 3D models. In the 2D
models, the dissipation is not related to the development of the
magnetic kink instability.  However, the current sheets are still
produced by the turbulent motion in the bulk of the nebula. We
conclude, that the combination of magnetic dissipation in the striped
wind zone and inside the nebula can already provide solution to the
$\sigma$-problem of PWN. Note that the spectral modelling of the Crab
nebula suggests that its synchrotron-emitting plasma is
particle-dominated, thus supporting the magnetic dissipation in the
nebula \citep{komissarov2013}. 

Although the integrated equations do not have terms describing
magnetic dissipation and hence all the dissipation is of numerical
origin, this does not mean that it should not be taken seriously. The
processes that create small scale current sheets in our simulations
and hence drive the magnetic energy flow from large to small scales,
where it dissipates, are properly accommodated in the code. On the
other hand, a rather high resolution is required in order to resolve
even the large wavelengths of the cascade. {\bf The statistical
analysis of our numerical solutions seem to indicate that the inertial 
range of magnetic turbulent spectrum is begining to emerge in our highest 
resolution 3D simulations (see Appendix~\ref{sec:turbulentSpectra}). 
The convergence study, though not entirely complete, also indicates 
convergence of our 3D solutions in the statistical sense 
(see Appendix~\ref{sec:convergence}).  However, further studies are required in 
order to settle the matter and at this stage we cannot rule out 
that the rate of magnetic dissipation in our simulations is still excessive. 
In fact, comparing the ratio of magnetic energy to the energy in relativistic 
particles obtained from simplified ``one-zone'' spectral models of 
the Crab nebula\footnote{See the calculations in \cite{komissarov2013} and references 
\citet{MH10,HA04,hillas-98}} with the ratio of magnetic to thermal energy 
in our simulations, we see that our value is lower by a factor of 
$\simeq 3$.  On the other hand, our simulations also show that the magnetic field 
strength is much higher in the inner part of the nebula, where most of 
the Crab Nebula X-ray and probably gamma-ray synchrotron emission is produced, 
compared to its outskirts, where most of the inverse-Compton emission 
is coming from. This raises doubts about the reliability of 
the one-zone model estimates. What is even more worrying is that the 
spectrum of the Crab Nebula can be fitted rather well even with the 1D 
Kennel-Coroniti model \citep{MH10}, which does not capture the PWN flow kinematics and 
predicts a completely different distribution of magnetic field, with the magnetic field 
increasing outwards.  Thus, the spectral data alone does not appear to be 
very informative.       
}

Given the similarities between the observed properties and theoretical 
models of PWN and other astrophysical phenomena involving relativistic 
flows, such as AGN and GRB jets, the kink instability and 
magnetic dissipation can play equally important roles in their dynamics 
as well.  In particular, this have been discussed in relation to the 
initial dynamics of magnetic bubbles in the magnetic model of long GRBs 
in \citet{LB03}. 
They speculated that while the expansion speed of the bubble is still 
below $0.1$c most of the magnetic energy supplied by the central engine 
is converted into heat, thus producing a fireball. The results of our 
simulations are in full agreement with this hypothesis.

In order to provide more direct tests against the observations, we
also attempted to model the synchrotron emission of the Crab
Nebula. In this modelling, we followed the assumption of
\cite{kennel1984} that the synchrotron electrons and positrons are
accelerated only at the termination shock.  However, given the recent
developments in the theory of particle acceleration at relativistic
magnetized shocks, we considered two cases, one where the particles
are only accelerated at the part of the shock that terminates the
striped wind zone and another where we make no distinction between the
striped and unstriped zone.  In both cases, the synthetic synchrotron
maps reproduce the torus, inner ring, wisps and sprite of the 
inner Crab Nebula. Moreover, the
visual comparison with the Hubble images of the Crab nebula favours
particle injection in the striped region only; otherwise a polar column of
excesively  strong emission is generated within the inner ring, in
conflict with the observations. However, our simple treatment of
relativistic particles produces an under-luminous X-ray jet, as
particles injected at the termination shock have cooled considerably
until entering the plume.  We suggest that additional particle
acceleration in the jet is necessary to reconcile MHD models with the
Chandra observations of the jet \citep[e.g.][]{weisskopf2000}.
Variability of the termination shock is connected to the generation of
wisps of increased intensity, moving outwards with approximate
velocity of $1/3 c$.  This is very similar to what is seen in the HST
observations of the Crab\footnote{A movie which demonstrates this
  behavior is a part of the on-line supplementing material.}.  In our
3D simulations, the wisps maintain their arc-like shape in the torus
region but become distorted further out.

The synthetic total intensity maps also show the bright extended inner
knot, which was discovered in the previous 2D simulations and
identified with HST knot 1 of the Crab Nebula. Thus, this feature is
very robust and not specific to 2D models only. This emission comes
from the immediate vicinity of the termination shock, it is highly
Doppler-beamed and originates in the high-speed part of the post-shock
flow.  We find a correlation between knot position and flux, such that
brighter states correspond to a smaller offset between knot and the
location of the pulsar.  The observed variability of our synthetic
knot is in excellent agreement with the recent optical observations of
Crab's knot 1 \citep{MoranShearer2013}.  Despite the large variability
of the knot flux, the optical polarisation signal of the knot is very
stable with a temporal mean polarisation degree of $0.58\pm0.01$,
again in excellent agreement with the value reported by
\cite{MoranShearer2013}.  Concerning the polarisation direction of the
knot, we report close alignment of the EVPA with the projected spin
direction of the pulsar with an angle of $-3.1^\circ\pm0.8^\circ$,
indicative of the dominating azimuthal magnetic field of the knot.  The
presence of this knot in the Crab images is the most direct evidence
so far of the particle acceleration taking place at the termination
shock of the Crab pulsar wind. 

The synthetic polarization maps show that in spite of the strong
disruption of the azimuthal magnetic field supplied by the pulsar in
our 3D simulations, the polarization remains substantial, particularly
in the inner part of the nebula.  The optical polarisation direction
on the scale of the torus clearly indicates an azimuthal field and
photon $b$-vectors appear to curve around with the torus.  The photon
$b$-field direction stays aligned with the wisps, even as the wisps
deform further outside.  Localised features in the intensity maps
(wisps, knot, jet) exhibit a strong polarisation degree $\gtrsim 0.5$.
This shows that the jet-torus region of the Crab pulsar is dominated
by the recently supplied plasma, which still keeps memory of the
highly organized magnetic field of the wind. 

{\bf 
One feature that has a rather different appearence in our synthetic synchrotron 
maps compared to  the images of the inner Crab Nebula is the inner 
ring\footnote{None of the synthetic maps of the nebula 
made by other researches have managed to reproduce this feature too.}. 
In X-rays, the real inner ring is very bright, knotty, 
and rather symmetric, and it does not seem have a counterpart in the optical 
images of the nebula \citep{weisskopf2000,hester2002}. 
In the synthetic images of 3D simulations, it is relatively weak, knotless, 
asymmetric and looks similar both in optical light and in X-rays.     
Its asymmetry is due to the Doppler beaming of the emission produced  
the fast post-termination shock flow. It is hard to see what could cancel 
this effect in X-rays.  Perhaps, here we need to go beyond the RMHD 
approximation in our search for answers. }
{\bf
The strong magnetic dissipation observed in our simulations forces us to question 
the very model we used to compute the synchrotron emission. Such dissipation 
would almost certainly have an impact on the emission, via  
acceleration of synchrotron-emitting electrons (and positrons) inside the nebula, whereas 
in our model we followed \citet{kennel1984} and assumed that they are accelerated 
only at the termination shock. 
The fact that our synthetic X-ray images do not show such a prominent 
jet-like feature as the Crab's X-ray jet already suggests that an additional 
channel of particle acceleration is needed to explain the observations.  
Other indicators are the excessively high polarization of the 
integral synthetic flux and the excessive brightness of the optical torus compared to 
the rest of the nebula (not shown). The polarization is so high because of the major 
contribution to the computed integral flux from the inner region (torus), where 
the magnetic field is still very much ordered. Provided the additional particle 
acceleration brightens up the outer nebula, both the polarization and the 
brightness contrast between the inner and outer nebula in optics may be reduced.   
These, however, are less convincing indicators because in the 3D simulations we 
have not reached the phase of self-similar expansion. Moreover, the RT filaments 
did not have enough time to develop and penetrate the synchrotron nebula, 
which would have an impact on its dynamics. This prohibits a direct comparison of 
our 3D models with the observations of the whole of Crab Nebula.  
A weaker magnetic dissipation may also 
help to reduce the total polarisation and the brightness contrast, via increasing 
the magnetic field and thus the emissivity in the outer parts of the synchrotron
nebula. These issues will have to be investigated in future studies.

On the other hand, our current model successfully reproduces the inner structure 
of the Crab Nebula, with the exception of its X-ray inner ring and jet. The success 
of the theory in the interpretation of Crab's inner knot is particularly impressive.   
This allows us to conclude that the shock acceleration is still an important 
contributor to the nebula emission.  PIC simulations of shocks in striped 
flow also speak in favour of such acceleration \citep{sironi2011}, though they 
do not seem to get the right spectrum\footnote{In this context, we consider the 
layer where the magnetic stripes dissipate as a part of the shock structure and 
any acceleration occurring in this layer as a kind of shock acceleration.}. 
The fact that our recipe B, where the acceleration occurs only in the striped 
part of the termination shock, provides synthetic maps which are more reminiscent 
of the inner Crab Nebula agrees with the PIC results. 
All these make a strong case for particle acceleration at the Crab's 
termination shock.    
}  

During a year of synthetic observation, the flux variability of the whole 
nebula in both soft X-ray and optical bands were on the $1\%$ level, which can be 
described as generally consistent with the observations 
\citep[][ and references therein]{W-H11}.

Concerning the origin of the gamma ray flares of the Crab Nebula, our results
restrict the possible location of the magnetic reconnection events that are 
believed to be responsible for the flares. In order to produce synchrotron photons
of the energy $\epsilon_\ind{ph}$ in the magnetic field of strength
$B$ after acceleration in the electric field of strength $E=\delta B$,
the magnetic field strength has to be about
\begin{align}
     B \simeq 3\times 10^{-3} \delta^{-2/3}
     \fracp{\epsilon_\ind{ph}}{1\,\mbox{GeV}}^{1/3}
     \fracp{L}{1\,\mbox{ld}}^{-2/3} \mbox{Gauss} \, ,
\end{align}
where $L$ is the length of the ``linear accelerator'' in light
days. Given that the variability time scale can be as short as few
hours \citep{balbo-11,buehler-12}, this length is unlikely to exceed
few light days and thus the magnetic field in the reconnection layer
should be well in excess of the mean magnetic field strength in the
nebula, which is only $\sim 0.1\times10^{-3}$Gauss \citep{MH10}. 
Figure \ref{fig:aniso} shows that in our numerical models, 
$B\simeq 10^{-3}$Gauss can be found only  
in the highly magnetized polar region, which is fed by the unstriped 
component of the polar wind. Elsewhere, including the vicinity of the 
equatorial current sheet, the magnetic field is significantly weaker. 
Thus, the polar region at the base of the Crab jet is the most likely 
location of the flares.

Simulations up to the current age of the Crab nebula can only be
reached with our axisymmetric models.  At a nebula age of $1000\ \rm
years$, we find the structure at the PWN-SNR interface in the
non-linear phase of the Rayleigh-Taylor instability and dense filaments
of SNR material protrude into the nebula with a length of up to $1/4$
nebula radii.  This is in good agreement with the prominent optical
filaments observed in the Crab nebula.  We obtain a power-law
expansion of the nebula $r_{\rm n}\propto t^{1.37}$ slightly faster
than the expectation in the self-similar regime with power-law index
of $6/5$.  Using the expansion law measured from the simulations, we
checked that the length of the filaments is consistent with the RT
scenario.  

{\bf \cite{bucciantini2004} carried out 2D RMHD 
simulations to study the development of Rayleigh-Taylor instability in 
PWN described by the 1D solution of \citet{kc84a}. 
Given the imposed symmetry, they could only consider perturbations whose 
wave vector was aligned with the magnetic field. 
They observed strong suppression of the perturbation growth in models where 
the magnetic pressure was about or exceeded the thermal pressure at the interface 
between the supernova shell and the relativistic bubble.  In the Kennel-Coroniti 
model of the Crab nebula such a strong interface magnetic field is achieved  
already for the wind magnetization as low as $\sigma=10^{-3}$. We observed RT 
instability in the model with $\sigma_0=0.01$.   
Since, the linear growth of perturbations with the wave vector orthogonal 
to the magnetic field is not influenced by the magnetic field, the development 
of RT-instability in our run is not that unexpected. However, the low 
magnetic field strength found in our 3D simulations near the interface  
implies that also the development of parallel modes would not be suppressed. 
In contrast to the Kennel-Coroniti model, we do not see an increase 
of the nebula magnetic field strength towards the interface, but rather the 
opposite.  Even in the high-$\sigma$ models, 
the magnetic dissipation ensures that the magnetic pressure is much lower 
than the thermal one near the interface (see Figure~\ref{fig:dissSlice}). 
Unfortunately, the computational cost of our 
3D simulations is too high to let us study the RT instability. Future  
specialised long-term 3D simulations are needed to explore these.}

\section{Conclusions}\label{sec:conclusions}

\begin{enumerate}
\item
The jet-torus morphology of the Crab Nebula, reproduced in the previous 2D 
relativistic MHD simulations, is a robust feature of MHD models, which reflects 
the anisotropic structure of pulsar winds. It is retained in our 3D simulations.  
\item
The jets are formed downstream of the termination shock where the magnetic hoop stress 
causes collimation of the flow lines that pass through the shock at intermediate 
latitudes.
\item
2D models exaggerate the jet power and collimation degree.  This leads
to artificially strong axial compression and unphysical ``jet
breakouts'' in models with higher wind magnetization. This result has implications 
for the magnetar model of GRB central engines as it questions the mechanism of 
jet formation.   
\item
As in the earlier 2D models, the variability of the termination shock is
caused by the intricate feedback mechanism between the shock and the nebula
flow. In 3D simulations, we obtain a somewhat lower
amplitude of the radial shock oscillations.  The inhomogeneities, formed in the
post-shock flow as a result of this variability, appear as wisps in our 
synthetic optical images of the Crab Nebula.
\item
Our results suggest that the magnetic dissipation inside PWN could be a key 
factor of their dynamics. Combined with the magnetic dissipation in the striped 
zone of the pulsar wind, it allows us to reconcile the observations of the 
Crab Nebula  with the expected high magnetisation of such winds.
In particular, the termination shock radius in our run with the highest
wind magnetisation ( $\sigma_0=3$ ), is very close to that of our 
lowest magnetisation run ( $\sigma_0=0.01$ ),  as well as to the prediction 
of the analytical model for unmagnetized wind. 
\item
The energetics of the simulated PWN is dominated by the thermal energy.  
This agrees with the ``one zone'' spectral modelling of the Crab Nebula. 
However, we also obtain a significant magnetic field variation in the nebula, 
which questions the validity of such simplistic spectral models. 
{\bf Although we cannot exclude that the magnetic dissipation rate in our simulations
is too high, the results of our limited convergence study suggest that it is not 
competely off the scale. The low magnetic field strength, found in our simulations 
in the outskirts of the nebula, agrees with the sub-equipartion values 
deduced from the observations.} 
\item 
The strong magnetic field, $B\sim 10^{-3}$Gauss, required in the magnetic reconnection 
models of the Crab gamma ray flares, can only be found at the base of the Crab jet.  
\item
Under the assumption that the synchrotron emitting particles are accelerated in 
the striped zone of the pulsar wind termination shock, our synthetic images 
reproduce the inner structure of the Crab Nebula, revealed by the Hubble and 
Chandra observations, very well indeed -- we obtain 
dynamical counterparts for its torus, inner ring, knot and sprite. 
\item
{\bf However, an additional particle acceleration, most likely inside the jet 
itself or at its base, is required ``to illuminate'' the jet, which is not as 
bright as the X-ray jet of the Crab Nebula. The inner X-ray ring of the nebula 
remains an enigma. }
\item
We have analysed the variability of the synthetic inner knot and obtained
excellent agreement with the recent Hubble data.  In particular, within one 
year the  luminosity of the knot increased by $35\%$ while its 
unresolved polarisation degree remained nearly constant at the level of
$0.58\pm0.01$.  Moreover, our simulations reveal a correlation between the knot
flux and position, with brighter states being found at the times of smaller 
separation  between the knot and the pulsar.
\item
The linear polarisation in the torus region is indicative of the
freshly injected azimuthal magnetic field. The photon b-field vectors
generally align with wisps, even as these deform further out. 
{\bf The overall polarisation degree of the nebula emission is significantly higher 
than the observed values, both in optical light and in X-rays. Given the 
short duration of our 3D runs, it is not clear how significant this disagreement
is. If confirmed by future longer runs, this may be considered as another
evidence for particle acceleration away from the termination 
shock. Alternatively, this could be an indication of excessive magnetic dissipation.}     
\item
Our long time 2D simulations show clear signs of the Rayleigh-Taylor
instability at the PWN--SNR interface.  At the current age of the
nebula, the longest filaments measure up to 1/4 of the nebula radius,
in overall good agreement with the observations of the Crab's optical filaments.
\item 
{\bf More work has to be done in order to clarify a number of issues raised 
by our simulations. In particular, it is important to run longer 3D simulations 
and reach the phase of approximate self-similar expansion of the nebula. This 
will allow more direct tests against the observations of the Crab Nebula. 
A more advanced convergence study is needed to settle the important issue of 
magnetic dissipation rate inside the nebula. An additional particle acceleration 
inside the nebula has to be incorporated in the model of its emission.    
}

\end{enumerate}

\section{Acknowledgments}
SSK and OP are supported by STFC under the standard grant
ST/I001816/1.  SSK acknowledges support by the Russian Ministry of
Education and Research under the state contract 14.B37.21.0915 for
Federal Target-Oriented Program.  RK acknowledges FWO-Vlaanderen,
grant G.0238.12, and BOF F+ financing related to EC FP7/2007-2013
grant agreement SWIFF (no.263340) and the Interuniversity Attraction
Poles Programme initiated by the Belgian Space Science Policy Office
(IAP P7/08 CHARM). {\bf The simulations were carried out on the Arc-1
cluster of the University of Leeds.}

\bibliographystyle{mn2e}
\bibliography{pwn,mypapers,lyubarsky,lyutikov,hea,astro,local,astro1}

\appendix

\section{Shape of the termination shock}\label{sec:shape}

The solution for the oblique shock is given in \cite{komissarov2011}. 
We utilize their (A12) for $\sigma_1 = a_1 = 0$, $\Gamma_1 \gg 1$, 
$\delta_1\gg 1/\Gamma_1$ and adopt the notation of the latter authors. This gives

\be 
   \beta_{2x}/\beta_{1x}=1/3, 
\ee 

\be 
    \Gamma_2= \sqrt{\frac{9}{8}} \frac{1}{\sin\delta_1}. 
\ee 
The normal component of the momentum equation reads 
\be 
   \rho_1c^2\Gamma_1^2\sin\delta_1^2 = p_2(4\Gamma_2^2\beta_{x2}^2 +1). 
\ee 
and after substitution 
\be 
  p_2 = \frac{2}{3}\rho_1c^2\Gamma_1^2 \sin^2\delta_1 \, .
\ee 

\begin{figure}
\includegraphics[width=80mm]{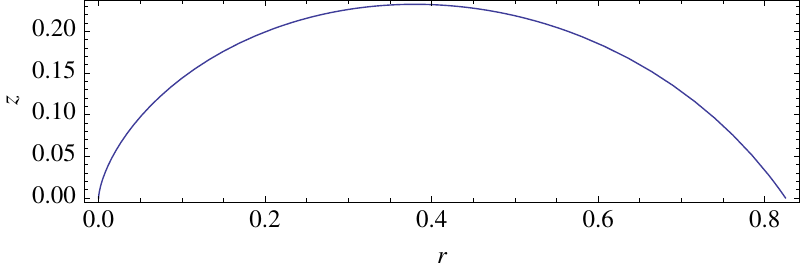}
\caption{ Numerical solution for the shock shape with $f_{\rm tot}\propto \sin^{2}\theta $. 
The unit of length is $r_0$.  }
\label{fig:hydroshock}
\end{figure}

We apply these results to the PW termination shock with energy flux density profile 
of the wind

\be 
   \rho c^3\Gamma^2 \propto \sin\theta^2 
\ee 
as in the monopole solution. If $p$ is the pressure in the nebula 
and $L$ is the wind luminosity then 

\be 
  p = \frac{L}{\textcolor{black}{4}\pi c}\frac{\sin^2\theta}{r^2} \sin^2\delta \, ,\label{eq:pneb}
\ee 
where we dropped the suffix for the obliqueness angle $\delta$ as well. 
Adopting $r(\theta)$ as the curve describing the shock shape, then 
the shock obliqueness is given by 
\be 
\sin^2\delta = \frac{r^2}{\dot{r}^2+r^2}\, .\label{eq:sin2delta}
\ee 
Introducing the length scale 
\be 
   r_0=\fracb{L}{4\pi pc}^{1/2} \, ,
\ee 
we have the dimensionless variable $\xi=r/r_{0}$ and equations (\ref{eq:pneb}) and (\ref{eq:sin2delta}) can be combined to 
\be 
    1 = \frac{\sin^2\theta}{\xi^2} \sin^2\delta \,
\ee 
whereby 
\be 
\sin^2\delta = \frac{\xi^2}{\dot{\xi}^2+\xi^2}\, .  
\ee 
and we end up with the non-linear differential equation describing the shape of the shock 
\be 
   \dot{\xi}^2+\xi^2 = \sin^2\theta \, .
\ee 
For $\theta\ll 1$ we can make the Ansatz $\xi=a\theta^b$ and obtain 
$b=2$ and $a=1/2$. Thus, 

\be 
\xi=\frac{1}{2}\theta^2, \quad\text{for}\ \theta\ll 1 \, .
\ee 
The full solution is found numerically and is presented in figure 
\ref{fig:hydroshock}. Its properties are 

\be 
   r_{\rm max} \simeq \textcolor{black}{0.82496}\, r_0\ \text{and}\ z_{\rm max} \simeq \textcolor{black}{0.232462}\, r_0 \, .
\label{rmax}
\ee 
Surely, the approximation $p=$const is not satisfied in reality and 
this is why the actual termination shock is actually a complex of 
shocks. However, we expect the basic shock dimensions to follow this estimate which is used for the analytical reference model described in section \ref{sec:analyt-refer-model}.    

As an aside, if we assume an isotropic spherical wind, we end up 
with the shock radius 
\be 
    r_{\rm s} = \textcolor{black}{\left(\frac{2}{3}\right)^{1/2}} r_0 \simeq \textcolor{black}{0.816497} r_0, 
\ee 
which is very close to the numerical solution for $r_{max}$. 
\textcolor{black}{Neglecting the downstream momentum we would arrive at a smaller value 
\be
	r_{\rm s} = \frac{1}{\sqrt{3}} r_{0} \simeq 0.58 r_{0}.  
\ee
}

\section{Notes on convergence}\label{sec:convergence}

\begin{figure*}
\begin{center}
\includegraphics[width=80mm]{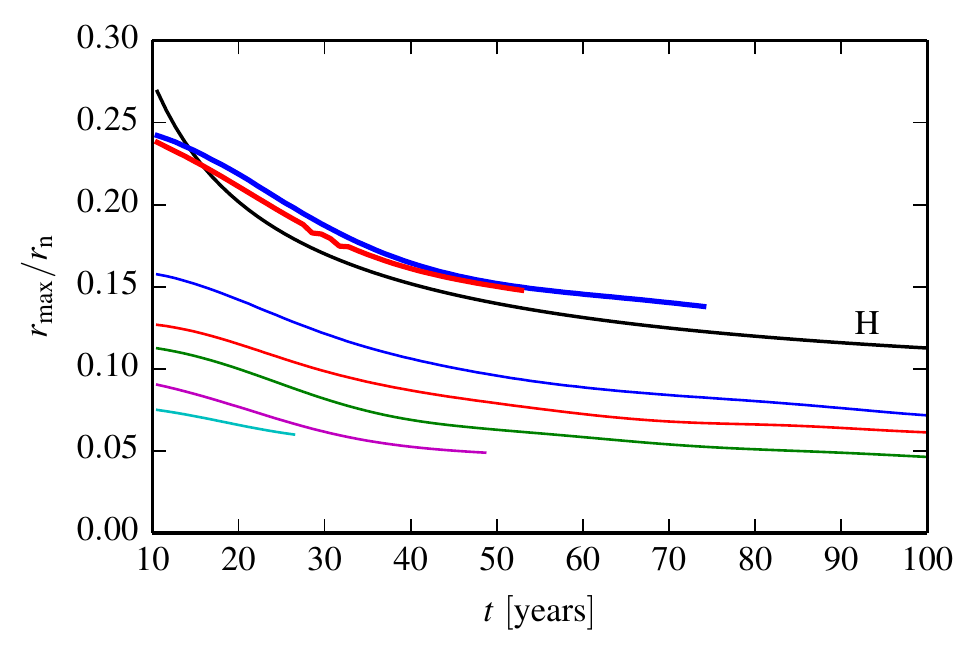}
\includegraphics[width=80mm]{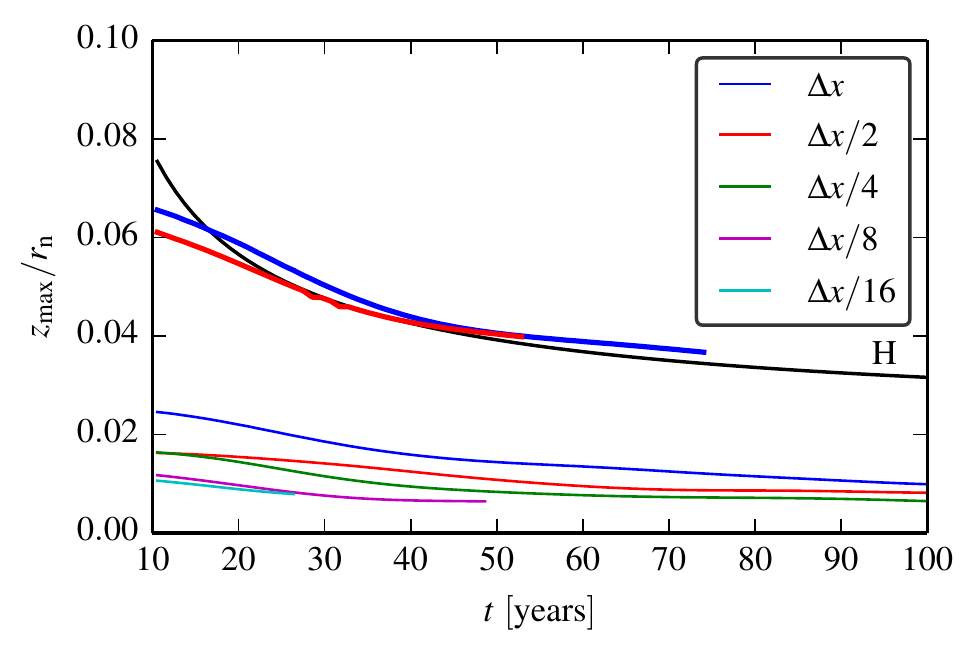}
\caption{Maximum horizontal and vertical sizes of the TS 
as functions of numerical resolution.  The thick and thin smooth lines 
show the low-pass filtered data for the 3D and 2D runs respectively.     
The 3D runs seem to display well converged shock size, whereas in 2D runs 
the convergence is not evident even at the resolution $\Delta x/16$.}
\label{fig:rmaxResolution}
\end{center}
\end{figure*}

Let us now consider how the ``observables'', $r_{\rm
  max}$ and $E_{\rm m}/E_{\rm t}$ depend on the resolution. 
As indicated by figure \ref{fig:rmaxResolution}, the 3D cases
show almost identical equatorial shock sizes $r_{\rm max}/r_{\rm n}$ when the resolution is doubled.  We note that the 2D runs are not as well converged.  Upon increasing the spatial
resolution, the shock size tends to decrease as a general trend.
Hence an increase of the resolution by a factor of 16 leads to a
reduced shock radius and height by roughly a factor of two in the axisymmetric
simulations.  As the nebula medium is dominated by discontinuous
shocks, the convergence rate is expected to be first order at best and
it is not surprising that convergence of a local quantity such as the
size of the termination shock is not fully reached.  
A likely explanation to the difference in convergence behavior for the 2D and 3D cases
can also be found in the presence of an artificial polar jet in axisymmetry that lacks
almost entirely in 3D.  
The emerging strong radial gradients in azimuthal magnetic field and thermal pressure become increasingly resolved, leading to higher compression of jet and termination shock.  
Regardless of the lack of local convergence, our high resolution simulations corroborate
that the shock sizes are systematically under-estimated in 2D, a trend
that even strengthens as the resolution is increased.  We stress once more that the shock radii in the 3D simulations on the other hand appear well converged.

\begin{figure}
\begin{center}
\includegraphics[width=80mm]{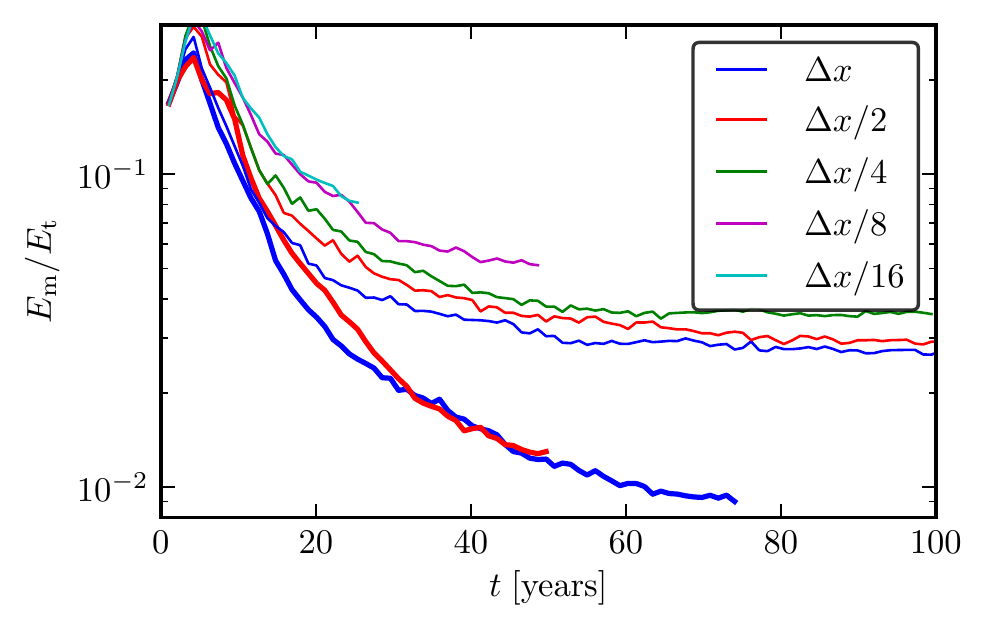}
\caption{Magnetic dissipation as a function of numerical resolution. The plot 
shows $E_{\rm m}/E_{\rm t}$ against the time. 
Like in figure~\ref{fig:rmaxResolution}, the thick lines show the 3D data 
and the thin lines the 2D data. The color-scheme of lines is also the same 
as in figure~\ref{fig:rmaxResolution}. In 2D, a convergence is noted only at 
the highest resolutions with $\Delta x /16$. In contrast, and 
in accord with our findings for the termination shock size,  the 3D 
simulations feature early convergence again. 
}
\label{fig:energeticsResolution}
\end{center}
\end{figure}

Turning to global (integral) quantities, we show the evolution of the
nebula energetics in figure \ref{fig:energeticsResolution}.  Only the
highest 2D resolutions indicate convergence of $E_{\rm m}/E_{\rm t}$
while the 3D realizations again show near identical evolution for $\Delta x$
and $\Delta x/2$.  

\section{Turbulent spectra}\label{sec:turbulentSpectra}

Periodic boundary conditions and  energy injection via ``forcing'' 
are two elements of current direct numerical simulations of 
magnetohydrodynamic turbulence
(e.g. \cite{kritsuk2009, beresnyak2009, lemaster2009} or
\cite{ZrakeMacFadyen2012} for the case of RMHD turbulence).  
In contrast, in the global simulations of PWN we do not use the periodic 
boundary conditions and do not introduce artificial force terms.  
The turbulence is produced via instability of the equatorial flow and
and polar jet as well as the unsteady termination shock, all taken into 
account in a self-consistent manner.  The pulsar wind injects strong
(azimuthal) guide field, a particularly interesting case of MHD
turbulence.  The turbulent flow is naturally confined by the expanding
supernova shell which acts like a reflecting wall for the waves generated 
inside the nebula. 
All these render the PWN an ideal host to study relativistic
magnetohydrodynamic turbulence\footnote{In addition, the MHD turbulence 
can be responsible, at least in part, for the particle
acceleration in the nebula proper via  wave-particle interactions or
turbulent shear acceleration \citep[e.g.][]{lazarian2012, ohira2013} }.

If a turbulent cascade develops in the nebula, the energy is transported
from the driving scale over the self-similar inertial range to the
dissipation scale, which in our case is determined by the numerical 
resolution. In order to properly account for the dissipation, the 
inertial range has to be resolved in the simulations.   
Since the dissipation of the magnetic energy
injected by the wind plays a key role in the dynamics of the PWN, 
this issue gains particular importance.  

To obtain magnetic- and velocity-power spectra, we select a sub-domain
containing the entire nebula with $x,y,z \in (-1.5\times
10^{18},1.5\times10^{18})$. Inside this sub-domain, the numerical 
solution is mapped onto a uniform grid with the same resolution as 
that used for the integration inside the nebula and away from the termination 
shock, where the resolution is higher. This corresponds to $150^3$ grid
cells in the fiducial case and $300^3$ cells for the high resolution
case.  Then we perform the 3D discrete Fourier transform

\begin{align}
\hat{X}(m,n,l) = \frac{L_x}{N_x} \frac{L_y}{N_y} \frac{L_z}{N_z}
\sum_{i=0}^{N_x-1} \sum_{j=0}^{N_x-1} \sum_{k=0}^{N_z-1}
X(i,j,k)\\ \exp{\left[-2\pi i \left( \frac{m i}{N_x} + \frac{n j}{N_y}
    + \frac{l k}{N_z}\right) \right]}
\end{align}
where $m\in [-N_x/2,N_x/2]$, $n\in [-N_y/2,N_y/2]$ and $l\in
[-N_z/2,N_z/2]$ are integers that span from the negative to the
positive Nyquist frequency.  The magnetic power in the $(m,n,l)$
spatial frequency domain is defined as

\begin{equation}
P_{\rm m} (m,n,l) = \frac{1}{8\pi} \left(\mathbf{\hat{B}}(m,n,l)\cdot
\mathbf{\hat{B}}^*(m,n,l)\right) \, ,
\end{equation}
where the superscript $^*$ indicates the complex conjugate. Next,
assuming isotropy in $k$-space, we average over spherical shells:

\begin{equation}
P_{\rm m} (k) = 4 \pi k^2 \langle P_{\rm m} (m,n,l)\rangle_{\rm S}(k)
= 4 \pi k^2 \frac{\sum_{k-\Delta k}^{k+\Delta k}P_{\rm m} (m,n,l)
}{\sum_{k-\Delta k}^{k+\Delta k} 1}
\end{equation}
whereby the sum is taken over all elements within the shell, $k(m,n,l)
= \sqrt{(m/L_x)^2 + (n/L_y)^2 + (l/L_z)^2} \in [k-\Delta k, k+\Delta
  k)$.  This yields the spectral energy density normalized such that
  the continuous limit results in

\begin{equation}
\frac{1}{8\pi} \iiint B^2(x,y,z) dx^3 = \int P_{\rm m}(k)
dk \label{eq:intb2}
\end{equation}
by Parseval's theorem.

To reduce the noise in the spectra, the temporal average
\begin{equation}
\bar{P}_{\rm m}(k) = \langle P_{\rm m}(k)\rangle_t.
\end{equation}
is performed over 11 consecutive snapshots starting at t=20 years with
$\Delta t = 1~\rm yr$.  

\begin{figure}
\begin{center}
\includegraphics[width=80mm]{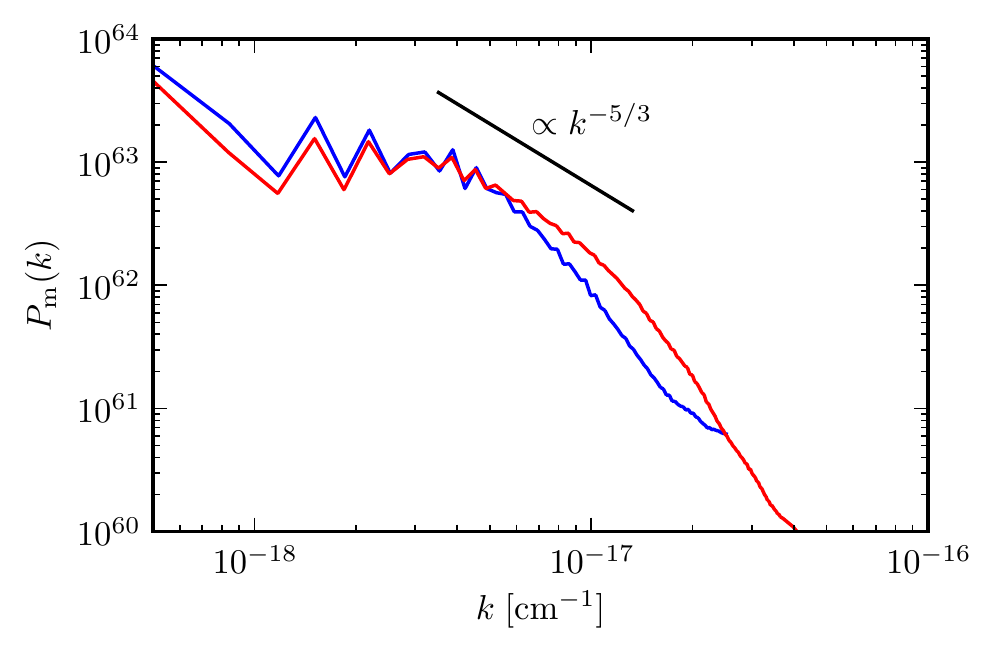}
\caption{Shell averaged power spectra of the nebula magnetic field in the 3D runs. The blue lines show the data for the run 
  B3D and the red ones for the run B3Dhr which has doubled resolution.    
  The magnetic energy spectra indicate emergence of   
  the inertial range in the higher resolution run. In this range, the magnetic 
  energy varies approximately $\propto k^{-5/3}$.  
}
\label{fig:turbulentSpectra3D}
\end{center}
\end{figure}

Figure \ref{fig:turbulentSpectra3D} shows the shell-averaged power spectra 
for the 3D runs  B3D and B3Dhr which differ only by their resolution 
( see table~\ref{tab:simulations} ). A quick inspection of the data shows 
that the large scale structure is well reproduced in the simulations. 
The magnetic energy spectra indicate the emergence, 
in the higher resolution run, of an  
inertial range with the spectral index $-5/3$. 
The corresponding  power spectra for our 2D runs are described 
in appendix \ref{sec:2d-power-spectra}. 

\section{2D power spectra}\label{sec:2d-power-spectra}

We consider a rectangular test volume in the shocked region of the turbulent nebula bubble with $r\in (3\times10^{17},9\times 10^{17})~\rm cm$ and  $z\in (-6\times10^{17},6\times10^{17})~\rm cm$ to exclude the wind zone and the polar jet.  

\begin{figure}
\begin{center}
\includegraphics[width=80mm]{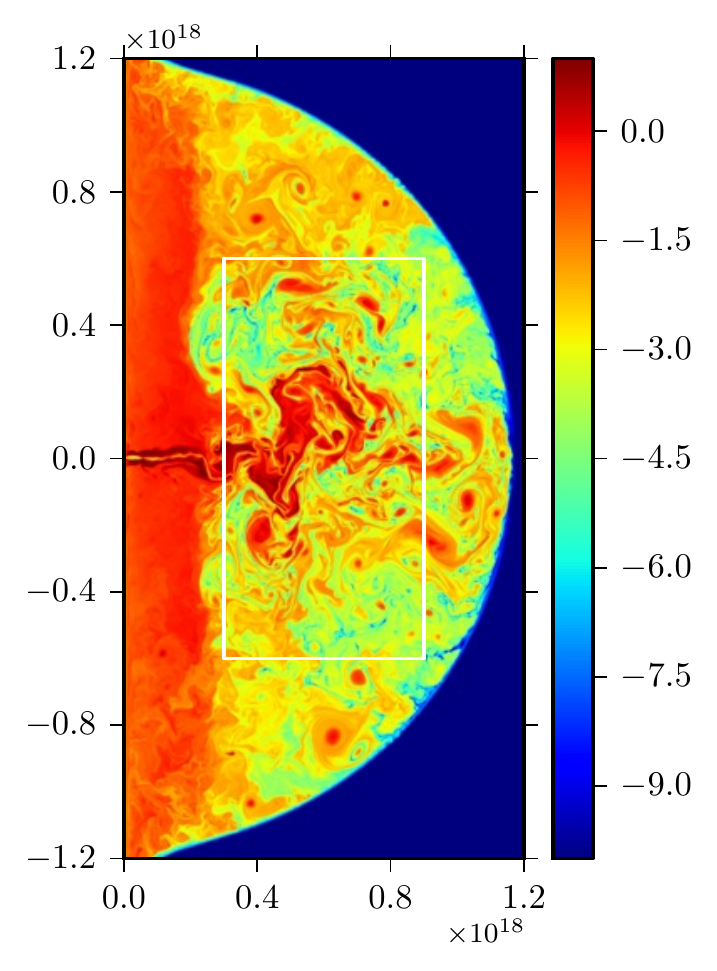}
\caption{Box of the turbulent region at t=28 years, shown is the magnetization $\log_{10}\sigma_{\rm s}$ in the highest resolution run B2Dehr.  }
\label{fig:sigmaBox}
\end{center}
\end{figure}
To obtain results independent of the interpolation
technique, the spectra are cut at the Nyquist frequency corresponding
to the original grid spacing in the nebula part of the domain.

Power spectra in 2D $(m,n)$ space are obtained analogue to the 3D case.  In 2D, the shell average becomes:
\begin{equation}
P_{\rm m} (k) = 2 \pi k \langle P_{\rm m} (m,n)\rangle_s(k) = 2 \pi k \frac{\sum_{k-\Delta k}^{k+\Delta k}P_{\rm m} (m,n) }{\sum_{k-\Delta k}^{k+\Delta k} 1}
\end{equation}
whereby the sum is taken over all elements within the shell, $k(m,n) =
\sqrt{(m/L_r)^2 + (n/L_z)^2} \in [k-\Delta k, k+\Delta k)$.

\begin{figure}
\begin{center}
\includegraphics[width=80mm]{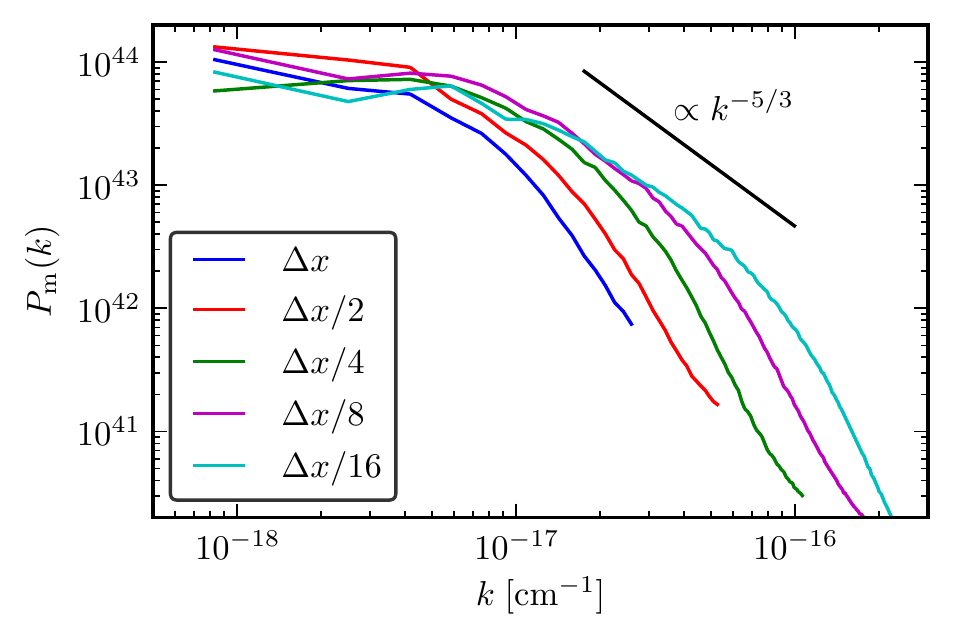}
\caption{Shell averaged power spectra of the magnetic field in the test-volume.  The magnetic power
  spectrum shows an approximate inertial range $\propto k^{-5/3}$.  
  In agreement with results shown in the previous
  section, no convergence is found for the 2D case leading to an
  increase of the area under the curve with increasing resolution}
\label{fig:turbulentSpectra}
\end{center}
\end{figure}

Figure \ref{fig:turbulentSpectra} shows the averaged power spectra of
magnetic field and four-velocity for the 2D simulations.  

Even with the highest resolution, the inertial range is barely resolved and indicates a scaling according to $k^{-5/3}$.  
This is consistent with the lack of convergence of the observables noted in
the previous section.

\end{document}